\documentclass[aps,prl,reprint,twocolumn,showpacs,amsmath,amssymb, amsfonts,10pt]{revtex4-2}

\usepackage[T1]{fontenc}
\usepackage{graphicx}
\usepackage{enumitem}
\usepackage{xcolor}
\usepackage{amsmath}

\usepackage{amsfonts}

\usepackage{amsmath}
\usepackage{amsfonts}
\usepackage{amssymb}
\usepackage{mwe}
\usepackage{graphicx}
\usepackage{color}
\usepackage{bm}
\usepackage[normalem]{ulem}
\usepackage{xcolor}

\usepackage[colorlinks,bookmarks=false,urlcolor=blue, citecolor=blue, linkcolor = blue]{hyperref}
\usepackage[capitalise]{cleveref}

\DeclareUnicodeCharacter{2212}{\textendash}
\begin{document}
\title{Emergent Fermi polarons in Dirac materials}

\author{Xin Chen$^{1}$, Eugen Dizer$^{1}$, Rafa\l{} O\l{}dziejewski$^{2}$, \\Kostya S. Novoselov$^{3,4}$, Marton Kan\'asz-Nagy$^{5}$, and Richard Schmidt$^{1}$}
\email{richard.schmidt@thphys.uni-heidelberg.de}

\affiliation{%
    $^1$Institut f\"ur Theoretische Physik, Universit\"at Heidelberg, D-69120 Heidelberg, Germany\\%
    $^2$Centre for Quantum Optical Technologies, Centre of New Technologies, University of Warsaw, Banacha 2c, 02-097 Warsaw, Poland\\
    $^3$Department of Materials Science and Engineering, National University of Singapore, Singapore\\
    $^4$Institute for Functional Intelligent Materials, National University of Singapore, Singapore\\
    $^5$Max-Planck-Institut f\"ur Quantenoptik, Hans-Kopfermann-Str. 1, D-85748 Garching, Germany
}

\begin{abstract}

We investigate band-structure effects on the absorption spectra of quantum impurities in Dirac materials. We uncover the formation of novel quasiparticles --- Dirac–Fermi polarons --- emerging from the dressing of impurities by excitations near the Dirac point. These quasiparticles are remarkably robust, persisting for both attractive and repulsive interactions, and across the full range of electron and hole doping. We show that their spectroscopic signature is a generic feature of Dirac materials, accessible with established techniques in both solid-state and ultracold atomic platforms. Our results establish polaron spectroscopy as a powerful probe of Dirac points at energies far from the Fermi surface, providing direct access to band-structure effects beyond conventional approaches.

\end{abstract}

\maketitle

The Fermi polaron is a paradigmatic phenomenon in quantum many-body physics~\cite{massignan:2025}. Despite its conceptual simplicity, it plays a central role in characterizing elementary excitations and probing quantum phases across a wide range of condensed matter systems~\cite{sidler2017fermi,Grusdt2018,Koepsell:2020qlf,smolenski2021,Colussi2023,Huang2023,Prichard2025}. Its physics is also essential for interpreting spectroscopic probes, ranging from X-ray absorption in solids~\cite{Mahan1967,Nozieres:1969,Benjamin2014} to radio-frequency spectroscopy in ultracold atomic gases~\cite{Gupta2003,zwierlein2009,kohstall2012metastability,koschorreck2012attractive}.
In strongly interacting two-dimensional semiconductors and Bose-Fermi mixtures~\cite{Cotle2016,vonMilczewski2022,Duda2023}, the formation of Fermi polarons serves as a powerful diagnostic tool for correlated electronic matter~\cite{xu2020correlated,Gao2025,Wang:2025,Zhang:2025,Liu:2026}. The study of Fermi polarons thus provides a unique window into quantum many-body physics, bridging fundamental questions and potential applications in quantum technologies.

Rapid advances in two-dimensional van der Waals materials~\cite{novoselov2004electric,Novoselov:2005kj,Bistritzer:2011rxr,mccann2013bilayer,cao2018unconventional,Xu2021,Dindorkar2023}, alongside the development of tunable optical lattices~\cite{Lee:2009hkd,Jo2012,Gross:2017ehn,Hofstetter:2018rwi,Chalopin:2024}, have  opened new avenues for experimentally studying Fermi polarons in non-trivial band structures.
Yet to date, most theoretical and experimental investigations have focused on regimes well-described by the effective mass model, where the bath exhibits a simple quadratic dispersion~\cite{Lobo2006,Fey:2020,Tan2020PRX}, and only recently have band-structure effects become a subject of theoretical interest~\cite{Hu2024PRA,Pimenov:2024,Amelio:2024ED,Amelio:2024Mott,Vashisht:2024}. Consequently, the interplay between non-trivial band structures and polaron formation, as well as the potential to utilize these quasiparticles as spectral probes has emerged as an active area of research.

\begin{figure*}[t]
    \centering
    \includegraphics[width=1.\linewidth]{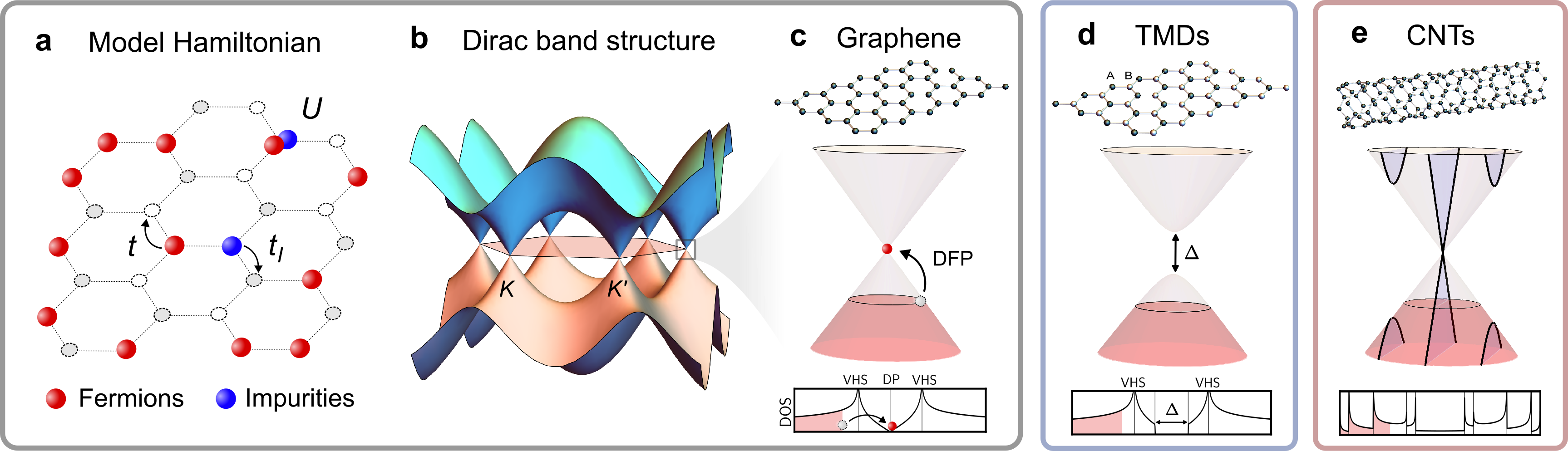}
    \caption{Setup and physical regimes. Massless Dirac regime (gray box): \textbf{a} Schematic realization, where majority fermions (red) and impurities (blue) populate a honeycomb lattice with hopping amplitudes $t$ and $t_I$, respectively, and interact via an on-site potential $U$. \textbf{b} Three-dimensional honeycomb band structure featuring two inequivalent Dirac points $K$ and $K'$ at the corners of the Brillouin zone. \textbf{c} Zoom into the linear Dirac cone with the corresponding gapless DOS (bottom), vanishing at the charge-neutrality point. The arrow illustrates the particle-hole excitation responsible for the emergence of the Dirac-Fermi polaron (DFP). Massive Dirac regime (blue box): \textbf{d} Massive Dirac cone with gap $\Delta$ opened by a sublattice energy detuning between A (black) and B (white) sites, as realized in transition metal dichalcogenides (TMDs), with the corresponding gapped DOS below. Metallic regime (red box): \textbf{e} Non-vanishing DOS, as realized in metallic carbon nanotubes (CNTs), with the corresponding band structure and finite DOS at the charge-neutrality point.}
    \label{fig:setup}
\end{figure*}

In this work, we investigate polaron formation in fermionic systems with effective honeycomb lattice structures~\cite{tarruell2012creating,Sorout_2020}, as relevant for graphene-like materials~\cite{Fan2023}, see~\cref{fig:setup}a-c. Analyzing the spectral response across the full bandwidth, we examine the contrasting roles of van Hove singularities (VHS) and Dirac points (DP), where the density of states (DOS) diverges and vanishes, respectively.

Our results reveal that a vanishing DOS, as realized by Dirac points~\cite{Wehling2009,Wehling2010,Ni2010,wehling2014dirac}, drives the emergence of a novel quasiparticle resonance that we term the Dirac–Fermi polaron (DFP). This branch appears as a distinct feature in the absorption spectrum and is intrinsically tied to the nature of the band structure. Counterintuitively, despite featuring a divergent density of states, VHS do not give rise to qualitatively new spectral resonances. 

The DFP resonance is strikingly robust, persisting from the local to the mobile impurity limit, in both massless and massive Dirac regimes. As a consequence, we predict that Dirac materials generically host three distinct Fermi polaron branches --- attractive, repulsive, and Dirac–Fermi polarons --- indicating that standard effective field theories developed for quadratic-band systems are generally insufficient for such systems.

We further extend our analysis to the case of gapped honeycomb lattices with a massive Dirac band structure, as well as highly asymmetric geometries such as relevant for metallic carbon nanotubes [see~\cref{fig:setup}d-e]. This reveals that the DFP is in fact part of a much broader class of polaron states that require strict zeros in the bath DOS.
We propose experimentally viable detection schemes for the DFP using absorption spectroscopy in transition-metal-dichalcogenide (TMD)–graphene heterostructures and radio-frequency spectroscopy with ultracold atoms in optical lattices.


\textit{Model.---} We consider a system defined on a honeycomb lattice~\cite{tarruell2012creating,Sorout_2020}, characterized by nearest-neighbor hopping amplitudes $t$ and $t_I$ for the bath fermions and the mobile impurity, respectively [see~\cref{fig:setup}a]. The Hamiltonian is given by
\begin{align}
    \hat{H} = -t \sum_{\langle i, j \rangle} \hat{c}_{i}^\dagger \hat{c}_{j}
    -t_I \sum_{\langle i, j \rangle} \hat{d}_{i}^\dagger \hat{d}_{j}
    +U \sum_{i} \hat{n}^c_{i} \hat{n}^d_{i} \,, \label{eq:H}
\end{align}
where the operators $\hat{c}_{i}^\dagger$ ($\hat{d}_{i}^\dagger$) create a bath fermion (impurity) at site $i$, and the corresponding number operators are $\hat{n}^c_i = \hat{c}_{i}^\dagger \hat{c}_{i}$ and $\hat{n}^d_i = \hat{d}_{i}^\dagger \hat{d}_{i}$.

Beyond the solid-state platforms illustrated in Fig.~\ref{fig:setup}, ultracold atoms in optical lattices provide a high degree of control over the system parameters~\cite{tarruell2012creating}. The hopping ratio $t_I/t$ can be adjusted via lattice detuning, while the fermion chemical potential $\mu$ can be tuned via the filling, from the bottom to the top of the band. Furthermore, the on-site impurity-bath interaction strength $U$ and its sign can be modified by tuning lattice parameters or using Feshbach resonances~\cite{bloch2008many,Bloch:2012uep,Schafer:2020ccc}.

\textit{Absorption spectrum.---} The absorption spectrum characterizes the response of the system to the sudden appearance of a quantum impurity~\cite{schmidt2018universal,massignan:2025}. Experimental realizations of this process include hyperfine transitions in ultracold atoms~\cite{cetina2016ultrafast}, core-hole creation in X-ray absorption~\cite{Nozieres:1969}, and exciton formation in 2D material heterostructures~\cite{wang2018tmdspectroscopy}.

In linear response theory, the absorption spectrum, ${\cal A}(\omega) = (1/\pi)\, \text{Re}\int_0^\infty d\tau \, e^{i\omega \tau} \, S(\tau)$, is determined by the Fourier transform of the Loschmidt echo~\cite{knap2012time},
\begin{align}
S(\tau) = \langle\text{FS}|
\hat{V} e^{-i \hat{H} \tau} e^{i \hat{H}_0 \tau} \hat{V}^\dagger 
|\text{FS}\rangle \,.
\label{eq:Loschmidt}
\end{align}
Here, $\hat{H}_0$ is the non-interacting Hamiltonian, and $\hat{V}^\dagger = \hat{d}_0^\dagger$ is the perturbation operator for the creation of a local impurity at site $i=0$~\cite{note}. The Fermi sea $|\text{FS}\rangle=\prod_{\epsilon_{\bf q}<\mu} \hat{c}^{\dagger}_{\bf q} |0\rangle$ is filled with all single-particle momentum states $\bf q$ below the chemical potential $\mu$, and $\epsilon_{\bf q}$ denotes the bath dispersion relation.

\begin{figure*}[t]
    \centering
    \includegraphics[width=\linewidth]{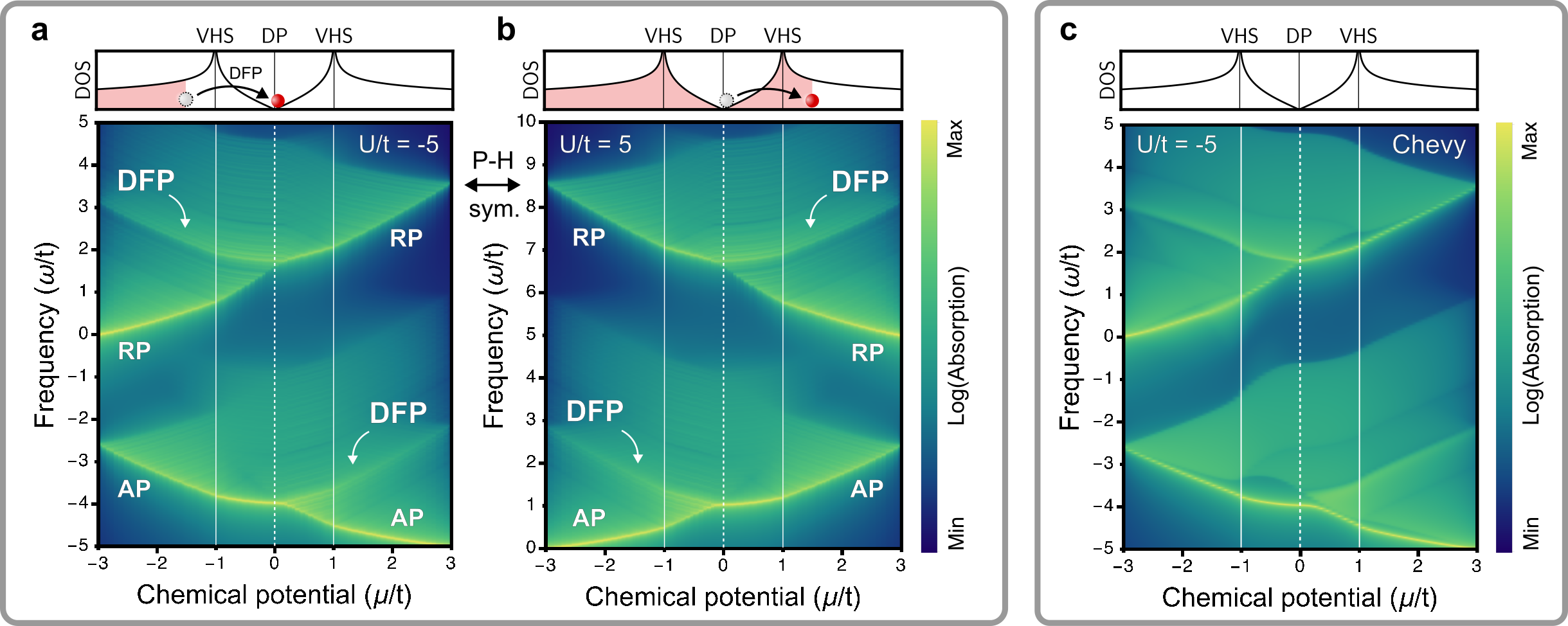}
    \caption{\textbf{a,b} Absorption spectrum of a static impurity ($t_I=0$) in the massless Dirac regime for attractive ($U/t=-5$) and repulsive ($U/t=5$) interactions, calculated via the functional determinant approach (FDA). In addition to the conventional attractive polaron (AP) and repulsive polaron (RP) branches, Dirac-Fermi polaron (DFP) resonances are visible. 
    The symmetry between panels \textbf{a} and \textbf{b} reflects the particle-hole symmetry of the honeycomb band structure. \textbf{c} The corresponding spectrum of a static impurity for attractive interactions obtained from the variational Chevy ansatz, Eq.~\eqref{eq:ChevyAnsatzWF}. Vertical dotted and solid lines mark the Dirac point (DP) and van Hove singularities (VHS), respectively, as also shown in the schematic density of states (DOS) above. Red shading illustrates occupation of states within the Fermi sea.}
    \label{fig:2DSpec}
\end{figure*}

To initiate our investigation, we consider the absorption spectrum of a static impurity ($t_I=0$) in the massless Dirac regime. In cold atoms, this limit can be realized using deep optical lattices for the impurity species, on-site optical potentials in quantum gas microscopes~\cite{Bakr2009,Koepsell2019,gross2021quantum,Sohmen2023}, or optical tweezers~\cite{Endres2016,Browaeys2020,kaufman2021quantum}. In solid-state settings, this scenario applies to excitations of defect sites~\cite{Klein:2019} or core-hole excitations~\cite{Benjamin2014,Schacherl2025}. For a static impurity, the Fermi sea dynamics is governed by a quadratic Hamiltonian. Hence it is exactly solvable, providing an ideal starting point before turning to the mobile impurity.

Strictly speaking, true quasiparticles do not form in the static limit ($t_I=0$). Anderson’s orthogonality catastrophe~\cite{anderson1967} leads to the creation of infinitely many low-energy particle–hole excitations and turns spectral peaks into power-law singularities (or `absorption wings'~\cite{Wang2023_FDA}), signaling the breakdown of quasiparticle behavior. However, these features evolve into robust quasiparticle peaks once the impurity becomes mobile. For consistency, we therefore retain the term `polaron' to describe the dominant absorption features even in the static limit.

We calculate the Loschmidt echo $S(\tau)$ [Eq.~\eqref{eq:Loschmidt}] within the functional determinant approach (FDA)~\cite{levitov,Levitov1996ElectronCS,Klich2003}, which yields
\begin{equation}
    S(\tau) = \det\left( \mathbb{I} - {\bf n}_F + {\bf n}_F \, e^{-i ({\bf h}_0 + {\bf u}) \tau} \, e^{i {\bf h}_0 \tau} \right) \,.
    \label{eq:Loschmidt_FDA}
\end{equation}
Here, ${\bf h}_0$ is the single-particle hopping matrix with elements $[{\bf h}_0]_{ij} = -t \, \delta_{\langle i, j \rangle}$, and ${\bf u}$ represents the localized impurity potential defined by $[{\bf u}]_{ij} = U \, \delta_{i 0} \delta_{j 0}$. The initial state is characterized by the Fermi occupation matrix ${\bf n}_F = \left(\mathbb{I} + e^{({\bf h}_0 - \mu \mathbb{I}) / T}\right)^{-1}$, where $\mu$ is the chemical potential and $\mathbb{I}$ is the identity matrix. The FDA reduces the exponentially large many-body Hilbert space onto $N \times N$ single-particle matrices for a system of $N$ sites. We have verified convergence of our results with respect to system size. For numerical details, see the Supplementary Material~\cite{Supplementary}.

The absorption spectra of the static impurity are shown in \cref{fig:2DSpec} for both attractive ($U/t=-5$) and repulsive ($U/t=5$) interactions. For attractive interactions ($U<0$) and low doping, i.~e. $\mu/t\gtrsim -3$ near the valence band minimum, the spectrum is dominated by the conventional attractive (AP) and repulsive (RP) polaron branches [see \cref{fig:2DSpec}a]. The high-energy tail of the AP is separated from the onset of the RP by the bound state energy $E_B/t=2.6$. This spectral gap arises because the upper edge of the AP corresponds to an attractive polaron accompanied by a particle-hole excitation from the band bottom to the Fermi level~\cite{schmidt2018universal}.

With increasing doping, a third absorption branch emerges, which we identify as the Dirac-Fermi polaron (DFP). Remarkably, the DFP persists down to very low filling ($\mu<0$), where it arises due to dressing of the repulsive polaron with a fermion excited from the Fermi surface to the Dirac point [see \cref{fig:setup}c]. All three spectral features remain discernible across all chemical potentials, except precisely at half-filling ($\mu=0$), where the Dirac point coincides with the Fermi level and the DFP merges with the main polaron branches [see \cref{fig:2DSpec}a,b]. As the chemical potential exceeds the Dirac point ($\mu>0$), a new DFP branch emerges from the attractive polaron, now originating from dressing the attractive polaron with a hole excitation from the Fermi level to the Dirac point.

For repulsive interactions ($U>0$), analogous spectral features appear. To clarify their physical nature, we relate them to the attractive case via a particle-hole mapping.
A comparison of \cref{fig:2DSpec}a,b makes evident that the spectra for attractive ($U/t=-5$) and repulsive ($U/t=5$) interactions are mirror images of each other under the transformation $\mu \leftrightarrow -\mu$, up to an overall energy shift. Indeed, due to the particle-hole symmetry of the graphene band structure, the Loschmidt echo $S_{U, \mu}(\tau)$ at interaction $U$ and chemical potential $\mu$ satisfies the symmetry relation $S_{-U, -\mu}(\tau) = S_{U, \infty}(\tau) \, S_{U, \mu}(\tau)$. Here, the prefactor $S_{U, \infty}(\tau) = \det( e^{-i ({\bf h}_0 + {\bf u})\tau} \, e^{i {\bf h}_0 \tau} )$ represents the Loschmidt echo of the completely filled band, whose Fourier transform produces the energy shift that places the repulsive absorption edge at zero energy relative to the band bottom. Consequently, this transformation shows that the repulsive peaks correspond to repulsively bound states~\cite{Winkler2006}, which can be viewed as the direct particle-hole analogues of the attractive bound states.

We have verified that the DFP is robust with respect to finite temperature, which can be readily incorporated into the FDA framework. We find that increasing temperature primarily broadens the spectral features, while the overall structure remains largely unaffected~\cite{Supplementary}.

\textit{Variational ansatz.---}
We now explore the case of a mobile impurity and study the effect of quantum coherent motion on the DFP excitation. To this end, we employ a truncated basis variational approach that describes the dressing of the impurity by particle-hole excitations of the Fermi sea~\cite{chevy2006universal,Parish2013}. For clarity, we suppress sublattice indices here; the full derivation including the complete two-band structure is provided in Ref.~\cite{Supplementary}. For a mobile polaron with total momentum $\bf Q$, the so-called `Chevy ansatz' takes the form
\begin{align}
    |\psi_{\bf Q}\rangle = \big(\phi_{0} \hat{d}_{\bf Q}^\dagger +\sum_{{\bf k}>,{\bf q}<} \phi_{{\bf k}{\bf q}} \hat{d}_{{\bf Q}-{\bf k}+{\bf q}}^\dagger \hat{c}_{\bf k}^\dagger \hat{c}_{\bf q}\big)|\text{FS}\rangle \,. \label{eq:ChevyAnsatzWF}
\end{align}
Here, the momentum sums are restricted to the occupied (empty) fermionic states $\mathbf{q}$ ($\mathbf{k}$), and the non-interacting Fermi sea is  given by $|\text{FS}\rangle=\prod_{\epsilon_{\bf q}<\mu} \hat{c}^{\dagger}_{\bf q} |0\rangle$. 
Projection of Eq.~\eqref{eq:ChevyAnsatzWF} onto the Schrödinger equation yields a set of coupled equations, which are solved iteratively to determine the mobile-polaron properties. In contrast to continuum polaron models~\cite{chevy2006universal}, hole scattering cannot be neglected in the lattice setting, where the relevant momenta are set by the Brillouin-zone scale and $k_F$ is not parametrically small compared to the ultraviolet cutoff~\cite{Amelio:2024Mott}.

As a benchmark of the variational approach, we first compute the spectra in the static limit ($t_I=0$) and compare to the exact functional determinant approach. This comparison is shown in~\cref{fig:2DSpec}a,c and demonstrates striking agreement between the variational calculation and the exact FDA results. This confirms that the variational framework ---but only when incorporating hole scattering--- accurately predicts both the positions and widths of the spectral peaks, including the DFP. The only notable difference arises in the spectral tails, where the truncated ansatz fails to reproduce the power-law decay characteristic of Anderson's orthogonality catastrophe~\cite{anderson1967}. We note that this limitation can be overcome within the mass-gap description of Fermi polarons~\cite{Chen:2025gdx}.

In the mobile-impurity case ($t_I\neq 0$), we find that the qualitative spectral structure established above is fully preserved (for a detailed discussion and numerical results, see Ref.~\cite{Supplementary}): the DFP remains clearly discernible across the full doping range, while the AP, RP and DFP energies are modified with increasing $t_I$. Importantly, the mobile impurity acquires a finite quasiparticle residue and lifetime, allowing the DFP to emerge as a genuine quasiparticle in the absorption spectrum. This confirms that the DFP is not an artifact of the static-impurity limit.

\textit{Massive Dirac fermions.---}
Having established the DFP as a robust spectral feature of the massless Dirac system, we now show that its emergence reflects a far more general principle. Indeed, we find that the physical mechanism underlying the DFP is not specific to the linear Dirac dispersion, but rather tied to the vanishing of the DOS at specific energies. This distinguishes it also fundamentally from the recently observed Wigner polarons~\cite{Wang:2025,Zhang:2025,Liu:2026}, which originate from vibrational excitations of the electron lattice~\cite{Adlong:2025}.

To analyze this mechanism, we first consider a massive Dirac band structure, realized by introducing a staggered on-site potential, $\Delta/2\sum_i (\hat{n}^A_i-\hat{n}^B_i)$. As a consequence,  the sublattice inversion symmetry is broken, opening a gap of size $\Delta$ [see~\cref{fig:setup}d].
In ultracold atom experiments with optical lattices, this can be achieved by detuning the A and B sublattice sites with an external potential, while in solid-state systems this scenario is naturally realized in TMDs or twisted bilayer graphene.

\begin{figure*}[t]
    \centering
    \includegraphics[width=1.\linewidth]{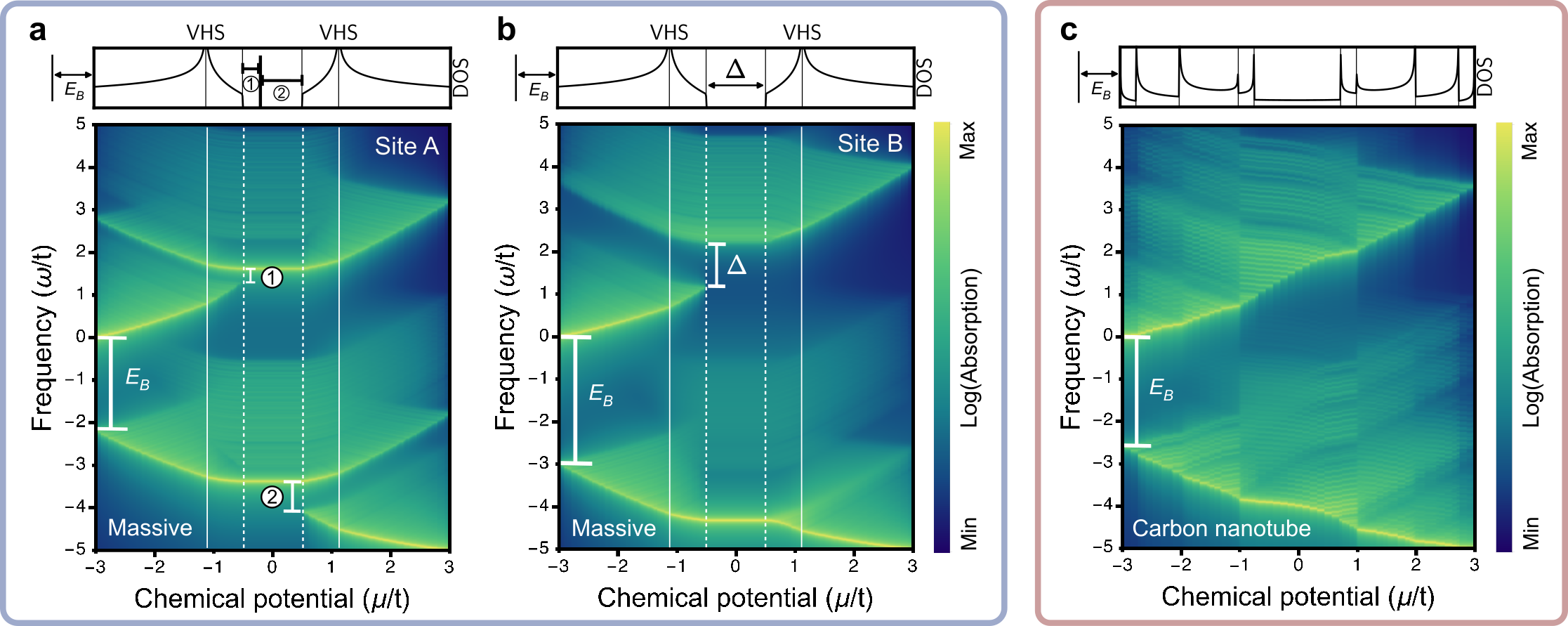}
    \caption{Absorption spectrum of a static impurity ($t_I = 0$) for attractive interactions ($U/t = -5$). \textbf{a,b} Massive Dirac regime (blue box, $\Delta = t$): FDA spectra for the impurity located at sublattice site A and B, with the schematic interacting DOS shown above each panel indicating bound state energies and band edges. The separation between the attractive and repulsive polaron branches is set by the bound state energy $E_B$. The distances from the in-gap bound state to the lower and upper band edges are marked by 1 and 2, respectively. For site A, the in-gap state resides near the centre of the gap, giving rise to two distinct spectral gaps. In contrast, for an impurity at site B, no in-gap state is present within the gap, resulting in a single gap $\Delta$ that separates the DFP branch from the RP. \textbf{c} Metallic carbon nanotube (red box): FDA spectrum showing the modified spectral structure arising from the non-vanishing DOS of this geometry. The calculation has been performed for a metallic nanotube ($6\times 30$) with open boundary conditions along the tube axis and periodic boundary conditions along the circumference.}
    \label{fig:gapped_vs_nogap}
\end{figure*}

The absorption spectra for a static impurity located at sublattice site A and B are shown in \cref{fig:gapped_vs_nogap}a,b. Despite the opening of the gap, the DFP resonance persists, confirming that a vanishing DOS ---rather than the linear dispersion--- is the key ingredient driving its emergence.
Opening the gap does, however, lead to additional features of the system.

For instance, the DFP is no longer smoothly connected to the AP and RP branches at half-filling. Instead, a gap opens between the branches, whose magnitude is set by the energy separation between the upper and lower band edges.
Furthermore, when the chemical potential lies within the gap, the absence of low-energy particle-hole excitations suppresses the orthogonality catastrophe: the polaron acquires a finite quasiparticle residue $Z$, even at $t_I = 0$, and the lineshape becomes truly Lorentzian, signalling a well-defined quasiparticle [see~\cref{fig:gapped_vs_nogap}a]. A similar mechanism is found for static impurities in BCS superfluids~\cite{Wang:2022BCSPRA,Wang:2022BCSPRL,Rodriguez:2025} and for mobile impurities in the mass-gap description~\cite{Chen:2025gdx}.

Additionally, the massive case can host an impurity bound state within the gap, whose appearance depends on both the sublattice position of the impurity and the interaction strength. For the interaction considered here ($U/t=-5$), such an in-gap state emerges when the impurity occupies site A, but is absent for site B [\cref{fig:gapped_vs_nogap}a,b]: for site A it resides near the centre of the gap, producing two distinct spectral gaps, whereas for site B a single gap separates the DFP branch from the RP. This sublattice sensitivity provides an additional handle for characterizing the underlying band structure experimentally.

The emergence of the DFP across these systems admits a unified interpretation rooted in the two-body problem of an impurity interacting with a single fermion near a band crossing or gap. When a true in-gap bound state forms [\cref{fig:gapped_vs_nogap}a], the DFP emerges as a distinct quasiparticle, also known as  Hopfield exciton~\cite{Hopfield1961}. When the interaction instead supports only a virtual bound state --- a pole of the scattering $T$-matrix embedded in the continuum [\cref{fig:gapped_vs_nogap}b] --- a DFP can still form, but broadens into a resonance as the virtual state detunes from the gap. The gapless Dirac regime realizes precisely this scenario, with the virtual state energy pinned near the Dirac point. The DFP is thus analogous to the repulsive polaron, which is likewise a quasiparticle embedded in the continuum~\cite{Scazza:2022bez}. For a detailed discussion, see the Supplementary Material~\cite{Supplementary}.
 
To further verify the role of a vanishing DOS in the formation of the DFP, we have considered the case of metallic carbon nanotubes (CNTs). Despite hosting linearly dispersing bands that touch at the Dirac point, metallic CNTs feature a constant DOS due to their quasi-one-dimensional character [see~\cref{fig:setup}e]. As shown in~\cref{fig:gapped_vs_nogap}c, the DFP is indeed strongly suppressed in this case.

Taken together, these results establish the absorption spectrum as a sensitive probe of the DOS: the behavior of the DFP branch and the presence or absence of a spectral gap directly distinguish the massless from the massive Dirac regime, suggesting that polaron spectroscopy can serve as a precision tool for band structure characterization in various platforms.

\textit{Experimental implementation.---}
The phenomena investigated in this work are accessible in a wide range of solid-state and synthetic Dirac materials. In the context of ultracold atoms, static impurities can be realized using optical tweezers, single-site potentials in quantum gas microscopes, or by tuning the optical lattice depth to suppress impurity tunneling ($t_I \to 0$) while maintaining bath mobility~\cite{bloch2008many}. Beyond standard frequency-domain absorption spectroscopy, the time-resolved Loschmidt echo [Eq.~\eqref{eq:Loschmidt}] can be measured using Ramsey interferometry~\cite{schmidt2018universal}. In both cold atoms and solid-state systems, pump-probe spectroscopy may help to further elucidate the nature of the DFP and its decay~\cite{DeSantis:2025lgp,Salvador:2025bkh}.
In solid-state systems, X-ray absorption spectroscopy provides an ideal probe for immobile impurities, where the photo-created core hole exerts a strong attractive potential.

Mobile impurities can be realized in solid-state heterostructures comprised of transition metal dichalcogenides and small-angle twisted bilayer graphene. In this setup, intralayer excitons could play the role of mobile impurities~\cite{lisi2021observation, popert2021optical}. Crucially, flat Moiré bands in twisted bilayer graphene suppress the electron kinetic energy, making it comparable to the electron-exciton interaction energy. Finally, we note that recent advances enable wide tunability of the chemical potential in Dirac materials, allowing experimental access to regimes across van Hove singularities, particularly in twisted bilayer graphene~\cite{rosenzweig2020overdoping}.

{\it Conclusion.--- }
Our work establishes polaron spectroscopy as a sensitive probe of band structure effects, with unique sensitivity to the density of states across the full bandwidth. The Dirac-Fermi polaron emerges as a direct spectroscopic fingerprint of a vanishing DOS, enabling a clear distinction between qualitatively different band structures: massless versus massive Dirac cones, and linear versus quadratic band crossings. Notably, the DFP resonance is absent when the DOS remains finite at the band crossing, providing a sharp diagnostic criterion. We anticipate that resonant inelastic X-ray scattering~\cite{ament2011xrayreview} could exploit this mechanism to achieve frequency- and momentum-resolved detection of critical band features at energy scales far from the Fermi surface. Such capabilities would further facilitate the study of exotic quantum materials, including nodal superconductors~\cite{zeroEnergyStates2002} and Weyl semimetals~\cite{armitage2018weylSemimetals}, where the precise mapping of nodal lines and band crossings remains an open experimental challenge. Finally, we propose that the phenomena described here are generic to any system in which charge carriers interact with a second particle species near a band crossing with suppressed DOS, suggesting that Dirac-Fermi polarons may arise in a broad class of quantum materials and synthetic quantum systems beyond the specific realizations considered here.

\begin{acknowledgments}
{\textbf{Acknowledgements.---}} We are grateful to M. Beringuer, E. Demler, T. Hilker, M. Salmhofer, T. Shi and J. Zaumseil for fruitful discussions. We acknowledge funding  by the DFG (German Research Foundation) – Project-ID 273811115 – SFB 1225 ISOQUANT, and under Germany's Excellence Strategy EXC 2181/1 - 390900948 (the Heidelberg STRUCTURES Excellence Cluster). M. K.-N. is supported by the Max Planck Society and the Deutsche Forschungsgemeinschaft (DFG, German Research Foundation) under Germany’s Excellence Strategy– EXC-2111–390814868. R.S. was supported within the DFG scientific network `A(E)MP - Appearance of the Effective Mass in Polaron Models' (Grant No. 569490025). R.O. was supported by the ``Quantum Optical Technologies'' (FENG.02.01-IP.05-0017/23) project which is being carried out within the Measure 2.1 International Research Agendas programme of the Foundation for Polish Science co-financed by the European Union under the European Funds for Smart Economy 2021-2027 (FENG). K.~S.~N. acknowledges support by the National Research Foundation, Singapore under its AI Singapore Programme (AISG Award No: AISG3-RP-2022-028), by the Ministry of Education, Singapore under Research Centre of Excellence award to the Institute for Functional Intelligent Materials, I-FIM (project No. EDUNC-33-18-279-V12) and by the Tier 3 program (MOE-MOET32024-0001).

{\textbf{Author contributions.---}} X.C., E.D. performed all calculations with initial input from M.K.-N. The project was conceived by R.S. with input from K.S.N. The results were analyzed and interpreted by X.C., E.D. and R.S. The manuscript was written by X.C., E.D., and R.S. with initial contributions from R.O. and M.K.-N.

{\textbf{Competing interests.---}} The authors declare no competing interests.

{\textbf{Data and materials availability.---}} The data is available from the authors upon request.
\end{acknowledgments}

\bibliographystyle{apsrev4-1}

\bibliographystyle{apsrev4-1}

\begin{thebibliography}{92}%
\makeatletter
\providecommand \@ifxundefined [1]{%
 \@ifx{#1\undefined}
}%
\providecommand \@ifnum [1]{%
 \ifnum #1\expandafter \@firstoftwo
 \else \expandafter \@secondoftwo
 \fi
}%
\providecommand \@ifx [1]{%
 \ifx #1\expandafter \@firstoftwo
 \else \expandafter \@secondoftwo
 \fi
}%
\providecommand \natexlab [1]{#1}%
\providecommand \enquote  [1]{``#1''}%
\providecommand \bibnamefont  [1]{#1}%
\providecommand \bibfnamefont [1]{#1}%
\providecommand \citenamefont [1]{#1}%
\providecommand \href@noop [0]{\@secondoftwo}%
\providecommand \href [0]{\begingroup \@sanitize@url \@href}%
\providecommand \@href[1]{\@@startlink{#1}\@@href}%
\providecommand \@@href[1]{\endgroup#1\@@endlink}%
\providecommand \@sanitize@url [0]{\catcode `\\12\catcode `\$12\catcode
  `\&12\catcode `\#12\catcode `\^12\catcode `\_12\catcode `\%12\relax}%
\providecommand \@@startlink[1]{}%
\providecommand \@@endlink[0]{}%
\providecommand \url  [0]{\begingroup\@sanitize@url \@url }%
\providecommand \@url [1]{\endgroup\@href {#1}{\urlprefix }}%
\providecommand \urlprefix  [0]{URL }%
\providecommand \Eprint [0]{\href }%
\providecommand \doibase [0]{http://dx.doi.org/}%
\providecommand \selectlanguage [0]{\@gobble}%
\providecommand \bibinfo  [0]{\@secondoftwo}%
\providecommand \bibfield  [0]{\@secondoftwo}%
\providecommand \translation [1]{[#1]}%
\providecommand \BibitemOpen [0]{}%
\providecommand \bibitemStop [0]{}%
\providecommand \bibitemNoStop [0]{.\EOS\space}%
\providecommand \EOS [0]{\spacefactor3000\relax}%
\providecommand \BibitemShut  [1]{\csname bibitem#1\endcsname}%
\let\auto@bib@innerbib\@empty
\bibitem [{\citenamefont {Massignan}\ \emph {et~al.}(2025)\citenamefont
  {Massignan}, \citenamefont {Schmidt}, \citenamefont {Astrakharchik},
  \citenamefont {Imamoglu}, \citenamefont {Zwierlein}, \citenamefont {Arlt},\
  and\ \citenamefont {Bruun}}]{massignan:2025}%
  \BibitemOpen
  \bibfield  {author} {\bibinfo {author} {\bibfnamefont {P.}~\bibnamefont
  {Massignan}}, \bibinfo {author} {\bibfnamefont {R.}~\bibnamefont {Schmidt}},
  \bibinfo {author} {\bibfnamefont {G.~E.}\ \bibnamefont {Astrakharchik}},
  \bibinfo {author} {\bibfnamefont {A.}~\bibnamefont {Imamoglu}}, \bibinfo
  {author} {\bibfnamefont {M.}~\bibnamefont {Zwierlein}}, \bibinfo {author}
  {\bibfnamefont {J.~J.}\ \bibnamefont {Arlt}}, \ and\ \bibinfo {author}
  {\bibfnamefont {G.~M.}\ \bibnamefont {Bruun}},\ }\href {\doibase
  10.48550/arXiv.2501.09618} {\  (\bibinfo {year} {2025}),\
  10.48550/arXiv.2501.09618}\BibitemShut {NoStop}%
\bibitem [{\citenamefont {Sidler}\ \emph {et~al.}(2017)\citenamefont {Sidler},
  \citenamefont {Back}, \citenamefont {Cotlet}, \citenamefont {Srivastava},
  \citenamefont {Fink}, \citenamefont {Kroner}, \citenamefont {Demler},\ and\
  \citenamefont {Imamoglu}}]{sidler2017fermi}%
  \BibitemOpen
  \bibfield  {author} {\bibinfo {author} {\bibfnamefont {M.}~\bibnamefont
  {Sidler}}, \bibinfo {author} {\bibfnamefont {P.}~\bibnamefont {Back}},
  \bibinfo {author} {\bibfnamefont {O.}~\bibnamefont {Cotlet}}, \bibinfo
  {author} {\bibfnamefont {A.}~\bibnamefont {Srivastava}}, \bibinfo {author}
  {\bibfnamefont {T.}~\bibnamefont {Fink}}, \bibinfo {author} {\bibfnamefont
  {M.}~\bibnamefont {Kroner}}, \bibinfo {author} {\bibfnamefont
  {E.}~\bibnamefont {Demler}}, \ and\ \bibinfo {author} {\bibfnamefont
  {A.}~\bibnamefont {Imamoglu}},\ }\href {\doibase 10.1038/nphys3949}
  {\bibfield  {journal} {\bibinfo  {journal} {Nat. Phys.}\ }\textbf {\bibinfo
  {volume} {13}},\ \bibinfo {pages} {255} (\bibinfo {year} {2017})}\BibitemShut
  {NoStop}%
\bibitem [{\citenamefont {{Grusdt, Fabian and K{\'a}nasz-Nagy, M{\'a}rton and
  Bohrdt, Annabelle and Chiu, Christie S. and Ji, Geoffrey and Greiner, Markus
  and Greif, Daniel and Demler, Eugene}}(2018)}]{Grusdt2018}%
  \BibitemOpen
  \bibfield  {author} {\bibinfo {author} {\bibnamefont {{Grusdt, Fabian and
  K{\'a}nasz-Nagy, M{\'a}rton and Bohrdt, Annabelle and Chiu, Christie S. and
  Ji, Geoffrey and Greiner, Markus and Greif, Daniel and Demler, Eugene}}},\
  }\href {\doibase 10.1103/PhysRevX.8.011046} {\bibfield  {journal} {\bibinfo
  {journal} {Phys. Rev. X}\ }\textbf {\bibinfo {volume} {8}},\ \bibinfo {pages}
  {011046} (\bibinfo {year} {2018})}\BibitemShut {NoStop}%
\bibitem [{\citenamefont {Koepsell}\ \emph {et~al.}(2020)\citenamefont
  {Koepsell}, \citenamefont {Bourgund}, \citenamefont {Sompet}, \citenamefont
  {Hirthe}, \citenamefont {Bohrdt}, \citenamefont {Wang}, \citenamefont
  {Grusdt}, \citenamefont {Demler}, \citenamefont {Salomon}, \citenamefont
  {Gross},\ and\ \citenamefont {Bloch}}]{Koepsell:2020qlf}%
  \BibitemOpen
  \bibfield  {author} {\bibinfo {author} {\bibfnamefont {J.}~\bibnamefont
  {Koepsell}}, \bibinfo {author} {\bibfnamefont {D.}~\bibnamefont {Bourgund}},
  \bibinfo {author} {\bibfnamefont {P.}~\bibnamefont {Sompet}}, \bibinfo
  {author} {\bibfnamefont {S.}~\bibnamefont {Hirthe}}, \bibinfo {author}
  {\bibfnamefont {A.}~\bibnamefont {Bohrdt}}, \bibinfo {author} {\bibfnamefont
  {Y.}~\bibnamefont {Wang}}, \bibinfo {author} {\bibfnamefont {F.}~\bibnamefont
  {Grusdt}}, \bibinfo {author} {\bibfnamefont {E.}~\bibnamefont {Demler}},
  \bibinfo {author} {\bibfnamefont {G.}~\bibnamefont {Salomon}}, \bibinfo
  {author} {\bibfnamefont {C.}~\bibnamefont {Gross}}, \ and\ \bibinfo {author}
  {\bibfnamefont {I.}~\bibnamefont {Bloch}},\ }\href {\doibase
  10.1126/science.abe7165} {\bibfield  {journal} {\bibinfo  {journal}
  {Science}\ }\textbf {\bibinfo {volume} {374}},\ \bibinfo {pages} {82}
  (\bibinfo {year} {2020})}\BibitemShut {NoStop}%
\bibitem [{\citenamefont {Smoleński}\ \emph {et~al.}(2021)\citenamefont
  {Smoleński}, \citenamefont {Dolgirev}, \citenamefont {Kuhlenkamp},
  \citenamefont {Popert}, \citenamefont {Shimazaki}, \citenamefont {Back},
  \citenamefont {Lu}, \citenamefont {Kroner}, \citenamefont {Watanabe},
  \citenamefont {Taniguchi}, \citenamefont {Esterlis}, \citenamefont {Demler},\
  and\ \citenamefont {Imamoğlu}}]{smolenski2021}%
  \BibitemOpen
  \bibfield  {author} {\bibinfo {author} {\bibfnamefont {T.}~\bibnamefont
  {Smoleński}}, \bibinfo {author} {\bibfnamefont {P.~E.}\ \bibnamefont
  {Dolgirev}}, \bibinfo {author} {\bibfnamefont {C.}~\bibnamefont
  {Kuhlenkamp}}, \bibinfo {author} {\bibfnamefont {A.}~\bibnamefont {Popert}},
  \bibinfo {author} {\bibfnamefont {Y.}~\bibnamefont {Shimazaki}}, \bibinfo
  {author} {\bibfnamefont {P.}~\bibnamefont {Back}}, \bibinfo {author}
  {\bibfnamefont {X.}~\bibnamefont {Lu}}, \bibinfo {author} {\bibfnamefont
  {M.}~\bibnamefont {Kroner}}, \bibinfo {author} {\bibfnamefont
  {K.}~\bibnamefont {Watanabe}}, \bibinfo {author} {\bibfnamefont
  {T.}~\bibnamefont {Taniguchi}}, \bibinfo {author} {\bibfnamefont
  {I.}~\bibnamefont {Esterlis}}, \bibinfo {author} {\bibfnamefont
  {E.}~\bibnamefont {Demler}}, \ and\ \bibinfo {author} {\bibfnamefont
  {A.}~\bibnamefont {Imamoğlu}},\ }\href {\doibase 10.1038/s41586-021-03590-4}
  {\bibfield  {journal} {\bibinfo  {journal} {Nature}\ }\textbf {\bibinfo
  {volume} {595}},\ \bibinfo {pages} {53} (\bibinfo {year} {2021})}\BibitemShut
  {NoStop}%
\bibitem [{\citenamefont {Colussi}\ \emph {et~al.}(2023)\citenamefont
  {Colussi}, \citenamefont {Caleffi}, \citenamefont {Menotti},\ and\
  \citenamefont {Recati}}]{Colussi2023}%
  \BibitemOpen
  \bibfield  {author} {\bibinfo {author} {\bibfnamefont {V.~E.}\ \bibnamefont
  {Colussi}}, \bibinfo {author} {\bibfnamefont {F.}~\bibnamefont {Caleffi}},
  \bibinfo {author} {\bibfnamefont {C.}~\bibnamefont {Menotti}}, \ and\
  \bibinfo {author} {\bibfnamefont {A.}~\bibnamefont {Recati}},\ }\href
  {\doibase 10.1103/PhysRevLett.130.173002} {\bibfield  {journal} {\bibinfo
  {journal} {Phys. Rev. Lett.}\ }\textbf {\bibinfo {volume} {130}},\ \bibinfo
  {pages} {173002} (\bibinfo {year} {2023})}\BibitemShut {NoStop}%
\bibitem [{\citenamefont {Huang}\ \emph {et~al.}(2023)\citenamefont {Huang},
  \citenamefont {Sampson}, \citenamefont {Ni}, \citenamefont {Liu},
  \citenamefont {Liang}, \citenamefont {Watanabe}, \citenamefont {Taniguchi},
  \citenamefont {Li}, \citenamefont {Martin}, \citenamefont {Levinsen},
  \citenamefont {Parish}, \citenamefont {Tutuc}, \citenamefont {Efimkin},\ and\
  \citenamefont {Li}}]{Huang2023}%
  \BibitemOpen
  \bibfield  {author} {\bibinfo {author} {\bibfnamefont {D.}~\bibnamefont
  {Huang}}, \bibinfo {author} {\bibfnamefont {K.}~\bibnamefont {Sampson}},
  \bibinfo {author} {\bibfnamefont {Y.}~\bibnamefont {Ni}}, \bibinfo {author}
  {\bibfnamefont {Z.}~\bibnamefont {Liu}}, \bibinfo {author} {\bibfnamefont
  {D.}~\bibnamefont {Liang}}, \bibinfo {author} {\bibfnamefont
  {K.}~\bibnamefont {Watanabe}}, \bibinfo {author} {\bibfnamefont
  {T.}~\bibnamefont {Taniguchi}}, \bibinfo {author} {\bibfnamefont
  {H.}~\bibnamefont {Li}}, \bibinfo {author} {\bibfnamefont {E.}~\bibnamefont
  {Martin}}, \bibinfo {author} {\bibfnamefont {J.}~\bibnamefont {Levinsen}},
  \bibinfo {author} {\bibfnamefont {M.~M.}\ \bibnamefont {Parish}}, \bibinfo
  {author} {\bibfnamefont {E.}~\bibnamefont {Tutuc}}, \bibinfo {author}
  {\bibfnamefont {D.~K.}\ \bibnamefont {Efimkin}}, \ and\ \bibinfo {author}
  {\bibfnamefont {X.}~\bibnamefont {Li}},\ }\href {\doibase
  10.1103/PhysRevX.13.011029} {\bibfield  {journal} {\bibinfo  {journal} {Phys.
  Rev. X}\ }\textbf {\bibinfo {volume} {13}},\ \bibinfo {pages} {011029}
  (\bibinfo {year} {2023})}\BibitemShut {NoStop}%
\bibitem [{\citenamefont {Prichard}\ \emph {et~al.}(2025)\citenamefont
  {Prichard}, \citenamefont {Ba}, \citenamefont {Morera}, \citenamefont {Spar},
  \citenamefont {Huse}, \citenamefont {Demler},\ and\ \citenamefont
  {Bakr}}]{Prichard2025}%
  \BibitemOpen
  \bibfield  {author} {\bibinfo {author} {\bibfnamefont {M.~L.}\ \bibnamefont
  {Prichard}}, \bibinfo {author} {\bibfnamefont {Z.}~\bibnamefont {Ba}},
  \bibinfo {author} {\bibfnamefont {I.}~\bibnamefont {Morera}}, \bibinfo
  {author} {\bibfnamefont {B.~M.}\ \bibnamefont {Spar}}, \bibinfo {author}
  {\bibfnamefont {D.~A.}\ \bibnamefont {Huse}}, \bibinfo {author}
  {\bibfnamefont {E.}~\bibnamefont {Demler}}, \ and\ \bibinfo {author}
  {\bibfnamefont {W.~S.}\ \bibnamefont {Bakr}},\ }\href {\doibase
  10.1038/s41567-025-03004-6} {\bibfield  {journal} {\bibinfo  {journal} {Nat.
  Phys.}\ }\textbf {\bibinfo {volume} {21}},\ \bibinfo {pages} {1548} (\bibinfo
  {year} {2025})}\BibitemShut {NoStop}%
\bibitem [{\citenamefont {Mahan}(1967)}]{Mahan1967}%
  \BibitemOpen
  \bibfield  {author} {\bibinfo {author} {\bibfnamefont {G.~D.}\ \bibnamefont
  {Mahan}},\ }\href {\doibase 10.1103/PhysRev.163.612} {\bibfield  {journal}
  {\bibinfo  {journal} {Phys. Rev.}\ }\textbf {\bibinfo {volume} {163}},\
  \bibinfo {pages} {612} (\bibinfo {year} {1967})}\BibitemShut {NoStop}%
\bibitem [{\citenamefont {Nozi\`eres}\ and\ \citenamefont
  {De~Dominicis}(1969)}]{Nozieres:1969}%
  \BibitemOpen
  \bibfield  {author} {\bibinfo {author} {\bibfnamefont {P.}~\bibnamefont
  {Nozi\`eres}}\ and\ \bibinfo {author} {\bibfnamefont {C.~T.}\ \bibnamefont
  {De~Dominicis}},\ }\href {\doibase 10.1103/PhysRev.178.1097} {\bibfield
  {journal} {\bibinfo  {journal} {Phys. Rev.}\ }\textbf {\bibinfo {volume}
  {178}},\ \bibinfo {pages} {1097} (\bibinfo {year} {1969})}\BibitemShut
  {NoStop}%
\bibitem [{\citenamefont {Benjamin}\ \emph {et~al.}(2014)\citenamefont
  {Benjamin}, \citenamefont {Klich},\ and\ \citenamefont
  {Demler}}]{Benjamin2014}%
  \BibitemOpen
  \bibfield  {author} {\bibinfo {author} {\bibfnamefont {D.}~\bibnamefont
  {Benjamin}}, \bibinfo {author} {\bibfnamefont {I.}~\bibnamefont {Klich}}, \
  and\ \bibinfo {author} {\bibfnamefont {E.}~\bibnamefont {Demler}},\ }\href
  {\doibase 10.1103/PhysRevLett.112.247002} {\bibfield  {journal} {\bibinfo
  {journal} {Phys. Rev. Lett.}\ }\textbf {\bibinfo {volume} {112}},\ \bibinfo
  {pages} {247002} (\bibinfo {year} {2014})}\BibitemShut {NoStop}%
\bibitem [{\citenamefont {Gupta}\ \emph {et~al.}(2003)\citenamefont {Gupta},
  \citenamefont {Hadzibabic}, \citenamefont {Zwierlein}, \citenamefont {Stan},
  \citenamefont {Dieckmann}, \citenamefont {Schunck}, \citenamefont {van
  Kempen}, \citenamefont {Verhaar},\ and\ \citenamefont
  {Ketterle}}]{Gupta2003}%
  \BibitemOpen
  \bibfield  {author} {\bibinfo {author} {\bibfnamefont {S.}~\bibnamefont
  {Gupta}}, \bibinfo {author} {\bibfnamefont {Z.}~\bibnamefont {Hadzibabic}},
  \bibinfo {author} {\bibfnamefont {M.~W.}\ \bibnamefont {Zwierlein}}, \bibinfo
  {author} {\bibfnamefont {C.~A.}\ \bibnamefont {Stan}}, \bibinfo {author}
  {\bibfnamefont {K.}~\bibnamefont {Dieckmann}}, \bibinfo {author}
  {\bibfnamefont {C.~H.}\ \bibnamefont {Schunck}}, \bibinfo {author}
  {\bibfnamefont {E.~G.~M.}\ \bibnamefont {van Kempen}}, \bibinfo {author}
  {\bibfnamefont {B.~J.}\ \bibnamefont {Verhaar}}, \ and\ \bibinfo {author}
  {\bibfnamefont {W.}~\bibnamefont {Ketterle}},\ }\href {\doibase
  10.1126/science.1085335} {\bibfield  {journal} {\bibinfo  {journal}
  {Science}\ }\textbf {\bibinfo {volume} {300}},\ \bibinfo {pages} {1723}
  (\bibinfo {year} {2003})}\BibitemShut {NoStop}%
\bibitem [{\citenamefont {Schirotzek}\ \emph {et~al.}(2009)\citenamefont
  {Schirotzek}, \citenamefont {Wu}, \citenamefont {Sommer},\ and\ \citenamefont
  {Zwierlein}}]{zwierlein2009}%
  \BibitemOpen
  \bibfield  {author} {\bibinfo {author} {\bibfnamefont {A.}~\bibnamefont
  {Schirotzek}}, \bibinfo {author} {\bibfnamefont {C.-H.}\ \bibnamefont {Wu}},
  \bibinfo {author} {\bibfnamefont {A.}~\bibnamefont {Sommer}}, \ and\ \bibinfo
  {author} {\bibfnamefont {M.~W.}\ \bibnamefont {Zwierlein}},\ }\href {\doibase
  10.1103/PhysRevLett.102.230402} {\bibfield  {journal} {\bibinfo  {journal}
  {Phys. Rev. Lett.}\ }\textbf {\bibinfo {volume} {102}},\ \bibinfo {pages}
  {230402} (\bibinfo {year} {2009})}\BibitemShut {NoStop}%
\bibitem [{\citenamefont {Kohstall}\ \emph {et~al.}(2012)\citenamefont
  {Kohstall}, \citenamefont {Zaccanti}, \citenamefont {Jag}, \citenamefont
  {Trenkwalder}, \citenamefont {Massignan}, \citenamefont {Bruun},
  \citenamefont {Schreck},\ and\ \citenamefont
  {Grimm}}]{kohstall2012metastability}%
  \BibitemOpen
  \bibfield  {author} {\bibinfo {author} {\bibfnamefont {C.}~\bibnamefont
  {Kohstall}}, \bibinfo {author} {\bibfnamefont {M.}~\bibnamefont {Zaccanti}},
  \bibinfo {author} {\bibfnamefont {M.}~\bibnamefont {Jag}}, \bibinfo {author}
  {\bibfnamefont {A.}~\bibnamefont {Trenkwalder}}, \bibinfo {author}
  {\bibfnamefont {P.}~\bibnamefont {Massignan}}, \bibinfo {author}
  {\bibfnamefont {G.~M.}\ \bibnamefont {Bruun}}, \bibinfo {author}
  {\bibfnamefont {F.}~\bibnamefont {Schreck}}, \ and\ \bibinfo {author}
  {\bibfnamefont {R.}~\bibnamefont {Grimm}},\ }\href {\doibase
  10.1038/nature11065} {\bibfield  {journal} {\bibinfo  {journal} {Nature}\
  }\textbf {\bibinfo {volume} {485}},\ \bibinfo {pages} {615} (\bibinfo {year}
  {2012})}\BibitemShut {NoStop}%
\bibitem [{\citenamefont {Koschorreck}\ \emph {et~al.}(2012)\citenamefont
  {Koschorreck}, \citenamefont {Pertot}, \citenamefont {Vogt}, \citenamefont
  {Fröhlich}, \citenamefont {Feld},\ and\ \citenamefont
  {Köhl}}]{koschorreck2012attractive}%
  \BibitemOpen
  \bibfield  {author} {\bibinfo {author} {\bibfnamefont {M.}~\bibnamefont
  {Koschorreck}}, \bibinfo {author} {\bibfnamefont {D.}~\bibnamefont {Pertot}},
  \bibinfo {author} {\bibfnamefont {E.}~\bibnamefont {Vogt}}, \bibinfo {author}
  {\bibfnamefont {B.}~\bibnamefont {Fröhlich}}, \bibinfo {author}
  {\bibfnamefont {M.}~\bibnamefont {Feld}}, \ and\ \bibinfo {author}
  {\bibfnamefont {M.}~\bibnamefont {Köhl}},\ }\href {\doibase
  10.1038/nature11151} {\bibfield  {journal} {\bibinfo  {journal} {Nature}\
  }\textbf {\bibinfo {volume} {485}},\ \bibinfo {pages} {619} (\bibinfo {year}
  {2012})}\BibitemShut {NoStop}%
\bibitem [{\citenamefont {Cotle\ifmmode~\mbox{\c{t}}\else \c{t}\fi{}}\ \emph
  {et~al.}(2016)\citenamefont {Cotle\ifmmode~\mbox{\c{t}}\else \c{t}\fi{}},
  \citenamefont {Zeytino\ifmmode~\check{g}\else \v{g}\fi{}lu}, \citenamefont
  {Sigrist}, \citenamefont {Demler},\ and\ \citenamefont
  {Imamo\ifmmode~\check{g}\else \v{g}\fi{}lu}}]{Cotle2016}%
  \BibitemOpen
  \bibfield  {author} {\bibinfo {author} {\bibfnamefont {O.}~\bibnamefont
  {Cotle\ifmmode~\mbox{\c{t}}\else \c{t}\fi{}}}, \bibinfo {author}
  {\bibfnamefont {S.}~\bibnamefont {Zeytino\ifmmode~\check{g}\else
  \v{g}\fi{}lu}}, \bibinfo {author} {\bibfnamefont {M.}~\bibnamefont
  {Sigrist}}, \bibinfo {author} {\bibfnamefont {E.}~\bibnamefont {Demler}}, \
  and\ \bibinfo {author} {\bibfnamefont {A.}~\bibnamefont
  {Imamo\ifmmode~\check{g}\else \v{g}\fi{}lu}},\ }\href {\doibase
  10.1103/PhysRevB.93.054510} {\bibfield  {journal} {\bibinfo  {journal} {Phys.
  Rev. B}\ }\textbf {\bibinfo {volume} {93}},\ \bibinfo {pages} {054510}
  (\bibinfo {year} {2016})}\BibitemShut {NoStop}%
\bibitem [{\citenamefont {von Milczewski}\ \emph {et~al.}(2022)\citenamefont
  {von Milczewski}, \citenamefont {Rose},\ and\ \citenamefont
  {Schmidt}}]{vonMilczewski2022}%
  \BibitemOpen
  \bibfield  {author} {\bibinfo {author} {\bibfnamefont {J.}~\bibnamefont {von
  Milczewski}}, \bibinfo {author} {\bibfnamefont {F.}~\bibnamefont {Rose}}, \
  and\ \bibinfo {author} {\bibfnamefont {R.}~\bibnamefont {Schmidt}},\ }\href
  {\doibase 10.1103/PhysRevA.105.013317} {\bibfield  {journal} {\bibinfo
  {journal} {Phys. Rev. A}\ }\textbf {\bibinfo {volume} {105}},\ \bibinfo
  {pages} {013317} (\bibinfo {year} {2022})}\BibitemShut {NoStop}%
\bibitem [{\citenamefont {Duda}\ \emph {et~al.}(2023)\citenamefont {Duda},
  \citenamefont {Chen}, \citenamefont {Schindewolf}, \citenamefont {Bause},
  \citenamefont {von Milczewski}, \citenamefont {Schmidt}, \citenamefont
  {Bloch},\ and\ \citenamefont {Luo}}]{Duda2023}%
  \BibitemOpen
  \bibfield  {author} {\bibinfo {author} {\bibfnamefont {M.}~\bibnamefont
  {Duda}}, \bibinfo {author} {\bibfnamefont {X.-Y.}\ \bibnamefont {Chen}},
  \bibinfo {author} {\bibfnamefont {A.}~\bibnamefont {Schindewolf}}, \bibinfo
  {author} {\bibfnamefont {R.}~\bibnamefont {Bause}}, \bibinfo {author}
  {\bibfnamefont {J.}~\bibnamefont {von Milczewski}}, \bibinfo {author}
  {\bibfnamefont {R.}~\bibnamefont {Schmidt}}, \bibinfo {author} {\bibfnamefont
  {I.}~\bibnamefont {Bloch}}, \ and\ \bibinfo {author} {\bibfnamefont {X.-Y.}\
  \bibnamefont {Luo}},\ }\href {\doibase 10.1038/s41567-023-01948-1} {\bibfield
   {journal} {\bibinfo  {journal} {Nat. Phys.}\ }\textbf {\bibinfo {volume}
  {19}},\ \bibinfo {pages} {720} (\bibinfo {year} {2023})}\BibitemShut
  {NoStop}%
\bibitem [{\citenamefont {Xu}\ \emph {et~al.}(2020)\citenamefont {Xu},
  \citenamefont {Liu}, \citenamefont {Rhodes}, \citenamefont {Watanabe},
  \citenamefont {Taniguchi}, \citenamefont {Hone}, \citenamefont {Elser},
  \citenamefont {Mak},\ and\ \citenamefont {Shan}}]{xu2020correlated}%
  \BibitemOpen
  \bibfield  {author} {\bibinfo {author} {\bibfnamefont {Y.}~\bibnamefont
  {Xu}}, \bibinfo {author} {\bibfnamefont {S.}~\bibnamefont {Liu}}, \bibinfo
  {author} {\bibfnamefont {D.~A.}\ \bibnamefont {Rhodes}}, \bibinfo {author}
  {\bibfnamefont {K.}~\bibnamefont {Watanabe}}, \bibinfo {author}
  {\bibfnamefont {T.}~\bibnamefont {Taniguchi}}, \bibinfo {author}
  {\bibfnamefont {J.}~\bibnamefont {Hone}}, \bibinfo {author} {\bibfnamefont
  {V.}~\bibnamefont {Elser}}, \bibinfo {author} {\bibfnamefont {K.~F.}\
  \bibnamefont {Mak}}, \ and\ \bibinfo {author} {\bibfnamefont
  {J.}~\bibnamefont {Shan}},\ }\href {\doibase 10.1038/s41586-020-2868-6}
  {\bibfield  {journal} {\bibinfo  {journal} {Nature}\ }\textbf {\bibinfo
  {volume} {587}},\ \bibinfo {pages} {214} (\bibinfo {year}
  {2020})}\BibitemShut {NoStop}%
\bibitem [{\citenamefont {Gao}\ \emph {et~al.}(2025)\citenamefont {Gao},
  \citenamefont {Ghafariasl}, \citenamefont {Jalali~Mehrabad}, \citenamefont
  {Huang}, \citenamefont {Zhang}, \citenamefont {Session}, \citenamefont
  {Upadhyay}, \citenamefont {Ma}, \citenamefont {Alshalan}, \citenamefont
  {Su{\'a}rez~Forero}, \citenamefont {Sarkar}, \citenamefont {Park},
  \citenamefont {Jang}, \citenamefont {Watanabe}, \citenamefont {Taniguchi},
  \citenamefont {Xie}, \citenamefont {Zhou},\ and\ \citenamefont
  {Hafezi}}]{Gao2025}%
  \BibitemOpen
  \bibfield  {author} {\bibinfo {author} {\bibfnamefont {B.}~\bibnamefont
  {Gao}}, \bibinfo {author} {\bibfnamefont {M.}~\bibnamefont {Ghafariasl}},
  \bibinfo {author} {\bibfnamefont {M.}~\bibnamefont {Jalali~Mehrabad}},
  \bibinfo {author} {\bibfnamefont {T.-S.}\ \bibnamefont {Huang}}, \bibinfo
  {author} {\bibfnamefont {L.}~\bibnamefont {Zhang}}, \bibinfo {author}
  {\bibfnamefont {D.}~\bibnamefont {Session}}, \bibinfo {author} {\bibfnamefont
  {P.}~\bibnamefont {Upadhyay}}, \bibinfo {author} {\bibfnamefont
  {R.}~\bibnamefont {Ma}}, \bibinfo {author} {\bibfnamefont {G.}~\bibnamefont
  {Alshalan}}, \bibinfo {author} {\bibfnamefont {D.~G.}\ \bibnamefont
  {Su{\'a}rez~Forero}}, \bibinfo {author} {\bibfnamefont {S.}~\bibnamefont
  {Sarkar}}, \bibinfo {author} {\bibfnamefont {S.}~\bibnamefont {Park}},
  \bibinfo {author} {\bibfnamefont {H.}~\bibnamefont {Jang}}, \bibinfo {author}
  {\bibfnamefont {K.}~\bibnamefont {Watanabe}}, \bibinfo {author}
  {\bibfnamefont {T.}~\bibnamefont {Taniguchi}}, \bibinfo {author}
  {\bibfnamefont {M.}~\bibnamefont {Xie}}, \bibinfo {author} {\bibfnamefont
  {Y.}~\bibnamefont {Zhou}}, \ and\ \bibinfo {author} {\bibfnamefont
  {M.}~\bibnamefont {Hafezi}},\ }\href@noop {} {\  (\bibinfo {year} {2025})},\
  \Eprint {http://arxiv.org/abs/2504.11530} {arXiv:2504.11530
  [cond-mat.str-el]} \BibitemShut {NoStop}%
\bibitem [{\citenamefont {Wang}\ \emph {et~al.}(2025)\citenamefont {Wang},
  \citenamefont {Menzel}, \citenamefont {Pichler}, \citenamefont {Knüppel},
  \citenamefont {Watanabe}, \citenamefont {Taniguchi}, \citenamefont {Knap},\
  and\ \citenamefont {Smoleński}}]{Wang:2025}%
  \BibitemOpen
  \bibfield  {author} {\bibinfo {author} {\bibfnamefont {L.}~\bibnamefont
  {Wang}}, \bibinfo {author} {\bibfnamefont {F.}~\bibnamefont {Menzel}},
  \bibinfo {author} {\bibfnamefont {F.}~\bibnamefont {Pichler}}, \bibinfo
  {author} {\bibfnamefont {P.}~\bibnamefont {Knüppel}}, \bibinfo {author}
  {\bibfnamefont {K.}~\bibnamefont {Watanabe}}, \bibinfo {author}
  {\bibfnamefont {T.}~\bibnamefont {Taniguchi}}, \bibinfo {author}
  {\bibfnamefont {M.}~\bibnamefont {Knap}}, \ and\ \bibinfo {author}
  {\bibfnamefont {T.}~\bibnamefont {Smoleński}},\ }\href@noop {} {\  (\bibinfo
  {year} {2025})},\ \Eprint {http://arxiv.org/abs/2512.16552} {arXiv:2512.16552
  [cond-mat.mes-hall]} \BibitemShut {NoStop}%
\bibitem [{\citenamefont {Zhang}\ \emph {et~al.}(2025)\citenamefont {Zhang},
  \citenamefont {Gu}, \citenamefont {Adlong}, \citenamefont {Christianen},
  \citenamefont {Dizer}, \citenamefont {Ni}, \citenamefont {Ma}, \citenamefont
  {Park}, \citenamefont {Jang}, \citenamefont {Taniguchi}, \citenamefont
  {Watanabe}, \citenamefont {Esterlis}, \citenamefont {Schmidt}, \citenamefont
  {Imamoglu},\ and\ \citenamefont {Zhou}}]{Zhang:2025}%
  \BibitemOpen
  \bibfield  {author} {\bibinfo {author} {\bibfnamefont {L.}~\bibnamefont
  {Zhang}}, \bibinfo {author} {\bibfnamefont {L.}~\bibnamefont {Gu}}, \bibinfo
  {author} {\bibfnamefont {H.~S.}\ \bibnamefont {Adlong}}, \bibinfo {author}
  {\bibfnamefont {A.}~\bibnamefont {Christianen}}, \bibinfo {author}
  {\bibfnamefont {E.}~\bibnamefont {Dizer}}, \bibinfo {author} {\bibfnamefont
  {R.}~\bibnamefont {Ni}}, \bibinfo {author} {\bibfnamefont {R.}~\bibnamefont
  {Ma}}, \bibinfo {author} {\bibfnamefont {S.}~\bibnamefont {Park}}, \bibinfo
  {author} {\bibfnamefont {H.}~\bibnamefont {Jang}}, \bibinfo {author}
  {\bibfnamefont {T.}~\bibnamefont {Taniguchi}}, \bibinfo {author}
  {\bibfnamefont {K.}~\bibnamefont {Watanabe}}, \bibinfo {author}
  {\bibfnamefont {I.}~\bibnamefont {Esterlis}}, \bibinfo {author}
  {\bibfnamefont {R.}~\bibnamefont {Schmidt}}, \bibinfo {author} {\bibfnamefont
  {A.}~\bibnamefont {Imamoglu}}, \ and\ \bibinfo {author} {\bibfnamefont
  {Y.}~\bibnamefont {Zhou}},\ }\href@noop {} {\  (\bibinfo {year} {2025})},\
  \Eprint {http://arxiv.org/abs/2512.16631} {arXiv:2512.16631
  [cond-mat.mes-hall]} \BibitemShut {NoStop}%
\bibitem [{\citenamefont {Liu}\ \emph {et~al.}(2026)\citenamefont {Liu},
  \citenamefont {Wilson}, \citenamefont {Hu}, \citenamefont {Zimmerman},
  \citenamefont {Mathew}, \citenamefont {Ouyang}, \citenamefont {Shi},
  \citenamefont {Taniguchi}, \citenamefont {Watanabe}, \citenamefont {Heinz},
  \citenamefont {Chang},\ and\ \citenamefont {Lui}}]{Liu:2026}%
  \BibitemOpen
  \bibfield  {author} {\bibinfo {author} {\bibfnamefont {E.}~\bibnamefont
  {Liu}}, \bibinfo {author} {\bibfnamefont {M.}~\bibnamefont {Wilson}},
  \bibinfo {author} {\bibfnamefont {J.}~\bibnamefont {Hu}}, \bibinfo {author}
  {\bibfnamefont {A.}~\bibnamefont {Zimmerman}}, \bibinfo {author}
  {\bibfnamefont {A.}~\bibnamefont {Mathew}}, \bibinfo {author} {\bibfnamefont
  {T.}~\bibnamefont {Ouyang}}, \bibinfo {author} {\bibfnamefont
  {A.}~\bibnamefont {Shi}}, \bibinfo {author} {\bibfnamefont {T.}~\bibnamefont
  {Taniguchi}}, \bibinfo {author} {\bibfnamefont {K.}~\bibnamefont {Watanabe}},
  \bibinfo {author} {\bibfnamefont {T.~F.}\ \bibnamefont {Heinz}}, \bibinfo
  {author} {\bibfnamefont {Y.-C.}\ \bibnamefont {Chang}}, \ and\ \bibinfo
  {author} {\bibfnamefont {C.~H.}\ \bibnamefont {Lui}},\ }\href@noop {} {\
  (\bibinfo {year} {2026})},\ \Eprint {http://arxiv.org/abs/2601.11914}
  {arXiv:2601.11914 [cond-mat.mes-hall]} \BibitemShut {NoStop}%
\bibitem [{\citenamefont {Novoselov}\ \emph {et~al.}(2004)\citenamefont
  {Novoselov}, \citenamefont {Geim}, \citenamefont {Morozov}, \citenamefont
  {Jiang}, \citenamefont {Zhang}, \citenamefont {Dubonos}, \citenamefont
  {Grigorieva},\ and\ \citenamefont {Firsov}}]{novoselov2004electric}%
  \BibitemOpen
  \bibfield  {author} {\bibinfo {author} {\bibfnamefont {K.~S.}\ \bibnamefont
  {Novoselov}}, \bibinfo {author} {\bibfnamefont {A.~K.}\ \bibnamefont {Geim}},
  \bibinfo {author} {\bibfnamefont {S.~V.}\ \bibnamefont {Morozov}}, \bibinfo
  {author} {\bibfnamefont {D.-e.}\ \bibnamefont {Jiang}}, \bibinfo {author}
  {\bibfnamefont {Y.}~\bibnamefont {Zhang}}, \bibinfo {author} {\bibfnamefont
  {S.~V.}\ \bibnamefont {Dubonos}}, \bibinfo {author} {\bibfnamefont {I.~V.}\
  \bibnamefont {Grigorieva}}, \ and\ \bibinfo {author} {\bibfnamefont {A.~A.}\
  \bibnamefont {Firsov}},\ }\href {\doibase 10.1126/science.1102896} {\bibfield
   {journal} {\bibinfo  {journal} {Science}\ }\textbf {\bibinfo {volume}
  {306}},\ \bibinfo {pages} {666} (\bibinfo {year} {2004})}\BibitemShut
  {NoStop}%
\bibitem [{\citenamefont {Novoselov}\ \emph {et~al.}(2005)\citenamefont
  {Novoselov}, \citenamefont {Geim}, \citenamefont {Morozov}, \citenamefont
  {Jiang}, \citenamefont {Katsnelson}, \citenamefont {Grigorieva},
  \citenamefont {Dubonos},\ and\ \citenamefont {Firsov}}]{Novoselov:2005kj}%
  \BibitemOpen
  \bibfield  {author} {\bibinfo {author} {\bibfnamefont {K.~S.}\ \bibnamefont
  {Novoselov}}, \bibinfo {author} {\bibfnamefont {A.~K.}\ \bibnamefont {Geim}},
  \bibinfo {author} {\bibfnamefont {S.~V.}\ \bibnamefont {Morozov}}, \bibinfo
  {author} {\bibfnamefont {D.}~\bibnamefont {Jiang}}, \bibinfo {author}
  {\bibfnamefont {M.~I.}\ \bibnamefont {Katsnelson}}, \bibinfo {author}
  {\bibfnamefont {I.~V.}\ \bibnamefont {Grigorieva}}, \bibinfo {author}
  {\bibfnamefont {S.~V.}\ \bibnamefont {Dubonos}}, \ and\ \bibinfo {author}
  {\bibfnamefont {A.~A.}\ \bibnamefont {Firsov}},\ }\href {\doibase
  10.1038/nature04233} {\bibfield  {journal} {\bibinfo  {journal} {Nature}\
  }\textbf {\bibinfo {volume} {438}},\ \bibinfo {pages} {197} (\bibinfo {year}
  {2005})}\BibitemShut {NoStop}%
\bibitem [{\citenamefont {Bistritzer}\ and\ \citenamefont
  {MacDonald}(2011)}]{Bistritzer:2011rxr}%
  \BibitemOpen
  \bibfield  {author} {\bibinfo {author} {\bibfnamefont {R.}~\bibnamefont
  {Bistritzer}}\ and\ \bibinfo {author} {\bibfnamefont {A.~H.}\ \bibnamefont
  {MacDonald}},\ }\href {\doibase 10.1073/pnas.1108174108} {\bibfield
  {journal} {\bibinfo  {journal} {Proc. Nat. Acad. Sci.}\ }\textbf {\bibinfo
  {volume} {108}},\ \bibinfo {pages} {12233} (\bibinfo {year}
  {2011})}\BibitemShut {NoStop}%
\bibitem [{\citenamefont {McCann}\ and\ \citenamefont
  {Koshino}(2013)}]{mccann2013bilayer}%
  \BibitemOpen
  \bibfield  {author} {\bibinfo {author} {\bibfnamefont {E.}~\bibnamefont
  {McCann}}\ and\ \bibinfo {author} {\bibfnamefont {M.}~\bibnamefont
  {Koshino}},\ }\href {\doibase 10.1088/0034-4885/76/5/056503} {\bibfield
  {journal} {\bibinfo  {journal} {Reports on Progress in Physics}\ }\textbf
  {\bibinfo {volume} {76}},\ \bibinfo {pages} {056503} (\bibinfo {year}
  {2013})}\BibitemShut {NoStop}%
\bibitem [{\citenamefont {Cao}\ \emph {et~al.}(2018)\citenamefont {Cao},
  \citenamefont {Fatemi}, \citenamefont {Fang}, \citenamefont {Watanabe},
  \citenamefont {Taniguchi}, \citenamefont {Kaxiras},\ and\ \citenamefont
  {Jarillo-Herrero}}]{cao2018unconventional}%
  \BibitemOpen
  \bibfield  {author} {\bibinfo {author} {\bibfnamefont {Y.}~\bibnamefont
  {Cao}}, \bibinfo {author} {\bibfnamefont {V.}~\bibnamefont {Fatemi}},
  \bibinfo {author} {\bibfnamefont {S.}~\bibnamefont {Fang}}, \bibinfo {author}
  {\bibfnamefont {K.}~\bibnamefont {Watanabe}}, \bibinfo {author}
  {\bibfnamefont {T.}~\bibnamefont {Taniguchi}}, \bibinfo {author}
  {\bibfnamefont {E.}~\bibnamefont {Kaxiras}}, \ and\ \bibinfo {author}
  {\bibfnamefont {P.}~\bibnamefont {Jarillo-Herrero}},\ }\href {\doibase
  10.1038/nature26160} {\bibfield  {journal} {\bibinfo  {journal} {Nature}\
  }\textbf {\bibinfo {volume} {556}},\ \bibinfo {pages} {43} (\bibinfo {year}
  {2018})}\BibitemShut {NoStop}%
\bibitem [{\citenamefont {Xu}\ \emph {et~al.}(2021)\citenamefont {Xu},
  \citenamefont {Al~Ezzi}, \citenamefont {Balakrishnan}, \citenamefont
  {Garcia-Ruiz}, \citenamefont {Tsim}, \citenamefont {Mullan}, \citenamefont
  {Barrier}, \citenamefont {Xin}, \citenamefont {Piot}, \citenamefont
  {Taniguchi}, \citenamefont {Katayama}, \citenamefont {Carvalho},
  \citenamefont {Mishchenko}, \citenamefont {Geim}, \citenamefont
  {Fal{\textquoteright}ko}, \citenamefont {Adam}, \citenamefont {Castro~Neto},
  \citenamefont {Novoselov},\ and\ \citenamefont {Shi}}]{Xu2021}%
  \BibitemOpen
  \bibfield  {author} {\bibinfo {author} {\bibfnamefont {S.}~\bibnamefont
  {Xu}}, \bibinfo {author} {\bibfnamefont {M.~M.}\ \bibnamefont {Al~Ezzi}},
  \bibinfo {author} {\bibfnamefont {N.}~\bibnamefont {Balakrishnan}}, \bibinfo
  {author} {\bibfnamefont {A.}~\bibnamefont {Garcia-Ruiz}}, \bibinfo {author}
  {\bibfnamefont {B.}~\bibnamefont {Tsim}}, \bibinfo {author} {\bibfnamefont
  {C.}~\bibnamefont {Mullan}}, \bibinfo {author} {\bibfnamefont
  {J.}~\bibnamefont {Barrier}}, \bibinfo {author} {\bibfnamefont
  {N.}~\bibnamefont {Xin}}, \bibinfo {author} {\bibfnamefont {B.~A.}\
  \bibnamefont {Piot}}, \bibinfo {author} {\bibfnamefont {T.}~\bibnamefont
  {Taniguchi}}, \bibinfo {author} {\bibfnamefont {K.}~\bibnamefont {Katayama}},
  \bibinfo {author} {\bibfnamefont {A.}~\bibnamefont {Carvalho}}, \bibinfo
  {author} {\bibfnamefont {A.}~\bibnamefont {Mishchenko}}, \bibinfo {author}
  {\bibfnamefont {A.}~\bibnamefont {Geim}}, \bibinfo {author} {\bibfnamefont
  {V.}~\bibnamefont {Fal{\textquoteright}ko}}, \bibinfo {author} {\bibfnamefont
  {S.}~\bibnamefont {Adam}}, \bibinfo {author} {\bibfnamefont {A.~H.}\
  \bibnamefont {Castro~Neto}}, \bibinfo {author} {\bibfnamefont
  {K.}~\bibnamefont {Novoselov}}, \ and\ \bibinfo {author} {\bibfnamefont
  {Y.}~\bibnamefont {Shi}},\ }\href {\doibase 10.1038/s41567-021-01172-9}
  {\bibfield  {journal} {\bibinfo  {journal} {Nat. Phys.}\ }\textbf {\bibinfo
  {volume} {17}},\ \bibinfo {pages} {619} (\bibinfo {year} {2021})}\BibitemShut
  {NoStop}%
\bibitem [{\citenamefont {Dindorkar}\ \emph {et~al.}(2023)\citenamefont
  {Dindorkar}, \citenamefont {Kurade},\ and\ \citenamefont
  {Shaikh}}]{Dindorkar2023}%
  \BibitemOpen
  \bibfield  {author} {\bibinfo {author} {\bibfnamefont {S.~S.}\ \bibnamefont
  {Dindorkar}}, \bibinfo {author} {\bibfnamefont {A.~S.}\ \bibnamefont
  {Kurade}}, \ and\ \bibinfo {author} {\bibfnamefont {A.~H.}\ \bibnamefont
  {Shaikh}},\ }\href {\doibase 10.1016/j.chphi.2023.100325} {\bibfield
  {journal} {\bibinfo  {journal} {Chemical Physics Impact}\ }\textbf {\bibinfo
  {volume} {7}},\ \bibinfo {pages} {100325} (\bibinfo {year}
  {2023})}\BibitemShut {NoStop}%
\bibitem [{\citenamefont {Lee}\ \emph {et~al.}(2009)\citenamefont {Lee},
  \citenamefont {Gremaud}, \citenamefont {Han}, \citenamefont {Englert},\ and\
  \citenamefont {Miniatura}}]{Lee:2009hkd}%
  \BibitemOpen
  \bibfield  {author} {\bibinfo {author} {\bibfnamefont {K.~L.}\ \bibnamefont
  {Lee}}, \bibinfo {author} {\bibfnamefont {B.}~\bibnamefont {Gremaud}},
  \bibinfo {author} {\bibfnamefont {R.}~\bibnamefont {Han}}, \bibinfo {author}
  {\bibfnamefont {B.-G.}\ \bibnamefont {Englert}}, \ and\ \bibinfo {author}
  {\bibfnamefont {C.}~\bibnamefont {Miniatura}},\ }\href {\doibase
  10.1103/PhysRevA.80.043411} {\bibfield  {journal} {\bibinfo  {journal} {Phys.
  Rev. A}\ }\textbf {\bibinfo {volume} {80}},\ \bibinfo {pages} {043411}
  (\bibinfo {year} {2009})}\BibitemShut {NoStop}%
\bibitem [{\citenamefont {Jo}\ \emph {et~al.}(2012)\citenamefont {Jo},
  \citenamefont {Guzman}, \citenamefont {Thomas}, \citenamefont {Hosur},
  \citenamefont {Vishwanath},\ and\ \citenamefont {Stamper-Kurn}}]{Jo2012}%
  \BibitemOpen
  \bibfield  {author} {\bibinfo {author} {\bibfnamefont {G.-B.}\ \bibnamefont
  {Jo}}, \bibinfo {author} {\bibfnamefont {J.}~\bibnamefont {Guzman}}, \bibinfo
  {author} {\bibfnamefont {C.~K.}\ \bibnamefont {Thomas}}, \bibinfo {author}
  {\bibfnamefont {P.}~\bibnamefont {Hosur}}, \bibinfo {author} {\bibfnamefont
  {A.}~\bibnamefont {Vishwanath}}, \ and\ \bibinfo {author} {\bibfnamefont
  {D.~M.}\ \bibnamefont {Stamper-Kurn}},\ }\href {\doibase
  10.1103/PhysRevLett.108.045305} {\bibfield  {journal} {\bibinfo  {journal}
  {Phys. Rev. Lett.}\ }\textbf {\bibinfo {volume} {108}},\ \bibinfo {pages}
  {045305} (\bibinfo {year} {2012})}\BibitemShut {NoStop}%
\bibitem [{\citenamefont {Gross}\ and\ \citenamefont
  {Bloch}(2017)}]{Gross:2017ehn}%
  \BibitemOpen
  \bibfield  {author} {\bibinfo {author} {\bibfnamefont {C.}~\bibnamefont
  {Gross}}\ and\ \bibinfo {author} {\bibfnamefont {I.}~\bibnamefont {Bloch}},\
  }\href {\doibase 10.1126/science.aal3837} {\bibfield  {journal} {\bibinfo
  {journal} {Science}\ }\textbf {\bibinfo {volume} {357}},\ \bibinfo {pages}
  {995} (\bibinfo {year} {2017})}\BibitemShut {NoStop}%
\bibitem [{\citenamefont {Hofstetter}\ and\ \citenamefont
  {Qin}(2018)}]{Hofstetter:2018rwi}%
  \BibitemOpen
  \bibfield  {author} {\bibinfo {author} {\bibfnamefont {W.}~\bibnamefont
  {Hofstetter}}\ and\ \bibinfo {author} {\bibfnamefont {T.}~\bibnamefont
  {Qin}},\ }\href {\doibase 10.1088/1361-6455/aaa31b} {\bibfield  {journal}
  {\bibinfo  {journal} {J. Phys. B}\ }\textbf {\bibinfo {volume} {51}},\
  \bibinfo {pages} {082001} (\bibinfo {year} {2018})}\BibitemShut {NoStop}%
\bibitem [{\citenamefont {Chalopin}\ \emph {et~al.}(2025)\citenamefont
  {Chalopin}, \citenamefont {Bojovi{\'c}}, \citenamefont {Bourgund},
  \citenamefont {Wang}, \citenamefont {Franz}, \citenamefont {Bloch},\ and\
  \citenamefont {Hilker}}]{Chalopin:2024}%
  \BibitemOpen
  \bibfield  {author} {\bibinfo {author} {\bibfnamefont {T.}~\bibnamefont
  {Chalopin}}, \bibinfo {author} {\bibfnamefont {P.}~\bibnamefont
  {Bojovi{\'c}}}, \bibinfo {author} {\bibfnamefont {D.}~\bibnamefont
  {Bourgund}}, \bibinfo {author} {\bibfnamefont {S.}~\bibnamefont {Wang}},
  \bibinfo {author} {\bibfnamefont {T.}~\bibnamefont {Franz}}, \bibinfo
  {author} {\bibfnamefont {I.}~\bibnamefont {Bloch}}, \ and\ \bibinfo {author}
  {\bibfnamefont {T.}~\bibnamefont {Hilker}},\ }\href {\doibase
  10.1103/PhysRevLett.134.053402} {\bibfield  {journal} {\bibinfo  {journal}
  {Phys. Rev. Lett.}\ }\textbf {\bibinfo {volume} {134}},\ \bibinfo {pages}
  {053402} (\bibinfo {year} {2025})}\BibitemShut {NoStop}%
\bibitem [{\citenamefont {Lobo}\ \emph {et~al.}(2006)\citenamefont {Lobo},
  \citenamefont {Recati}, \citenamefont {Giorgini},\ and\ \citenamefont
  {Stringari}}]{Lobo2006}%
  \BibitemOpen
  \bibfield  {author} {\bibinfo {author} {\bibfnamefont {C.}~\bibnamefont
  {Lobo}}, \bibinfo {author} {\bibfnamefont {A.}~\bibnamefont {Recati}},
  \bibinfo {author} {\bibfnamefont {S.}~\bibnamefont {Giorgini}}, \ and\
  \bibinfo {author} {\bibfnamefont {S.}~\bibnamefont {Stringari}},\ }\href
  {\doibase 10.1103/PhysRevLett.97.200403} {\bibfield  {journal} {\bibinfo
  {journal} {Phys. Rev. Lett.}\ }\textbf {\bibinfo {volume} {97}},\ \bibinfo
  {pages} {200403} (\bibinfo {year} {2006})}\BibitemShut {NoStop}%
\bibitem [{\citenamefont {Fey}\ \emph {et~al.}(2020)\citenamefont {Fey},
  \citenamefont {Schmelcher}, \citenamefont {Imamoglu},\ and\ \citenamefont
  {Schmidt}}]{Fey:2020}%
  \BibitemOpen
  \bibfield  {author} {\bibinfo {author} {\bibfnamefont {C.}~\bibnamefont
  {Fey}}, \bibinfo {author} {\bibfnamefont {P.}~\bibnamefont {Schmelcher}},
  \bibinfo {author} {\bibfnamefont {A.}~\bibnamefont {Imamoglu}}, \ and\
  \bibinfo {author} {\bibfnamefont {R.}~\bibnamefont {Schmidt}},\ }\href
  {\doibase 10.1103/PhysRevB.101.195417} {\bibfield  {journal} {\bibinfo
  {journal} {Phys. Rev. B}\ }\textbf {\bibinfo {volume} {101}},\ \bibinfo
  {pages} {195417} (\bibinfo {year} {2020})}\BibitemShut {NoStop}%
\bibitem [{\citenamefont {Tan}\ \emph {et~al.}(2020)\citenamefont {Tan},
  \citenamefont {Cotlet}, \citenamefont {Bergschneider}, \citenamefont
  {Schmidt}, \citenamefont {Back}, \citenamefont {Shimazaki}, \citenamefont
  {Kroner},\ and\ \citenamefont {Imamoglu}}]{Tan2020PRX}%
  \BibitemOpen
  \bibfield  {author} {\bibinfo {author} {\bibfnamefont {L.~B.}\ \bibnamefont
  {Tan}}, \bibinfo {author} {\bibfnamefont {O.}~\bibnamefont {Cotlet}},
  \bibinfo {author} {\bibfnamefont {A.}~\bibnamefont {Bergschneider}}, \bibinfo
  {author} {\bibfnamefont {R.}~\bibnamefont {Schmidt}}, \bibinfo {author}
  {\bibfnamefont {P.}~\bibnamefont {Back}}, \bibinfo {author} {\bibfnamefont
  {Y.}~\bibnamefont {Shimazaki}}, \bibinfo {author} {\bibfnamefont
  {M.}~\bibnamefont {Kroner}}, \ and\ \bibinfo {author} {\bibfnamefont
  {A.}~\bibnamefont {Imamoglu}},\ }\href {\doibase 10.1103/PhysRevX.10.021011}
  {\bibfield  {journal} {\bibinfo  {journal} {Phys. Rev. X}\ }\textbf {\bibinfo
  {volume} {10}},\ \bibinfo {pages} {021011} (\bibinfo {year}
  {2020})}\BibitemShut {NoStop}%
\bibitem [{\citenamefont {Hu}\ \emph {et~al.}(2024)\citenamefont {Hu},
  \citenamefont {Wang},\ and\ \citenamefont {Liu}}]{Hu2024PRA}%
  \BibitemOpen
  \bibfield  {author} {\bibinfo {author} {\bibfnamefont {H.}~\bibnamefont
  {Hu}}, \bibinfo {author} {\bibfnamefont {J.}~\bibnamefont {Wang}}, \ and\
  \bibinfo {author} {\bibfnamefont {X.-J.}\ \bibnamefont {Liu}},\ }\href
  {\doibase 10.1103/PhysRevA.110.023314} {\bibfield  {journal} {\bibinfo
  {journal} {Phys. Rev. A}\ }\textbf {\bibinfo {volume} {110}},\ \bibinfo
  {pages} {023314} (\bibinfo {year} {2024})}\BibitemShut {NoStop}%
\bibitem [{\citenamefont {Pimenov}(2024)}]{Pimenov:2024}%
  \BibitemOpen
  \bibfield  {author} {\bibinfo {author} {\bibfnamefont {D.}~\bibnamefont
  {Pimenov}},\ }\href {\doibase 10.1103/PhysRevB.109.195153} {\bibfield
  {journal} {\bibinfo  {journal} {Phys. Rev. B}\ }\textbf {\bibinfo {volume}
  {109}},\ \bibinfo {pages} {195153} (\bibinfo {year} {2024})}\BibitemShut
  {NoStop}%
\bibitem [{\citenamefont {Amelio}\ and\ \citenamefont
  {Goldman}(2024)}]{Amelio:2024ED}%
  \BibitemOpen
  \bibfield  {author} {\bibinfo {author} {\bibfnamefont {I.}~\bibnamefont
  {Amelio}}\ and\ \bibinfo {author} {\bibfnamefont {N.}~\bibnamefont
  {Goldman}},\ }\href {\doibase 10.21468/SciPostPhys.16.2.056} {\bibfield
  {journal} {\bibinfo  {journal} {SciPost Phys.}\ }\textbf {\bibinfo {volume}
  {16}},\ \bibinfo {pages} {056} (\bibinfo {year} {2024})}\BibitemShut
  {NoStop}%
\bibitem [{\citenamefont {Amelio}\ \emph {et~al.}(2024)\citenamefont {Amelio},
  \citenamefont {Mazza},\ and\ \citenamefont {Goldman}}]{Amelio:2024Mott}%
  \BibitemOpen
  \bibfield  {author} {\bibinfo {author} {\bibfnamefont {I.}~\bibnamefont
  {Amelio}}, \bibinfo {author} {\bibfnamefont {G.}~\bibnamefont {Mazza}}, \
  and\ \bibinfo {author} {\bibfnamefont {N.}~\bibnamefont {Goldman}},\ }\href
  {\doibase 10.1103/PhysRevB.110.235302} {\bibfield  {journal} {\bibinfo
  {journal} {Phys. Rev. B}\ }\textbf {\bibinfo {volume} {110}},\ \bibinfo
  {pages} {235302} (\bibinfo {year} {2024})}\BibitemShut {NoStop}%
\bibitem [{\citenamefont {Vashisht}\ \emph {et~al.}(2025)\citenamefont
  {Vashisht}, \citenamefont {Amelio}, \citenamefont {Vanderstraeten},
  \citenamefont {Bruun}, \citenamefont {Diessel},\ and\ \citenamefont
  {Goldman}}]{Vashisht:2024}%
  \BibitemOpen
  \bibfield  {author} {\bibinfo {author} {\bibfnamefont {A.}~\bibnamefont
  {Vashisht}}, \bibinfo {author} {\bibfnamefont {I.}~\bibnamefont {Amelio}},
  \bibinfo {author} {\bibfnamefont {L.}~\bibnamefont {Vanderstraeten}},
  \bibinfo {author} {\bibfnamefont {G.~M.}\ \bibnamefont {Bruun}}, \bibinfo
  {author} {\bibfnamefont {O.~K.}\ \bibnamefont {Diessel}}, \ and\ \bibinfo
  {author} {\bibfnamefont {N.}~\bibnamefont {Goldman}},\ }\href {\doibase
  10.1038/s41467-025-60166-w} {\bibfield  {journal} {\bibinfo  {journal}
  {Nature Commun.}\ }\textbf {\bibinfo {volume} {16}},\ \bibinfo {pages} {4918}
  (\bibinfo {year} {2025})}\BibitemShut {NoStop}%
\bibitem [{\citenamefont {Tarruell}\ \emph {et~al.}(2012)\citenamefont
  {Tarruell}, \citenamefont {Greif}, \citenamefont {Uehlinger}, \citenamefont
  {Jotzu},\ and\ \citenamefont {Esslinger}}]{tarruell2012creating}%
  \BibitemOpen
  \bibfield  {author} {\bibinfo {author} {\bibfnamefont {L.}~\bibnamefont
  {Tarruell}}, \bibinfo {author} {\bibfnamefont {D.}~\bibnamefont {Greif}},
  \bibinfo {author} {\bibfnamefont {T.}~\bibnamefont {Uehlinger}}, \bibinfo
  {author} {\bibfnamefont {G.}~\bibnamefont {Jotzu}}, \ and\ \bibinfo {author}
  {\bibfnamefont {T.}~\bibnamefont {Esslinger}},\ }\href {\doibase
  10.1038/nature10871} {\bibfield  {journal} {\bibinfo  {journal} {Nature}\
  }\textbf {\bibinfo {volume} {483}},\ \bibinfo {pages} {302} (\bibinfo {year}
  {2012})}\BibitemShut {NoStop}%
\bibitem [{\citenamefont {Sorout}\ \emph {et~al.}(2020)\citenamefont {Sorout},
  \citenamefont {Sarkar},\ and\ \citenamefont {Gangadharaiah}}]{Sorout_2020}%
  \BibitemOpen
  \bibfield  {author} {\bibinfo {author} {\bibfnamefont {A.~K.}\ \bibnamefont
  {Sorout}}, \bibinfo {author} {\bibfnamefont {S.}~\bibnamefont {Sarkar}}, \
  and\ \bibinfo {author} {\bibfnamefont {S.}~\bibnamefont {Gangadharaiah}},\
  }\href {\doibase 10.1088/1361-648X/ab9d4d} {\bibfield  {journal} {\bibinfo
  {journal} {Journal of Physics: Condensed Matter}\ }\textbf {\bibinfo {volume}
  {32}},\ \bibinfo {pages} {415604} (\bibinfo {year} {2020})}\BibitemShut
  {NoStop}%
\bibitem [{\citenamefont {Fan}\ \emph {et~al.}(2023)\citenamefont {Fan},
  \citenamefont {Sun}, \citenamefont {Shao}, \citenamefont {Li},\ and\
  \citenamefont {Zhao}}]{Fan2023}%
  \BibitemOpen
  \bibfield  {author} {\bibinfo {author} {\bibfnamefont {R.}~\bibnamefont
  {Fan}}, \bibinfo {author} {\bibfnamefont {L.}~\bibnamefont {Sun}}, \bibinfo
  {author} {\bibfnamefont {X.}~\bibnamefont {Shao}}, \bibinfo {author}
  {\bibfnamefont {Y.}~\bibnamefont {Li}}, \ and\ \bibinfo {author}
  {\bibfnamefont {M.}~\bibnamefont {Zhao}},\ }\href {\doibase
  10.1016/j.chphma.2022.04.009} {\bibfield  {journal} {\bibinfo  {journal}
  {ChemPhysMater}\ }\textbf {\bibinfo {volume} {2}},\ \bibinfo {pages} {30}
  (\bibinfo {year} {2023})}\BibitemShut {NoStop}%
\bibitem [{\citenamefont {Wehling}\ \emph {et~al.}(2009)\citenamefont
  {Wehling}, \citenamefont {Katsnelson},\ and\ \citenamefont
  {Lichtenstein}}]{Wehling2009}%
  \BibitemOpen
  \bibfield  {author} {\bibinfo {author} {\bibfnamefont {T.~O.}\ \bibnamefont
  {Wehling}}, \bibinfo {author} {\bibfnamefont {M.~I.}\ \bibnamefont
  {Katsnelson}}, \ and\ \bibinfo {author} {\bibfnamefont {A.~I.}\ \bibnamefont
  {Lichtenstein}},\ }\href {\doibase 10.1103/PhysRevB.80.085428} {\bibfield
  {journal} {\bibinfo  {journal} {Phys. Rev. B}\ }\textbf {\bibinfo {volume}
  {80}},\ \bibinfo {pages} {085428} (\bibinfo {year} {2009})}\BibitemShut
  {NoStop}%
\bibitem [{\citenamefont {Wehling}\ \emph {et~al.}(2010)\citenamefont
  {Wehling}, \citenamefont {Yuan}, \citenamefont {Lichtenstein}, \citenamefont
  {Geim},\ and\ \citenamefont {Katsnelson}}]{Wehling2010}%
  \BibitemOpen
  \bibfield  {author} {\bibinfo {author} {\bibfnamefont {T.~O.}\ \bibnamefont
  {Wehling}}, \bibinfo {author} {\bibfnamefont {S.}~\bibnamefont {Yuan}},
  \bibinfo {author} {\bibfnamefont {A.~I.}\ \bibnamefont {Lichtenstein}},
  \bibinfo {author} {\bibfnamefont {A.~K.}\ \bibnamefont {Geim}}, \ and\
  \bibinfo {author} {\bibfnamefont {M.~I.}\ \bibnamefont {Katsnelson}},\ }\href
  {\doibase 10.1103/PhysRevLett.105.056802} {\bibfield  {journal} {\bibinfo
  {journal} {Phys. Rev. Lett.}\ }\textbf {\bibinfo {volume} {105}},\ \bibinfo
  {pages} {056802} (\bibinfo {year} {2010})}\BibitemShut {NoStop}%
\bibitem [{\citenamefont {Ni}\ \emph {et~al.}(2010)\citenamefont {Ni},
  \citenamefont {Ponomarenko}, \citenamefont {Nair}, \citenamefont {Yang},
  \citenamefont {Anissimova}, \citenamefont {Grigorieva}, \citenamefont
  {Schedin}, \citenamefont {Blake}, \citenamefont {Shen}, \citenamefont {Hill},
  \citenamefont {Novoselov},\ and\ \citenamefont {Geim}}]{Ni2010}%
  \BibitemOpen
  \bibfield  {author} {\bibinfo {author} {\bibfnamefont {Z.~H.}\ \bibnamefont
  {Ni}}, \bibinfo {author} {\bibfnamefont {L.~A.}\ \bibnamefont {Ponomarenko}},
  \bibinfo {author} {\bibfnamefont {R.~R.}\ \bibnamefont {Nair}}, \bibinfo
  {author} {\bibfnamefont {R.}~\bibnamefont {Yang}}, \bibinfo {author}
  {\bibfnamefont {S.}~\bibnamefont {Anissimova}}, \bibinfo {author}
  {\bibfnamefont {I.~V.}\ \bibnamefont {Grigorieva}}, \bibinfo {author}
  {\bibfnamefont {F.}~\bibnamefont {Schedin}}, \bibinfo {author} {\bibfnamefont
  {P.}~\bibnamefont {Blake}}, \bibinfo {author} {\bibfnamefont {Z.~X.}\
  \bibnamefont {Shen}}, \bibinfo {author} {\bibfnamefont {E.~H.}\ \bibnamefont
  {Hill}}, \bibinfo {author} {\bibfnamefont {K.~S.}\ \bibnamefont {Novoselov}},
  \ and\ \bibinfo {author} {\bibfnamefont {A.~K.}\ \bibnamefont {Geim}},\
  }\href {\doibase 10.1021/nl101399r} {\bibfield  {journal} {\bibinfo
  {journal} {Nano Letters}\ }\textbf {\bibinfo {volume} {10}},\ \bibinfo
  {pages} {3868} (\bibinfo {year} {2010})}\BibitemShut {NoStop}%
\bibitem [{\citenamefont {Wehling}\ \emph {et~al.}(2014)\citenamefont
  {Wehling}, \citenamefont {Black-Schaffer},\ and\ \citenamefont
  {Balatsky}}]{wehling2014dirac}%
  \BibitemOpen
  \bibfield  {author} {\bibinfo {author} {\bibfnamefont {T.~O.}\ \bibnamefont
  {Wehling}}, \bibinfo {author} {\bibfnamefont {A.~M.}\ \bibnamefont
  {Black-Schaffer}}, \ and\ \bibinfo {author} {\bibfnamefont {A.~V.}\
  \bibnamefont {Balatsky}},\ }\href {\doibase 10.1080/00018732.2014.927109}
  {\bibfield  {journal} {\bibinfo  {journal} {Advances in Physics}\ }\textbf
  {\bibinfo {volume} {63}},\ \bibinfo {pages} {1} (\bibinfo {year}
  {2014})}\BibitemShut {NoStop}%
\bibitem [{\citenamefont {Bloch}\ \emph {et~al.}(2008)\citenamefont {Bloch},
  \citenamefont {Dalibard},\ and\ \citenamefont {Zwerger}}]{bloch2008many}%
  \BibitemOpen
  \bibfield  {author} {\bibinfo {author} {\bibfnamefont {I.}~\bibnamefont
  {Bloch}}, \bibinfo {author} {\bibfnamefont {J.}~\bibnamefont {Dalibard}}, \
  and\ \bibinfo {author} {\bibfnamefont {W.}~\bibnamefont {Zwerger}},\ }\href
  {\doibase 10.1103/RevModPhys.80.885} {\bibfield  {journal} {\bibinfo
  {journal} {Rev. Mod. Phys.}\ }\textbf {\bibinfo {volume} {80}},\ \bibinfo
  {pages} {885} (\bibinfo {year} {2008})}\BibitemShut {NoStop}%
\bibitem [{\citenamefont {Bloch}\ \emph {et~al.}(2012)\citenamefont {Bloch},
  \citenamefont {Dalibard},\ and\ \citenamefont
  {Nascimb{\`e}ne}}]{Bloch:2012uep}%
  \BibitemOpen
  \bibfield  {author} {\bibinfo {author} {\bibfnamefont {I.}~\bibnamefont
  {Bloch}}, \bibinfo {author} {\bibfnamefont {J.}~\bibnamefont {Dalibard}}, \
  and\ \bibinfo {author} {\bibfnamefont {S.}~\bibnamefont {Nascimb{\`e}ne}},\
  }\href {\doibase 10.1038/nphys2259} {\bibfield  {journal} {\bibinfo
  {journal} {Nature Phys.}\ }\textbf {\bibinfo {volume} {8}},\ \bibinfo {pages}
  {267} (\bibinfo {year} {2012})}\BibitemShut {NoStop}%
\bibitem [{\citenamefont {Sch{\"a}fer}\ \emph {et~al.}(2020)\citenamefont
  {Sch{\"a}fer}, \citenamefont {Fukuhara}, \citenamefont {Sugawa},
  \citenamefont {Takasu},\ and\ \citenamefont {Takahashi}}]{Schafer:2020ccc}%
  \BibitemOpen
  \bibfield  {author} {\bibinfo {author} {\bibfnamefont {F.}~\bibnamefont
  {Sch{\"a}fer}}, \bibinfo {author} {\bibfnamefont {T.}~\bibnamefont
  {Fukuhara}}, \bibinfo {author} {\bibfnamefont {S.}~\bibnamefont {Sugawa}},
  \bibinfo {author} {\bibfnamefont {Y.}~\bibnamefont {Takasu}}, \ and\ \bibinfo
  {author} {\bibfnamefont {Y.}~\bibnamefont {Takahashi}},\ }\href {\doibase
  10.1038/s42254-020-0195-3} {\bibfield  {journal} {\bibinfo  {journal} {Nature
  Rev. Phys.}\ }\textbf {\bibinfo {volume} {2}},\ \bibinfo {pages} {411}
  (\bibinfo {year} {2020})}\BibitemShut {NoStop}%
\bibitem [{\citenamefont {Schmidt}\ \emph {et~al.}(2018)\citenamefont
  {Schmidt}, \citenamefont {Knap}, \citenamefont {Ivanov}, \citenamefont {You},
  \citenamefont {Cetina},\ and\ \citenamefont {Demler}}]{schmidt2018universal}%
  \BibitemOpen
  \bibfield  {author} {\bibinfo {author} {\bibfnamefont {R.}~\bibnamefont
  {Schmidt}}, \bibinfo {author} {\bibfnamefont {M.}~\bibnamefont {Knap}},
  \bibinfo {author} {\bibfnamefont {D.~A.}\ \bibnamefont {Ivanov}}, \bibinfo
  {author} {\bibfnamefont {J.-S.}\ \bibnamefont {You}}, \bibinfo {author}
  {\bibfnamefont {M.}~\bibnamefont {Cetina}}, \ and\ \bibinfo {author}
  {\bibfnamefont {E.}~\bibnamefont {Demler}},\ }\href {\doibase
  10.1088/1361-6633/aa9593} {\bibfield  {journal} {\bibinfo  {journal} {Rep.
  Prog. Phys.}\ }\textbf {\bibinfo {volume} {81}},\ \bibinfo {pages} {024401}
  (\bibinfo {year} {2018})}\BibitemShut {NoStop}%
\bibitem [{\citenamefont {Cetina}\ \emph {et~al.}(2016)\citenamefont {Cetina},
  \citenamefont {Jag}, \citenamefont {Lous}, \citenamefont {Fritsche},
  \citenamefont {Walraven}, \citenamefont {Grimm}, \citenamefont {Levinsen},
  \citenamefont {Parish}, \citenamefont {Schmidt}, \citenamefont {Knap},\ and\
  \citenamefont {Demler}}]{cetina2016ultrafast}%
  \BibitemOpen
  \bibfield  {author} {\bibinfo {author} {\bibfnamefont {M.}~\bibnamefont
  {Cetina}}, \bibinfo {author} {\bibfnamefont {M.}~\bibnamefont {Jag}},
  \bibinfo {author} {\bibfnamefont {R.~S.}\ \bibnamefont {Lous}}, \bibinfo
  {author} {\bibfnamefont {I.}~\bibnamefont {Fritsche}}, \bibinfo {author}
  {\bibfnamefont {J.~T.~M.}\ \bibnamefont {Walraven}}, \bibinfo {author}
  {\bibfnamefont {R.}~\bibnamefont {Grimm}}, \bibinfo {author} {\bibfnamefont
  {J.}~\bibnamefont {Levinsen}}, \bibinfo {author} {\bibfnamefont {M.~M.}\
  \bibnamefont {Parish}}, \bibinfo {author} {\bibfnamefont {R.}~\bibnamefont
  {Schmidt}}, \bibinfo {author} {\bibfnamefont {M.}~\bibnamefont {Knap}}, \
  and\ \bibinfo {author} {\bibfnamefont {E.}~\bibnamefont {Demler}},\ }\href
  {\doibase 10.1126/science.aaf5134} {\bibfield  {journal} {\bibinfo  {journal}
  {Science}\ }\textbf {\bibinfo {volume} {354}},\ \bibinfo {pages} {96}
  (\bibinfo {year} {2016})}\BibitemShut {NoStop}%
\bibitem [{\citenamefont {Wang}\ \emph {et~al.}(2018)\citenamefont {Wang},
  \citenamefont {Chernikov}, \citenamefont {Glazov}, \citenamefont {Heinz},
  \citenamefont {Marie}, \citenamefont {Amand},\ and\ \citenamefont
  {Urbaszek}}]{wang2018tmdspectroscopy}%
  \BibitemOpen
  \bibfield  {author} {\bibinfo {author} {\bibfnamefont {G.}~\bibnamefont
  {Wang}}, \bibinfo {author} {\bibfnamefont {A.}~\bibnamefont {Chernikov}},
  \bibinfo {author} {\bibfnamefont {M.~M.}\ \bibnamefont {Glazov}}, \bibinfo
  {author} {\bibfnamefont {T.~F.}\ \bibnamefont {Heinz}}, \bibinfo {author}
  {\bibfnamefont {X.}~\bibnamefont {Marie}}, \bibinfo {author} {\bibfnamefont
  {T.}~\bibnamefont {Amand}}, \ and\ \bibinfo {author} {\bibfnamefont
  {B.}~\bibnamefont {Urbaszek}},\ }\href {\doibase
  10.1103/RevModPhys.90.021001} {\bibfield  {journal} {\bibinfo  {journal}
  {Rev. Mod. Phys.}\ }\textbf {\bibinfo {volume} {90}},\ \bibinfo {pages}
  {021001} (\bibinfo {year} {2018})}\BibitemShut {NoStop}%
\bibitem [{\citenamefont {Knap}\ \emph {et~al.}(2012)\citenamefont {Knap},
  \citenamefont {Shashi}, \citenamefont {Nishida}, \citenamefont {Imambekov},
  \citenamefont {Abanin},\ and\ \citenamefont {Demler}}]{knap2012time}%
  \BibitemOpen
  \bibfield  {author} {\bibinfo {author} {\bibfnamefont {M.}~\bibnamefont
  {Knap}}, \bibinfo {author} {\bibfnamefont {A.}~\bibnamefont {Shashi}},
  \bibinfo {author} {\bibfnamefont {Y.}~\bibnamefont {Nishida}}, \bibinfo
  {author} {\bibfnamefont {A.}~\bibnamefont {Imambekov}}, \bibinfo {author}
  {\bibfnamefont {D.~A.}\ \bibnamefont {Abanin}}, \ and\ \bibinfo {author}
  {\bibfnamefont {E.}~\bibnamefont {Demler}},\ }\href {\doibase
  10.1103/PhysRevX.2.041020} {\bibfield  {journal} {\bibinfo  {journal} {Phys.
  Rev. X}\ }\textbf {\bibinfo {volume} {2}},\ \bibinfo {pages} {041020}
  (\bibinfo {year} {2012})}\BibitemShut {NoStop}%
\bibitem [{not()}]{note}%
  \BibitemOpen
  \href@noop {} {\ }\bibinfo {note} {{For a mobile impurity, the perturbation
  operator is given by $\hat{V}^\dagger = \hat{d}^\dagger_{\mathbf{q}}$,
  creating an impurity with momentum $\mathbf{q}$.}}\BibitemShut {Stop}%
\bibitem [{\citenamefont {Bakr}\ \emph {et~al.}(2009)\citenamefont {Bakr},
  \citenamefont {Gillen}, \citenamefont {Peng}, \citenamefont {F{\"o}lling},\
  and\ \citenamefont {Greiner}}]{Bakr2009}%
  \BibitemOpen
  \bibfield  {author} {\bibinfo {author} {\bibfnamefont {W.~S.}\ \bibnamefont
  {Bakr}}, \bibinfo {author} {\bibfnamefont {J.~I.}\ \bibnamefont {Gillen}},
  \bibinfo {author} {\bibfnamefont {A.}~\bibnamefont {Peng}}, \bibinfo {author}
  {\bibfnamefont {S.}~\bibnamefont {F{\"o}lling}}, \ and\ \bibinfo {author}
  {\bibfnamefont {M.}~\bibnamefont {Greiner}},\ }\href {\doibase
  10.1038/nature08482} {\bibfield  {journal} {\bibinfo  {journal} {Nature}\
  }\textbf {\bibinfo {volume} {462}},\ \bibinfo {pages} {74} (\bibinfo {year}
  {2009})}\BibitemShut {NoStop}%
\bibitem [{\citenamefont {Koepsell}\ \emph {et~al.}(2019)\citenamefont
  {Koepsell}, \citenamefont {Vijayan}, \citenamefont {Sompet}, \citenamefont
  {Grusdt}, \citenamefont {Hilker}, \citenamefont {Demler}, \citenamefont
  {Salomon}, \citenamefont {Bloch},\ and\ \citenamefont
  {Gross}}]{Koepsell2019}%
  \BibitemOpen
  \bibfield  {author} {\bibinfo {author} {\bibfnamefont {J.}~\bibnamefont
  {Koepsell}}, \bibinfo {author} {\bibfnamefont {J.}~\bibnamefont {Vijayan}},
  \bibinfo {author} {\bibfnamefont {P.}~\bibnamefont {Sompet}}, \bibinfo
  {author} {\bibfnamefont {F.}~\bibnamefont {Grusdt}}, \bibinfo {author}
  {\bibfnamefont {T.~A.}\ \bibnamefont {Hilker}}, \bibinfo {author}
  {\bibfnamefont {E.}~\bibnamefont {Demler}}, \bibinfo {author} {\bibfnamefont
  {G.}~\bibnamefont {Salomon}}, \bibinfo {author} {\bibfnamefont
  {I.}~\bibnamefont {Bloch}}, \ and\ \bibinfo {author} {\bibfnamefont
  {C.}~\bibnamefont {Gross}},\ }\href {\doibase 10.1038/s41586-019-1463-1}
  {\bibfield  {journal} {\bibinfo  {journal} {Nature}\ }\textbf {\bibinfo
  {volume} {572}},\ \bibinfo {pages} {358} (\bibinfo {year}
  {2019})}\BibitemShut {NoStop}%
\bibitem [{\citenamefont {Gross}\ and\ \citenamefont
  {Bakr}(2021)}]{gross2021quantum}%
  \BibitemOpen
  \bibfield  {author} {\bibinfo {author} {\bibfnamefont {C.}~\bibnamefont
  {Gross}}\ and\ \bibinfo {author} {\bibfnamefont {W.~S.}\ \bibnamefont
  {Bakr}},\ }\href {\doibase 10.1038/s41567-021-01370-5} {\bibfield  {journal}
  {\bibinfo  {journal} {Nat. Phys.}\ }\textbf {\bibinfo {volume} {17}},\
  \bibinfo {pages} {1316} (\bibinfo {year} {2021})}\BibitemShut {NoStop}%
\bibitem [{\citenamefont {Sohmen}\ \emph {et~al.}(2023)\citenamefont {Sohmen},
  \citenamefont {Mark}, \citenamefont {Greiner},\ and\ \citenamefont
  {Ferlaino}}]{Sohmen2023}%
  \BibitemOpen
  \bibfield  {author} {\bibinfo {author} {\bibfnamefont {M.}~\bibnamefont
  {Sohmen}}, \bibinfo {author} {\bibfnamefont {M.~J.}\ \bibnamefont {Mark}},
  \bibinfo {author} {\bibfnamefont {M.}~\bibnamefont {Greiner}}, \ and\
  \bibinfo {author} {\bibfnamefont {F.}~\bibnamefont {Ferlaino}},\ }\href
  {\doibase 10.21468/SciPostPhys.15.5.182} {\bibfield  {journal} {\bibinfo
  {journal} {SciPost Phys.}\ }\textbf {\bibinfo {volume} {15}},\ \bibinfo
  {pages} {182} (\bibinfo {year} {2023})}\BibitemShut {NoStop}%
\bibitem [{\citenamefont {Endres}\ \emph {et~al.}(2016)\citenamefont {Endres},
  \citenamefont {Bernien}, \citenamefont {Keesling}, \citenamefont {Levine},
  \citenamefont {Anschuetz}, \citenamefont {Krajenbrink}, \citenamefont
  {Senko}, \citenamefont {Vuletic}, \citenamefont {Greiner},\ and\
  \citenamefont {Lukin}}]{Endres2016}%
  \BibitemOpen
  \bibfield  {author} {\bibinfo {author} {\bibfnamefont {M.}~\bibnamefont
  {Endres}}, \bibinfo {author} {\bibfnamefont {H.}~\bibnamefont {Bernien}},
  \bibinfo {author} {\bibfnamefont {A.}~\bibnamefont {Keesling}}, \bibinfo
  {author} {\bibfnamefont {H.}~\bibnamefont {Levine}}, \bibinfo {author}
  {\bibfnamefont {E.~R.}\ \bibnamefont {Anschuetz}}, \bibinfo {author}
  {\bibfnamefont {A.}~\bibnamefont {Krajenbrink}}, \bibinfo {author}
  {\bibfnamefont {C.}~\bibnamefont {Senko}}, \bibinfo {author} {\bibfnamefont
  {V.}~\bibnamefont {Vuletic}}, \bibinfo {author} {\bibfnamefont
  {M.}~\bibnamefont {Greiner}}, \ and\ \bibinfo {author} {\bibfnamefont
  {M.~D.}\ \bibnamefont {Lukin}},\ }\href {\doibase 10.1126/science.aah3752}
  {\bibfield  {journal} {\bibinfo  {journal} {Science}\ }\textbf {\bibinfo
  {volume} {354}},\ \bibinfo {pages} {1024} (\bibinfo {year}
  {2016})}\BibitemShut {NoStop}%
\bibitem [{\citenamefont {Browaeys}\ and\ \citenamefont
  {Lahaye}(2020)}]{Browaeys2020}%
  \BibitemOpen
  \bibfield  {author} {\bibinfo {author} {\bibfnamefont {A.}~\bibnamefont
  {Browaeys}}\ and\ \bibinfo {author} {\bibfnamefont {T.}~\bibnamefont
  {Lahaye}},\ }\href {\doibase 10.1038/s41567-019-0733-z} {\bibfield  {journal}
  {\bibinfo  {journal} {Nat. Phys.}\ }\textbf {\bibinfo {volume} {16}},\
  \bibinfo {pages} {132} (\bibinfo {year} {2020})}\BibitemShut {NoStop}%
\bibitem [{\citenamefont {Kaufman}\ and\ \citenamefont
  {Ni}(2021)}]{kaufman2021quantum}%
  \BibitemOpen
  \bibfield  {author} {\bibinfo {author} {\bibfnamefont {A.~M.}\ \bibnamefont
  {Kaufman}}\ and\ \bibinfo {author} {\bibfnamefont {K.-K.}\ \bibnamefont
  {Ni}},\ }\href {\doibase 10.1038/s41567-021-01357-2} {\bibfield  {journal}
  {\bibinfo  {journal} {Nat. Phys.}\ }\textbf {\bibinfo {volume} {17}},\
  \bibinfo {pages} {1324} (\bibinfo {year} {2021})}\BibitemShut {NoStop}%
\bibitem [{\citenamefont {Klein}\ \emph {et~al.}(2019)\citenamefont {Klein},
  \citenamefont {Lorke}, \citenamefont {Florian}, \citenamefont {Sigger},
  \citenamefont {Sigl}, \citenamefont {Rey}, \citenamefont {Wierzbowski},
  \citenamefont {Cerne}, \citenamefont {M{\"u}ller}, \citenamefont
  {Mitterreiter}, \citenamefont {Zimmermann}, \citenamefont {Taniguchi},
  \citenamefont {Watanabe}, \citenamefont {Wurstbauer}, \citenamefont
  {Kaniber}, \citenamefont {Knap}, \citenamefont {Schmidt}, \citenamefont
  {Finley},\ and\ \citenamefont {Holleitner}}]{Klein:2019}%
  \BibitemOpen
  \bibfield  {author} {\bibinfo {author} {\bibfnamefont {J.}~\bibnamefont
  {Klein}}, \bibinfo {author} {\bibfnamefont {M.}~\bibnamefont {Lorke}},
  \bibinfo {author} {\bibfnamefont {M.}~\bibnamefont {Florian}}, \bibinfo
  {author} {\bibfnamefont {F.}~\bibnamefont {Sigger}}, \bibinfo {author}
  {\bibfnamefont {L.}~\bibnamefont {Sigl}}, \bibinfo {author} {\bibfnamefont
  {S.}~\bibnamefont {Rey}}, \bibinfo {author} {\bibfnamefont {J.}~\bibnamefont
  {Wierzbowski}}, \bibinfo {author} {\bibfnamefont {J.}~\bibnamefont {Cerne}},
  \bibinfo {author} {\bibfnamefont {K.}~\bibnamefont {M{\"u}ller}}, \bibinfo
  {author} {\bibfnamefont {E.}~\bibnamefont {Mitterreiter}}, \bibinfo {author}
  {\bibfnamefont {P.}~\bibnamefont {Zimmermann}}, \bibinfo {author}
  {\bibfnamefont {T.}~\bibnamefont {Taniguchi}}, \bibinfo {author}
  {\bibfnamefont {K.}~\bibnamefont {Watanabe}}, \bibinfo {author}
  {\bibfnamefont {U.}~\bibnamefont {Wurstbauer}}, \bibinfo {author}
  {\bibfnamefont {M.}~\bibnamefont {Kaniber}}, \bibinfo {author} {\bibfnamefont
  {M.}~\bibnamefont {Knap}}, \bibinfo {author} {\bibfnamefont {R.}~\bibnamefont
  {Schmidt}}, \bibinfo {author} {\bibfnamefont {J.~J.}\ \bibnamefont {Finley}},
  \ and\ \bibinfo {author} {\bibfnamefont {A.~W.}\ \bibnamefont {Holleitner}},\
  }\href {\doibase 10.1038/s41467-019-10632-z} {\bibfield  {journal} {\bibinfo
  {journal} {Nature Commun.}\ }\textbf {\bibinfo {volume} {10}},\ \bibinfo
  {pages} {2755} (\bibinfo {year} {2019})}\BibitemShut {NoStop}%
\bibitem [{\citenamefont {Schacherl}\ \emph {et~al.}(2025)\citenamefont
  {Schacherl}, \citenamefont {Tagliavini}, \citenamefont {Kaufmann-Heimeshoff},
  \citenamefont {G{\"o}ttlicher}, \citenamefont {Mazzanti}, \citenamefont
  {Popa}, \citenamefont {Walter}, \citenamefont {Pruessmann}, \citenamefont
  {Vollmer}, \citenamefont {Beck}, \citenamefont {Ekanayake}, \citenamefont
  {Branson}, \citenamefont {Neill}, \citenamefont {Fellhauer}, \citenamefont
  {Reitz}, \citenamefont {Schild}, \citenamefont {Brager}, \citenamefont
  {Cahill}, \citenamefont {Windorff}, \citenamefont {Sittel}, \citenamefont
  {Ramanantoanina}, \citenamefont {Haverkort},\ and\ \citenamefont
  {Vitova}}]{Schacherl2025}%
  \BibitemOpen
  \bibfield  {author} {\bibinfo {author} {\bibfnamefont {B.}~\bibnamefont
  {Schacherl}}, \bibinfo {author} {\bibfnamefont {M.}~\bibnamefont
  {Tagliavini}}, \bibinfo {author} {\bibfnamefont {H.}~\bibnamefont
  {Kaufmann-Heimeshoff}}, \bibinfo {author} {\bibfnamefont {J.}~\bibnamefont
  {G{\"o}ttlicher}}, \bibinfo {author} {\bibfnamefont {M.}~\bibnamefont
  {Mazzanti}}, \bibinfo {author} {\bibfnamefont {K.}~\bibnamefont {Popa}},
  \bibinfo {author} {\bibfnamefont {O.}~\bibnamefont {Walter}}, \bibinfo
  {author} {\bibfnamefont {T.}~\bibnamefont {Pruessmann}}, \bibinfo {author}
  {\bibfnamefont {C.}~\bibnamefont {Vollmer}}, \bibinfo {author} {\bibfnamefont
  {A.}~\bibnamefont {Beck}}, \bibinfo {author} {\bibfnamefont {R.~S.~K.}\
  \bibnamefont {Ekanayake}}, \bibinfo {author} {\bibfnamefont {J.~A.}\
  \bibnamefont {Branson}}, \bibinfo {author} {\bibfnamefont {T.}~\bibnamefont
  {Neill}}, \bibinfo {author} {\bibfnamefont {D.}~\bibnamefont {Fellhauer}},
  \bibinfo {author} {\bibfnamefont {C.}~\bibnamefont {Reitz}}, \bibinfo
  {author} {\bibfnamefont {D.}~\bibnamefont {Schild}}, \bibinfo {author}
  {\bibfnamefont {D.}~\bibnamefont {Brager}}, \bibinfo {author} {\bibfnamefont
  {C.}~\bibnamefont {Cahill}}, \bibinfo {author} {\bibfnamefont
  {C.}~\bibnamefont {Windorff}}, \bibinfo {author} {\bibfnamefont
  {T.}~\bibnamefont {Sittel}}, \bibinfo {author} {\bibfnamefont
  {H.}~\bibnamefont {Ramanantoanina}}, \bibinfo {author} {\bibfnamefont
  {M.~W.}\ \bibnamefont {Haverkort}}, \ and\ \bibinfo {author} {\bibfnamefont
  {T.}~\bibnamefont {Vitova}},\ }\href {\doibase 10.1038/s41467-024-54574-7}
  {\bibfield  {journal} {\bibinfo  {journal} {Nat. Comm.}\ }\textbf {\bibinfo
  {volume} {16}},\ \bibinfo {pages} {1221} (\bibinfo {year}
  {2025})}\BibitemShut {NoStop}%
\bibitem [{\citenamefont {Anderson}(1967)}]{anderson1967}%
  \BibitemOpen
  \bibfield  {author} {\bibinfo {author} {\bibfnamefont {P.~W.}\ \bibnamefont
  {Anderson}},\ }\href {\doibase 10.1103/PhysRevLett.18.1049} {\bibfield
  {journal} {\bibinfo  {journal} {Phys. Rev. Lett.}\ }\textbf {\bibinfo
  {volume} {18}},\ \bibinfo {pages} {1049} (\bibinfo {year}
  {1967})}\BibitemShut {NoStop}%
\bibitem [{\citenamefont {Wang}(2023)}]{Wang2023_FDA}%
  \BibitemOpen
  \bibfield  {author} {\bibinfo {author} {\bibfnamefont {J.}~\bibnamefont
  {Wang}},\ }\href {\doibase 10.1007/s43673-023-00092-5} {\bibfield  {journal}
  {\bibinfo  {journal} {AAPPS Bull.}\ }\textbf {\bibinfo {volume} {33}},\
  \bibinfo {pages} {20} (\bibinfo {year} {2023})}\BibitemShut {NoStop}%
\bibitem [{\citenamefont {{Levitov}}\ and\ \citenamefont
  {{Lesovik}}(1993)}]{levitov}%
  \BibitemOpen
  \bibfield  {author} {\bibinfo {author} {\bibfnamefont {L.~S.}\ \bibnamefont
  {{Levitov}}}\ and\ \bibinfo {author} {\bibfnamefont {G.~B.}\ \bibnamefont
  {{Lesovik}}},\ }\href@noop {} {\bibfield  {journal} {\bibinfo  {journal}
  {ZhETF Pisma Redaktsiiu}\ }\textbf {\bibinfo {volume} {58}},\ \bibinfo
  {pages} {225} (\bibinfo {year} {1993})}\BibitemShut {NoStop}%
\bibitem [{\citenamefont {Levitov}\ \emph {et~al.}(1996)\citenamefont
  {Levitov}, \citenamefont {Lee},\ and\ \citenamefont
  {Lesovik}}]{Levitov1996ElectronCS}%
  \BibitemOpen
  \bibfield  {author} {\bibinfo {author} {\bibfnamefont {L.~S.}\ \bibnamefont
  {Levitov}}, \bibinfo {author} {\bibfnamefont {H.-W.}\ \bibnamefont {Lee}}, \
  and\ \bibinfo {author} {\bibfnamefont {G.~B.}\ \bibnamefont {Lesovik}},\
  }\href@noop {} {\bibfield  {journal} {\bibinfo  {journal} {Journal of
  Mathematical Physics}\ }\textbf {\bibinfo {volume} {37}},\ \bibinfo {pages}
  {4845} (\bibinfo {year} {1996})}\BibitemShut {NoStop}%
\bibitem [{\citenamefont {Klich}(2003)}]{Klich2003}%
  \BibitemOpen
  \bibfield  {author} {\bibinfo {author} {\bibfnamefont {I.}~\bibnamefont
  {Klich}},\ }\enquote {\bibinfo {title} {{An Elementary Derivation of
  Levitov's Formula}},}\ in\ \href {\doibase 10.1007/978-94-010-0089-5_19}
  {\emph {\bibinfo {booktitle} {Quantum Noise in Mesoscopic Physics}}},\
  \bibinfo {editor} {edited by\ \bibinfo {editor} {\bibfnamefont {Y.~V.}\
  \bibnamefont {Nazarov}}}\ (\bibinfo  {publisher} {Springer Netherlands},\
  \bibinfo {address} {Dordrecht},\ \bibinfo {year} {2003})\ pp.\ \bibinfo
  {pages} {397--402}\BibitemShut {NoStop}%
\bibitem [{Sup()}]{Supplementary}%
  \BibitemOpen
  \href@noop {} {\ }\bibinfo {note} {{See Supplementary Material for details on
  the numerical implementation, the variational treatment of mobile impurities
  and the quasiparticle nature of the DFP; the Supplementary Material includes
  Refs.
  \cite{levitov,Levitov1996ElectronCS,Klich2003,MG_FDA,chevy2006universal,Scazza:2022bez}}}\BibitemShut
  {NoStop}%
\bibitem [{\citenamefont {Winkler}\ \emph {et~al.}(2006)\citenamefont
  {Winkler}, \citenamefont {Thalhammer}, \citenamefont {Lang}, \citenamefont
  {Grimm}, \citenamefont {Hecker~Denschlag}, \citenamefont {Daley},
  \citenamefont {Kantian}, \citenamefont {Büchler},\ and\ \citenamefont
  {Zoller}}]{Winkler2006}%
  \BibitemOpen
  \bibfield  {author} {\bibinfo {author} {\bibfnamefont {K.}~\bibnamefont
  {Winkler}}, \bibinfo {author} {\bibfnamefont {G.}~\bibnamefont {Thalhammer}},
  \bibinfo {author} {\bibfnamefont {F.}~\bibnamefont {Lang}}, \bibinfo {author}
  {\bibfnamefont {R.}~\bibnamefont {Grimm}}, \bibinfo {author} {\bibfnamefont
  {J.}~\bibnamefont {Hecker~Denschlag}}, \bibinfo {author} {\bibfnamefont
  {A.~J.}\ \bibnamefont {Daley}}, \bibinfo {author} {\bibfnamefont
  {A.}~\bibnamefont {Kantian}}, \bibinfo {author} {\bibfnamefont {H.~P.}\
  \bibnamefont {Büchler}}, \ and\ \bibinfo {author} {\bibfnamefont
  {P.}~\bibnamefont {Zoller}},\ }\href {\doibase 10.1038/nature04918}
  {\bibfield  {journal} {\bibinfo  {journal} {Nature}\ }\textbf {\bibinfo
  {volume} {441}},\ \bibinfo {pages} {853–856} (\bibinfo {year}
  {2006})}\BibitemShut {NoStop}%
\bibitem [{\citenamefont {Chevy}(2006)}]{chevy2006universal}%
  \BibitemOpen
  \bibfield  {author} {\bibinfo {author} {\bibfnamefont {F.}~\bibnamefont
  {Chevy}},\ }\href {\doibase 10.1103/PhysRevA.74.063628} {\bibfield  {journal}
  {\bibinfo  {journal} {Phys. Rev. A}\ }\textbf {\bibinfo {volume} {74}},\
  \bibinfo {pages} {063628} (\bibinfo {year} {2006})}\BibitemShut {NoStop}%
\bibitem [{\citenamefont {Parish}\ and\ \citenamefont
  {Levinsen}(2013)}]{Parish2013}%
  \BibitemOpen
  \bibfield  {author} {\bibinfo {author} {\bibfnamefont {M.~M.}\ \bibnamefont
  {Parish}}\ and\ \bibinfo {author} {\bibfnamefont {J.}~\bibnamefont
  {Levinsen}},\ }\href {\doibase 10.1103/PhysRevA.87.033616} {\bibfield
  {journal} {\bibinfo  {journal} {Phys. Rev. A}\ }\textbf {\bibinfo {volume}
  {87}},\ \bibinfo {pages} {033616} (\bibinfo {year} {2013})}\BibitemShut
  {NoStop}%
\bibitem [{\citenamefont {Chen}\ \emph {et~al.}(2025)\citenamefont {Chen},
  \citenamefont {Dizer}, \citenamefont {Rodr{\'\i}{\-}guez},\ and\
  \citenamefont {Schmidt}}]{Chen:2025gdx}%
  \BibitemOpen
  \bibfield  {author} {\bibinfo {author} {\bibfnamefont {X.}~\bibnamefont
  {Chen}}, \bibinfo {author} {\bibfnamefont {E.}~\bibnamefont {Dizer}},
  \bibinfo {author} {\bibfnamefont {E.~R.}\ \bibnamefont {Rodr{\'\i}{\-}guez}},
  \ and\ \bibinfo {author} {\bibfnamefont {R.}~\bibnamefont {Schmidt}},\ }\href
  {\doibase 10.1103/h2f7-dhjh} {\bibfield  {journal} {\bibinfo  {journal}
  {Phys. Rev. Lett.}\ }\textbf {\bibinfo {volume} {135}},\ \bibinfo {pages}
  {193401} (\bibinfo {year} {2025})}\BibitemShut {NoStop}%
\bibitem [{\citenamefont {Adlong}\ \emph {et~al.}(2025)\citenamefont {Adlong},
  \citenamefont {Dizer}, \citenamefont {Schmidt}, \citenamefont {Imamoglu},\
  and\ \citenamefont {Christianen}}]{Adlong:2025}%
  \BibitemOpen
  \bibfield  {author} {\bibinfo {author} {\bibfnamefont {H.~S.}\ \bibnamefont
  {Adlong}}, \bibinfo {author} {\bibfnamefont {E.}~\bibnamefont {Dizer}},
  \bibinfo {author} {\bibfnamefont {R.}~\bibnamefont {Schmidt}}, \bibinfo
  {author} {\bibfnamefont {A.}~\bibnamefont {Imamoglu}}, \ and\ \bibinfo
  {author} {\bibfnamefont {A.}~\bibnamefont {Christianen}},\ }\href@noop {} {\
  (\bibinfo {year} {2025})},\ \Eprint {http://arxiv.org/abs/2512.16651}
  {arXiv:2512.16651 [cond-mat.str-el]} \BibitemShut {NoStop}%
\bibitem [{\citenamefont {Wang}\ \emph
  {et~al.}(2022{\natexlab{a}})\citenamefont {Wang}, \citenamefont {Liu},\ and\
  \citenamefont {Hu}}]{Wang:2022BCSPRA}%
  \BibitemOpen
  \bibfield  {author} {\bibinfo {author} {\bibfnamefont {J.}~\bibnamefont
  {Wang}}, \bibinfo {author} {\bibfnamefont {X.-J.}\ \bibnamefont {Liu}}, \
  and\ \bibinfo {author} {\bibfnamefont {H.}~\bibnamefont {Hu}},\ }\href
  {\doibase 10.1103/PhysRevA.105.043320} {\bibfield  {journal} {\bibinfo
  {journal} {Phys. Rev. A}\ }\textbf {\bibinfo {volume} {105}},\ \bibinfo
  {pages} {043320} (\bibinfo {year} {2022}{\natexlab{a}})}\BibitemShut
  {NoStop}%
\bibitem [{\citenamefont {Wang}\ \emph
  {et~al.}(2022{\natexlab{b}})\citenamefont {Wang}, \citenamefont {Liu},\ and\
  \citenamefont {Hu}}]{Wang:2022BCSPRL}%
  \BibitemOpen
  \bibfield  {author} {\bibinfo {author} {\bibfnamefont {J.}~\bibnamefont
  {Wang}}, \bibinfo {author} {\bibfnamefont {X.-J.}\ \bibnamefont {Liu}}, \
  and\ \bibinfo {author} {\bibfnamefont {H.}~\bibnamefont {Hu}},\ }\href
  {\doibase 10.1103/PhysRevLett.128.175301} {\bibfield  {journal} {\bibinfo
  {journal} {Phys. Rev. Lett.}\ }\textbf {\bibinfo {volume} {128}},\ \bibinfo
  {pages} {175301} (\bibinfo {year} {2022}{\natexlab{b}})}\BibitemShut
  {NoStop}%
\bibitem [{\citenamefont {Rodr{\'\i}{\-}guez}\ \emph
  {et~al.}(2025)\citenamefont {Rodr{\'\i}{\-}guez}, \citenamefont {Gievers},\
  and\ \citenamefont {Schmidt}}]{Rodriguez:2025}%
  \BibitemOpen
  \bibfield  {author} {\bibinfo {author} {\bibfnamefont {E.~R.}\ \bibnamefont
  {Rodr{\'\i}{\-}guez}}, \bibinfo {author} {\bibfnamefont {M.}~\bibnamefont
  {Gievers}}, \ and\ \bibinfo {author} {\bibfnamefont {R.}~\bibnamefont
  {Schmidt}},\ }\href {https://arxiv.org/abs/2511.19191} {\bibfield  {journal}
  {\bibinfo  {journal} {arXiv:2511.19191}\ } (\bibinfo {year}
  {2025})}\BibitemShut {NoStop}%
\bibitem [{\citenamefont {Thomas}\ and\ \citenamefont
  {Hopfield}(1961)}]{Hopfield1961}%
  \BibitemOpen
  \bibfield  {author} {\bibinfo {author} {\bibfnamefont {D.~G.}\ \bibnamefont
  {Thomas}}\ and\ \bibinfo {author} {\bibfnamefont {J.~J.}\ \bibnamefont
  {Hopfield}},\ }\href {\doibase 10.1103/PhysRevLett.7.316} {\bibfield
  {journal} {\bibinfo  {journal} {Phys. Rev. Lett.}\ }\textbf {\bibinfo
  {volume} {7}},\ \bibinfo {pages} {316} (\bibinfo {year} {1961})}\BibitemShut
  {NoStop}%
\bibitem [{\citenamefont {Scazza}\ \emph {et~al.}(2022)\citenamefont {Scazza},
  \citenamefont {Zaccanti}, \citenamefont {Massignan}, \citenamefont {Parish},\
  and\ \citenamefont {Levinsen}}]{Scazza:2022bez}%
  \BibitemOpen
  \bibfield  {author} {\bibinfo {author} {\bibfnamefont {F.}~\bibnamefont
  {Scazza}}, \bibinfo {author} {\bibfnamefont {M.}~\bibnamefont {Zaccanti}},
  \bibinfo {author} {\bibfnamefont {P.}~\bibnamefont {Massignan}}, \bibinfo
  {author} {\bibfnamefont {M.~M.}\ \bibnamefont {Parish}}, \ and\ \bibinfo
  {author} {\bibfnamefont {J.}~\bibnamefont {Levinsen}},\ }\href {\doibase
  10.3390/atoms10020055} {\bibfield  {journal} {\bibinfo  {journal} {Atoms}\
  }\textbf {\bibinfo {volume} {10}},\ \bibinfo {pages} {55} (\bibinfo {year}
  {2022})}\BibitemShut {NoStop}%
\bibitem [{\citenamefont {De~Santis}\ \emph {et~al.}(2025)\citenamefont
  {De~Santis}, \citenamefont {Salvador}, \citenamefont {Bazhan}, \citenamefont
  {Erne}, \citenamefont {Pr{\"u}fer}, \citenamefont {Guarcello}, \citenamefont
  {Valenti}, \citenamefont {Schmiedmayer},\ and\ \citenamefont
  {Demler}}]{DeSantis:2025lgp}%
  \BibitemOpen
  \bibfield  {author} {\bibinfo {author} {\bibfnamefont {D.}~\bibnamefont
  {De~Santis}}, \bibinfo {author} {\bibfnamefont {A.~G.}\ \bibnamefont
  {Salvador}}, \bibinfo {author} {\bibfnamefont {N.}~\bibnamefont {Bazhan}},
  \bibinfo {author} {\bibfnamefont {S.}~\bibnamefont {Erne}}, \bibinfo {author}
  {\bibfnamefont {M.}~\bibnamefont {Pr{\"u}fer}}, \bibinfo {author}
  {\bibfnamefont {C.}~\bibnamefont {Guarcello}}, \bibinfo {author}
  {\bibfnamefont {D.}~\bibnamefont {Valenti}}, \bibinfo {author} {\bibfnamefont
  {J.}~\bibnamefont {Schmiedmayer}}, \ and\ \bibinfo {author} {\bibfnamefont
  {E.}~\bibnamefont {Demler}},\ }\href@noop {} {\  (\bibinfo {year} {2025})},\
  \Eprint {http://arxiv.org/abs/2509.25147} {arXiv:2509.25147
  [cond-mat.quant-gas]} \BibitemShut {NoStop}%
\bibitem [{\citenamefont {Salvador}\ \emph {et~al.}(2025)\citenamefont
  {Salvador}, \citenamefont {Morera}, \citenamefont {Michael}, \citenamefont
  {Dolgirev}, \citenamefont {Pavicevic}, \citenamefont {Liu}, \citenamefont
  {Cavalleri},\ and\ \citenamefont {Demler}}]{Salvador:2025bkh}%
  \BibitemOpen
  \bibfield  {author} {\bibinfo {author} {\bibfnamefont {A.~G.}\ \bibnamefont
  {Salvador}}, \bibinfo {author} {\bibfnamefont {I.}~\bibnamefont {Morera}},
  \bibinfo {author} {\bibfnamefont {M.~H.}\ \bibnamefont {Michael}}, \bibinfo
  {author} {\bibfnamefont {P.~E.}\ \bibnamefont {Dolgirev}}, \bibinfo {author}
  {\bibfnamefont {D.}~\bibnamefont {Pavicevic}}, \bibinfo {author}
  {\bibfnamefont {A.}~\bibnamefont {Liu}}, \bibinfo {author} {\bibfnamefont
  {A.}~\bibnamefont {Cavalleri}}, \ and\ \bibinfo {author} {\bibfnamefont
  {E.}~\bibnamefont {Demler}},\ }\href@noop {} {\  (\bibinfo {year} {2025})},\
  \Eprint {http://arxiv.org/abs/2501.16856} {arXiv:2501.16856
  [cond-mat.str-el]} \BibitemShut {NoStop}%
\bibitem [{\citenamefont {Lisi}\ \emph {et~al.}(2021)\citenamefont {Lisi},
  \citenamefont {Lu}, \citenamefont {Benschop}, \citenamefont {de~Jong},
  \citenamefont {Stepanov}, \citenamefont {Duran}, \citenamefont {Margot},
  \citenamefont {Cucchi}, \citenamefont {Cappelli}, \citenamefont {Hunter}
  \emph {et~al.}}]{lisi2021observation}%
  \BibitemOpen
  \bibfield  {author} {\bibinfo {author} {\bibfnamefont {S.}~\bibnamefont
  {Lisi}}, \bibinfo {author} {\bibfnamefont {X.}~\bibnamefont {Lu}}, \bibinfo
  {author} {\bibfnamefont {T.}~\bibnamefont {Benschop}}, \bibinfo {author}
  {\bibfnamefont {T.~A.}\ \bibnamefont {de~Jong}}, \bibinfo {author}
  {\bibfnamefont {P.}~\bibnamefont {Stepanov}}, \bibinfo {author}
  {\bibfnamefont {J.~R.}\ \bibnamefont {Duran}}, \bibinfo {author}
  {\bibfnamefont {F.}~\bibnamefont {Margot}}, \bibinfo {author} {\bibfnamefont
  {I.}~\bibnamefont {Cucchi}}, \bibinfo {author} {\bibfnamefont
  {E.}~\bibnamefont {Cappelli}}, \bibinfo {author} {\bibfnamefont
  {A.}~\bibnamefont {Hunter}},  \emph {et~al.},\ }\href {\doibase
  10.1038/s41567-020-01041-x} {\bibfield  {journal} {\bibinfo  {journal} {Nat.
  Phys.}\ }\textbf {\bibinfo {volume} {17}},\ \bibinfo {pages} {189} (\bibinfo
  {year} {2021})}\BibitemShut {NoStop}%
\bibitem [{\citenamefont {Popert}\ \emph {et~al.}(2022)\citenamefont {Popert},
  \citenamefont {Shimazaki}, \citenamefont {Kroner}, \citenamefont {Watanabe},
  \citenamefont {Taniguchi}, \citenamefont {Imamoğlu},\ and\ \citenamefont
  {Smoleński}}]{popert2021optical}%
  \BibitemOpen
  \bibfield  {author} {\bibinfo {author} {\bibfnamefont {A.}~\bibnamefont
  {Popert}}, \bibinfo {author} {\bibfnamefont {Y.}~\bibnamefont {Shimazaki}},
  \bibinfo {author} {\bibfnamefont {M.}~\bibnamefont {Kroner}}, \bibinfo
  {author} {\bibfnamefont {K.}~\bibnamefont {Watanabe}}, \bibinfo {author}
  {\bibfnamefont {T.}~\bibnamefont {Taniguchi}}, \bibinfo {author}
  {\bibfnamefont {A.}~\bibnamefont {Imamoğlu}}, \ and\ \bibinfo {author}
  {\bibfnamefont {T.}~\bibnamefont {Smoleński}},\ }\href {\doibase
  10.1021/acs.nanolett.2c02000} {\bibfield  {journal} {\bibinfo  {journal}
  {Nano Letters}\ }\textbf {\bibinfo {volume} {22}},\ \bibinfo {pages} {7363}
  (\bibinfo {year} {2022})}\BibitemShut {NoStop}%
\bibitem [{\citenamefont {Rosenzweig}\ \emph {et~al.}(2020)\citenamefont
  {Rosenzweig}, \citenamefont {Karakachian}, \citenamefont {Marchenko},
  \citenamefont {K\"uster},\ and\ \citenamefont
  {Starke}}]{rosenzweig2020overdoping}%
  \BibitemOpen
  \bibfield  {author} {\bibinfo {author} {\bibfnamefont {P.}~\bibnamefont
  {Rosenzweig}}, \bibinfo {author} {\bibfnamefont {H.}~\bibnamefont
  {Karakachian}}, \bibinfo {author} {\bibfnamefont {D.}~\bibnamefont
  {Marchenko}}, \bibinfo {author} {\bibfnamefont {K.}~\bibnamefont {K\"uster}},
  \ and\ \bibinfo {author} {\bibfnamefont {U.}~\bibnamefont {Starke}},\ }\href
  {\doibase 10.1103/PhysRevLett.125.176403} {\bibfield  {journal} {\bibinfo
  {journal} {Phys. Rev. Lett.}\ }\textbf {\bibinfo {volume} {125}},\ \bibinfo
  {pages} {176403} (\bibinfo {year} {2020})}\BibitemShut {NoStop}%
\bibitem [{\citenamefont {Ament}\ \emph {et~al.}(2011)\citenamefont {Ament},
  \citenamefont {van Veenendaal}, \citenamefont {Devereaux}, \citenamefont
  {Hill},\ and\ \citenamefont {van~den Brink}}]{ament2011xrayreview}%
  \BibitemOpen
  \bibfield  {author} {\bibinfo {author} {\bibfnamefont {L.~J.~P.}\
  \bibnamefont {Ament}}, \bibinfo {author} {\bibfnamefont {M.}~\bibnamefont
  {van Veenendaal}}, \bibinfo {author} {\bibfnamefont {T.~P.}\ \bibnamefont
  {Devereaux}}, \bibinfo {author} {\bibfnamefont {J.~P.}\ \bibnamefont {Hill}},
  \ and\ \bibinfo {author} {\bibfnamefont {J.}~\bibnamefont {van~den Brink}},\
  }\href {\doibase 10.1103/RevModPhys.83.705} {\bibfield  {journal} {\bibinfo
  {journal} {Rev. Mod. Phys.}\ }\textbf {\bibinfo {volume} {83}},\ \bibinfo
  {pages} {705} (\bibinfo {year} {2011})}\BibitemShut {NoStop}%
\bibitem [{\citenamefont {Ryu}\ and\ \citenamefont
  {Hatsugai}(2002)}]{zeroEnergyStates2002}%
  \BibitemOpen
  \bibfield  {author} {\bibinfo {author} {\bibfnamefont {S.}~\bibnamefont
  {Ryu}}\ and\ \bibinfo {author} {\bibfnamefont {Y.}~\bibnamefont {Hatsugai}},\
  }\href {\doibase 10.1103/PhysRevLett.89.077002} {\bibfield  {journal}
  {\bibinfo  {journal} {Phys. Rev. Lett.}\ }\textbf {\bibinfo {volume} {89}},\
  \bibinfo {pages} {077002} (\bibinfo {year} {2002})}\BibitemShut {NoStop}%
\bibitem [{\citenamefont {Armitage}\ \emph {et~al.}(2018)\citenamefont
  {Armitage}, \citenamefont {Mele},\ and\ \citenamefont
  {Vishwanath}}]{armitage2018weylSemimetals}%
  \BibitemOpen
  \bibfield  {author} {\bibinfo {author} {\bibfnamefont {N.~P.}\ \bibnamefont
  {Armitage}}, \bibinfo {author} {\bibfnamefont {E.~J.}\ \bibnamefont {Mele}},
  \ and\ \bibinfo {author} {\bibfnamefont {A.}~\bibnamefont {Vishwanath}},\
  }\href {\doibase 10.1103/RevModPhys.90.015001} {\bibfield  {journal}
  {\bibinfo  {journal} {Rev. Mod. Phys.}\ }\textbf {\bibinfo {volume} {90}},\
  \bibinfo {pages} {015001} (\bibinfo {year} {2018})}\BibitemShut {NoStop}%
\bibitem [{\citenamefont {Gievers}\ \emph {et~al.}(2024)\citenamefont
  {Gievers}, \citenamefont {Wagner},\ and\ \citenamefont {Schmidt}}]{MG_FDA}%
  \BibitemOpen
  \bibfield  {author} {\bibinfo {author} {\bibfnamefont {M.}~\bibnamefont
  {Gievers}}, \bibinfo {author} {\bibfnamefont {M.}~\bibnamefont {Wagner}}, \
  and\ \bibinfo {author} {\bibfnamefont {R.}~\bibnamefont {Schmidt}},\ }\href
  {\doibase 10.1103/PhysRevLett.132.053401} {\bibfield  {journal} {\bibinfo
  {journal} {Phys. Rev. Lett.}\ }\textbf {\bibinfo {volume} {132}},\ \bibinfo
  {pages} {053401} (\bibinfo {year} {2024})}\BibitemShut {NoStop}%
\end{thebibliography}

\end{document}


\title{Supplementary Material for ``Emergent Fermi polarons in Dirac materials''}

\author{Xin Chen$^{1}$, Eugen Dizer$^{1}$, Rafa\l{} O\l{}dziejewski$^{2}$, \\Kostya S. Novoselov$^{3,4}$, Marton Kan\'asz-Nagy$^{5}$, and Richard Schmidt$^{1}$}

\affiliation{%
    $^1$Institut f\"ur Theoretische Physik, Universit\"at Heidelberg, D-69120 Heidelberg, Germany\\%
    $^2$Centre for Quantum Optical Technologies, Centre of New Technologies, University of Warsaw, Banacha 2c, 02-097 Warsaw, Poland\\
    $^3$Department of Materials Science and Engineering, National University of Singapore, Singapore\\
    $^4$Institute for Functional Intelligent Materials, National University of Singapore, Singapore\\
    $^5$Max-Planck-Institut f\"ur Quantenoptik, Hans-Kopfermann-Str. 1, D-85748 Garching, Germany
}

\maketitle
In this Supplementary Material, we provide detailed derivations and extended analyses supporting the results presented in the main text. First, we outline the Functional Determinant Approach (FDA) applied to the single-layer honeycomb lattice, demonstrating the convergence of the absorption spectra with respect to system size and confirming the robustness of the Dirac-Fermi polaron (DFP) at finite temperatures. We then elaborate on the variational approach, incorporating the full two-band structure of the graphene lattice to derive the complete self-consistency equations for a mobile impurity. Utilizing this multi-band framework, we quantitatively characterize the quasiparticle residues and intrinsic lifetimes of the distinct polaron branches. Finally, we analyze the microscopic origin of the DFP, elucidating its fundamental formation mechanism represented in the analytic structure of the impurity self-energy and the two-body $T$-matrix scattering dynamics.

\section{Functional Determinant Approach}
\label{sec:FDA}

The absorption spectrum of an impurity immersed in a fermionic bath is
determined by the Loschmidt echo~\cite{levitov,Levitov1996ElectronCS,Klich2003,MG_FDA},
which encodes the overlap between the initial and time-evolved many-body
states.  For a bath described by a general single-particle Hamiltonian
$\hat{H}_0 = \sum_{ij} [\mathbf{h}_0]_{ij}\,\hat{c}_i^\dagger \hat{c}_j$ and
a localized impurity potential $\hat{V} = \sum_{ij}
[\mathbf{v}]_{ij}\,\hat{c}_i^\dagger \hat{c}_j$, the Loschmidt echo reduces
exactly to an $N \times N$ determinant~\cite{levitov,Levitov1996ElectronCS,Klich2003,MG_FDA},
\begin{equation}
    S(\tau) = \det\!\left(
        \mathbb{I} - \mathbf{n}_F
        + \mathbf{n}_F\,
          e^{-i(\mathbf{h}_0 + \mathbf{v})\tau}\,
          e^{+i\mathbf{h}_0\tau}
    \right) \,,
    \label{eq:S_FDA}
\end{equation}
where $N$ is the number of lattice sites, $\mathbf{h}_0$ and $\mathbf{v}$ are
the $N\times N$ single-particle matrices of the bath Hamiltonian and the
impurity potential, respectively, and
\begin{equation}
    \mathbf{n}_F = \left(
        \mathbb{I} + e^{(\mathbf{h}_0 - \mu\,\mathbb{I})/T}
    \right)^{-1}
    \label{eq:nF}
\end{equation}
is the Fermi occupation matrix at chemical potential $\mu$ and temperature
$T$.  The FDA thus reduces the exponentially large many-body problem to an
$N\times N$ linear-algebra problem, which is exact for any quadratic bath
Hamiltonian and impurity potential.
 
The absorption spectrum is obtained from the retarded impurity Green's
function $G^R(\omega) = -i\int_0^\infty S(\tau)\,e^{i(\omega+i\eta)\tau}\,
\mathrm{d}\tau$, yielding
\begin{equation}
    A(\omega)
    = -\frac{1}{\pi}\,\mathrm{Im}\,G^R(\omega)
    = \frac{1}{\pi}\,\mathrm{Re}
      \int_0^\infty S(\tau)\,
      e^{i(\omega + i\eta)\tau}\,\mathrm{d}\tau \,,
    \label{eq:A_omega}
\end{equation}
where $\eta > 0$ is a Lorentzian broadening that damps the long-$\tau$ tail and broadens each spectral peak to a Lorentzian of half-width $\eta$.

\section{Single-layer honeycomb lattice}
\label{sec:monolayer}

The single-layer system is described by the Hamiltonian, 
\begin{equation}
    \hat{H} = -t\!\sum_{\langle i,j\rangle}\!\hat{c}_i^\dagger\hat{c}_j
              + \frac{\Delta}{2}\!\sum_i\!\left(\hat{n}^A_i - \hat{n}^B_i\right)
              + U\,\hat{n}^{\alpha}_{i_0} \,,
    \label{eq:H_monolayer_SM}
\end{equation}
where $t$ is the nearest-neighbour hopping amplitude, $\Delta/2$ is the
sublattice mass that assigns on-site energies $+\Delta/2$ to sublattice A and
$-\Delta/2$ to sublattice B, and $U\,\hat{n}^\alpha_{i_0}$ is the impurity potential
at the sublattice $\alpha\in\{A,B\}$ of  unit cell with index $i_0$. Setting $\Delta = 0$ recovers the massless Dirac case
discussed in the main text, while $\Delta$ opens a gap at the charge-neutrality point. 

\begin{figure}[ht!]
\includegraphics[width = 0.35\linewidth]{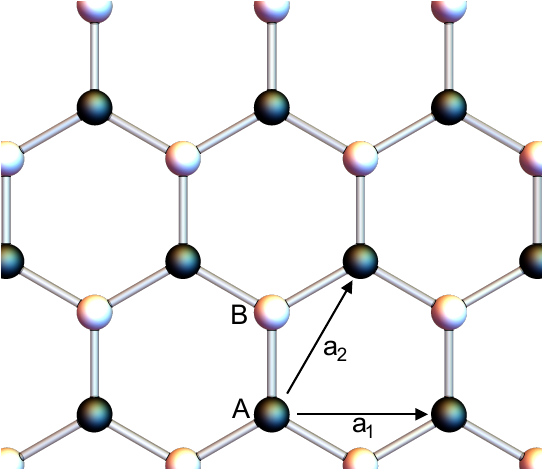}
\caption{ The honeycomb lattice is constructed
with lattice vectors $\mathbf{a}_1 = \sqrt{3}\,\hat{x}$ and $\mathbf{a}_2 =
(\sqrt{3}/2)\,\hat{x} + (3/2)\,\hat{y}$ (in units of the bond length), with
a two-site unit cell and a B-site offset $\bm{\tau}_B = \hat{y}$.  An
$L_x\times L_y$ supercell contains $N = 2L_xL_y$ sites, and boundary
conditions are applied independently along $\mathbf{a}_1$ and $\mathbf{a}_2$. \label{fig:grapheneLatt}}
\end{figure}
 
\section{Finite-size and temperature effects}
\label{sec:temperature}

\begin{figure}[ht!]
\includegraphics[width = \linewidth]{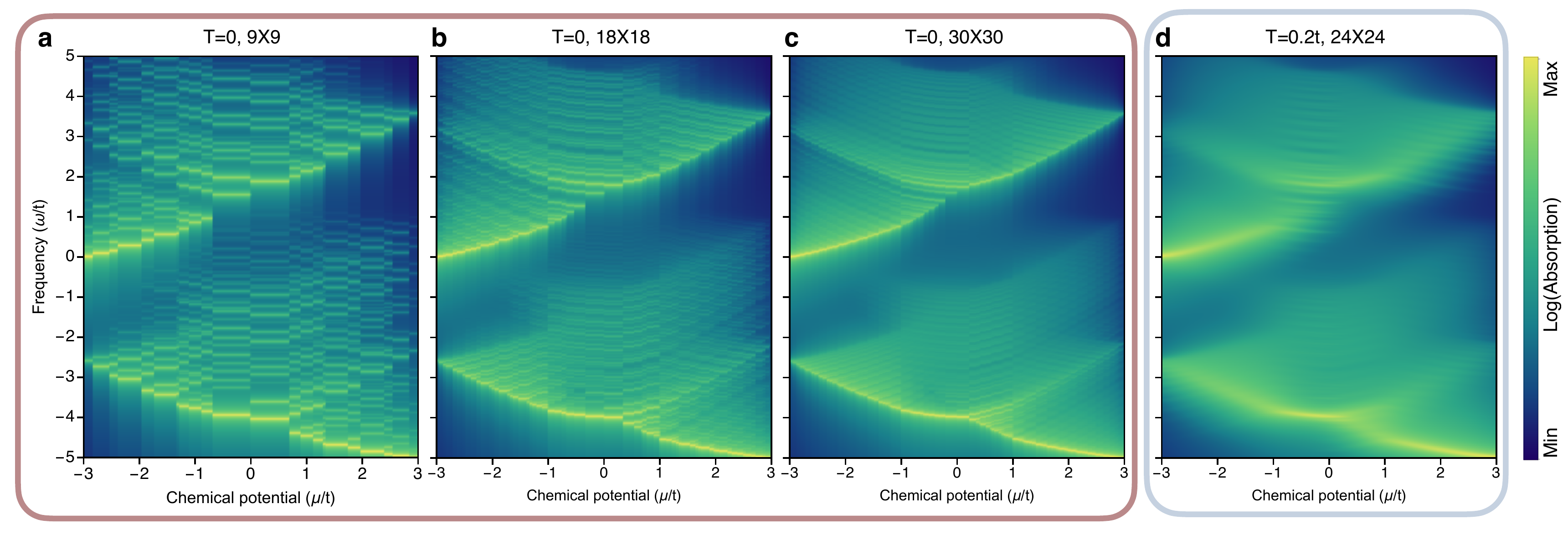}
\caption{FDA absorption spectra for an interaction strength of $U=-5\,t$ with periodic boundary conditions. {\bf a}–{\bf c} Spectra at zero temperature ($T=0$) for progressively larger lattice sizes: {\bf a} $9\times9$, {\bf b} $18\times 18$, and {\bf c} $30\times 30$. {\bf d} The corresponding finite-temperature absorption spectrum evaluated at $T=0.2\,t$ on a $24\times 24$ lattice.\label{fig:fnSize-Temp}}
\end{figure}

A key virtue of the FDA is that it can be applied to finite systems with arbitrary boundary conditions, allowing for the direct assessment of finite-size effects. We have systematically computed the absorption spectrum for system sizes ranging from $9\times9$ ($N=162$) to $30\times 30$ ($N=1800$) under periodic boundary conditions (PBC). The results, shown in \cref{fig:fnSize-Temp}{\bf a}–{\bf c}, demonstrate that the discrete single-particle spectrum converges rapidly to the thermodynamic density of states. For lattices of $18\times 18$ and larger, the positions and widths of all spectral features—including the AP, RP, and DFP branches—are virtually indistinguishable within the artificial broadening $\eta=0.02\,t$ used throughout. 

At small system sizes, the primary qualitative difference between open boundary conditions (OBC) and PBC arises from the zero-energy modes localized at zigzag edges under OBC. By employing PBC, the edgeless torus geometry circumvents this issue, ensuring that the discrete single-particle spectrum converges uniformly to the thermodynamic density of states. While all results presented in the main text are calculated using OBC with $L = 24$, they closely resemble the PBC results discussed here in \cref{fig:fnSize-Temp}{\bf c}. Finally, we note that the minor ripples observed in \cref{fig:fnSize-Temp}{\bf c} originate from the commensurate energy levels intrinsic to the finite-size PBC geometry.

Finite temperature enters the FDA exclusively through the Fermi occupation matrix $\mathbf{n}_F$ in \cref{eq:nF}, which smoothly interpolates between the zero-temperature step function and a broad thermal distribution. The effect of finite temperature on the absorption spectrum is illustrated in \cref{fig:fnSize-Temp}{\bf d} for an interaction strength of $U=-5\,t$ at $T=0.2\,t$. We find that finite temperature primarily broadens the spectral features while leaving the overall structure of the spectrum—particularly the DFP branch—qualitatively intact. The main quantitative effects are: (i) a thermal broadening of the sharp onset edges of the AP and RP, (ii) a reduction in spectral contrast between the polaron peaks and the incoherent background, and (iii) a slight shift of spectral weight from the DFP toward the continuum as thermally excited particle-hole pairs populate the region near the Dirac point. At $T = 0.2\,t$, all three branches remain clearly resolvable, confirming the robustness of the DFP against thermal fluctuations at experimentally relevant temperatures for cold-atom and solid-state platforms.

\newpage
\section{Variational ansatz}
To analyze the details of the microscopic mechanism underlying the absorption features as well as to describe the case of a mobile impurity, we consider a variational approach to the Fermi polaron. Here the state of the system is described by the `Chevy ansatz' \cite{chevy2006universal}:  
\begin{align}
    |\psi\rangle = (\phi_{0,E} d_{\bf Q}^\dagger +\sum_{{\bf k},{\bf q}} \phi_{{\bf k},{\bf q},E}d_{{\bf Q}-{\bf k}+{\bf q}}^\dagger c_{\bf k}^\dagger c_{\bf q})|\text{FS}\rangle,\label{eq:ChevyAnsatzWF}
\end{align}
representing a mobile polaron of total momentum $\bf Q$ with up to one particle-hole excitation. In contrast to the continuum model with unbounded dispersion~\cite{chevy2006universal}, hole scattering cannot be neglected as usually done in the continuum Chevy approach. In principle, the index $\bf Q$ is not enough to account for all degrees of freedom in a multi-band structure; however, we simplify the problem with a single-band Hamiltonian 
\begin{align}
    H_\text{single}=\sum_{\bf k} \epsilon_{\bf k}c_{\bf k}^\dagger c_{\bf k}+\sum_{\bf k} \epsilon_{\bf k}^\text{I} d_{\bf k}^\dagger d_{\bf k}+U\sum_{{\bf P},{\bf k},{\bf k}'}\varphi_{\mathbf{P},\mathbf{k}}^*\varphi_{\mathbf{P},\mathbf{k}'} d^\dagger_{{\bf P}-{\bf k}}d_{{\bf P}-{\bf k}'}c^\dagger_{{\bf k}}c_{{\bf k}'},\label{eq:H_singleband}
\end{align}
to elucidate features of a impurity-particle-hole trimer scattering variational ansatz. For the multiband case with a separable potential, the momentum summation in \cref{eq:H_singleband} generalizes straightforwardly to a joint summation over momentum and band index. The next section extends this framework to a two-band model of impurities and fermions on a graphene lattice, where the interaction is no longer a single separable potential but rather a sum of two separable potentials, reflecting the two inequivalent sublattice sites at which scattering can occur.
In Eq.~\eqref{eq:H_singleband}, $\varphi_{\mathbf{P},\mathbf{k}}$ is the form factor of the separable interaction. Solving the eigenstate equation leads to the self-consistency equations
\begin{subequations}
\begin{align}
    & \eta_{E,{\bf k}} = \sum_{\epsilon_{\bf q}<\mu}\frac{ {U}\,|\varphi_{\mathbf{Q}-\mathbf{k},\mathbf{q}}|^2\,(\chi_{E,{\bf q}} -  \eta_{E,{\bf k}})\,}{E - \epsilon_{{\bf Q}-{\bf k}+{\bf q}}^\text{I} - \epsilon_{\bf k} + \epsilon_{\bf q} -  {U} \, n_0\, +i0^+},\label{eq:chi}\\
    & \chi_{E,{\bf q}} =\varphi_{{\bf Q},{\bf q}} \phi_{0,E} + \sum_{\epsilon_{\bf k}>\mu}\frac{{U}\,|\varphi_{\mathbf{Q}+\mathbf{q},\mathbf{k}}|^2\,( \chi_{E,{\bf q}} - \eta_{E,{\bf k}})\,}{ E- \epsilon_{{\bf Q}-{\bf k}+{\bf q}}^\text{I} - \epsilon_{\bf k} + \epsilon_{\bf q} - {U}\, n_0\, +i0^+ },\label{eq:eta}\\
    &\Big[G_\text{I}^\text{R}\Big]^{-1} =E- \epsilon_{{\bf Q}}^\text{I} + i0^+ - {U} \sum_{\epsilon_{\bf q}<\mu}\frac{\varphi_{\mathbf{Q},\mathbf{q}}^*\,\chi_{E,{\bf q}}}{ \phi_{0,E}},
    \label{eq:SelfEngEq}
\end{align}\label{eq:chevySubEqns}
\end{subequations}
where $U\,n_0=U\,\sum_{\epsilon_{\bf q}<\mu}|\varphi_{{\bf Q},{\bf q}}|^2$ describes the mean-field potential induced by the bath fermions. The terms $\epsilon_{\bf k}$ and $\epsilon^\text{I}_{\bf k}$ are the dispersion energies for bath fermions and impurity. $G^\text{R}_{\text{I}}$ is the retarded Green's function of the single impurity. $\chi_{E,{\bf q}} = \varphi_{{\bf Q},{\bf q}}\phi_{0,E} + \sum_{\epsilon_{\bf k}>\mu}\varphi_{{\bf Q},{\bf k}}\phi_{{\bf k}, {\bf q},E}$, and $\eta_{E,{\bf k}} = \sum_{\epsilon_{\bf q}<\mu}\varphi^*_{{\bf Q},{\bf q}} \phi_{{\bf k}, {\bf q},E}$  can be interpreted as the scattering amplitudes from the non-interacting impurity state to the state of particle-dimer scattering with hole excitation and the state of hole-dimer scattering with particle excitation, respectively.

\section{Variational Ansatz for a Mobile Impurity in Graphene}
To investigate the mobile impurity variational ansatz within a model featuring  multiple bands, we expand the bath operators in the basis of the graphene Bloch functions $|\psi^{u/d}_{\bf k}\rangle=(\begin{array}{cc} \varphi^{u/d}_{{\bf k},a}, & \varphi^{u/d}_{{\bf k},b}\end{array})^T$ for the upper ($u$) and lower ($d$) bands. The Hamiltonian is given by
\begin{equation}
    H = \sum_{{\bf k}, \lambda} \epsilon_{\bf k}^\lambda c^\dagger_{{\bf k}\lambda} c_{{\bf k}\lambda} + \sum_{{\bf k}, \lambda} \epsilon_{\bf k}^{I,\lambda} d^\dagger_{{\bf k}\lambda} d_{{\bf k}\lambda} + H_{\text{int}},
\end{equation}
where $c^\dagger$ ($c$) and $d^\dagger$ ($d$) are the creation (annihilation) operators for bath fermions and the impurity, respectively, and $\lambda \in \{u,d\}$ is the band index.

$H_\text{int}$ describes the short-range on-site interaction between bath fermions and the impurity. Since the interaction is local to each of the two sublattices, $H_\text{int}$ decomposes into a sum of separable terms when expressed in the Bloch basis $|\psi^{u/d}_{\bf k}\rangle$:
\begin{align}
    H_\text{int} = U \sum_{\stackrel{{\bf P},{\bf k}, {\bf k}'}{\lambda_I,\lambda_I',\lambda,\lambda'=u,d}} \Big(\sum_{s={ a},{ b}}(\varphi_{{\bf P}/2-{\bf k}',s}^{\lambda_I'})^* \varphi_{{\bf P}/2-{\bf k},s}^{\lambda_I} (\varphi_{{\bf P}/2+{\bf k}',s}^{\lambda'})^* \varphi_{{\bf P}/2+{\bf k},s}^{\lambda} \Big)d_{{\bf P}/2-{\bf k}',\lambda_I'}^\dagger d_{{\bf P}/2-{\bf k},\lambda_I} c_{{\bf P}/2+{\bf k}',\lambda'}^\dagger c_{{\bf P}/2+{\bf k},\lambda} .
\end{align}
Here, the summation index $s\in \{a,b\}$ denotes the specific sublattice on which the on-site interaction occurs. We employ a generalized Chevy ansatz for the polaron with total momentum ${\bf Q}$:
\begin{equation}
    |\psi_{\bf Q}\rangle =  \sum_{\lambda_I=u,d}\Big( \alpha_0^{\lambda_I} d^\dagger_{{\bf Q},\lambda_I} + \sum_{{E^{\lambda_p}_{\bf k}>\mu,E^{\lambda_h}_{\bf q}<\mu}} \alpha_{{\bf k}{\bf q}}^{\lambda_I,\lambda_p,\lambda_h} d^\dagger_{{\bf Q}-{\bf k}+{\bf q},\lambda_I} c^\dagger_{{\bf k},\lambda_p} c_{{\bf q},\lambda_h} \Big) |\text{FS}\rangle.
\end{equation}

Minimizing the energy expectation value $\langle \psi | H - E | \psi \rangle = 0$ yields a system of coupled linear equations for the variational parameters. Although the multi-band nature of graphene introduces significant algebraic complexity, the resulting self-consistency equations retain the same structure as in the static impurity case. They can be expressed compactly in terms of generalized $T$-matrices for electron-impurity and hole-impurity scattering:
\begin{subequations}
    \begin{align}
    &\boldsymbol{\chi}(\mathbf{Q},E) = \mathbf{T}^e(\mathbf{Q},E) \cdot \left( \mathbf{\Psi}^\mathbf{Q}_0 +  \mathbf{G}(\mathbf{Q},E+i\varepsilon) \cdot \mathbf{T}^h (\mathbf{Q},E)\cdot  \mathbf{G}^\dagger(\mathbf{Q},E-i\varepsilon) \cdot \boldsymbol{\chi}(\mathbf{Q},E) \right),\label{eq:mobChevyChi}\\
    & (E-\epsilon^{I,\lambda_I}_{\bf Q}) \alpha_0^{\lambda_I} = (\varphi_{{\bf Q},s}^{\lambda_I} \varphi_{{\bf q}, s}^{\lambda_h})^*_{(s,{\bf q}, \lambda_h)}\cdot (\boldsymbol{\chi}(\mathbf{Q},E))_{(s,{\bf q},\lambda_h)} \label{eq:mobilChevyImpGF}
    \end{align}
\end{subequations}
where $\boldsymbol{\chi}$ represents the scattering amplitude, and $\mathbf{G}$ contains the propagators connecting the particle and hole scattering sectors.  The explicit definitions for the vector and matrix components appearing in the self-consistency equations are:
\begin{subequations}
    \begin{align}
        &\boldsymbol{\Psi}^\mathbf{Q}_0\equiv \left[\boldsymbol{\Psi}^\mathbf{Q}_0\right]_{(s,{\bf q}, \lambda_h)}=\sum_{\lambda_I=u,d}\alpha_0^{\lambda_I}\varphi_{{\bf Q},s}^{\lambda_I}\varphi_{{\bf q},s}^{\lambda_h},\\
        &\boldsymbol{\chi}\equiv[\boldsymbol{\chi}(\mathbf{Q},E)]_{(s,{\bf q},\lambda_h)} = \sum_{\lambda_I=u,d}\Big(\alpha_0^{\lambda_I} \varphi_{{\bf Q},s}^{\lambda_I}\varphi_{{\bf q},s}^{\lambda_h}+\sum_{{E^{\lambda_p}_{\bf k}>\mu}} {\varphi^{\lambda_I}_{{\bf P}_I,s}\varphi^{\lambda_p}_{{\bf k},s}} {\alpha_{{\bf k}{\bf q}}^{\lambda_I,\lambda_p,\lambda_h}}\Big),\\
        &{\bf G}\equiv\left[{\bf G}(\mathbf{Q},E)\right]_{(s,{\bf q},\lambda_h);(s',{\bf k},\lambda_p)}=\sum_{\lambda_I=u,d} \frac{\varphi^{\lambda_I}_{{\bf P}_I,s}(\varphi^{\lambda_h}_{{\bf q},s}\varphi^{\lambda_I}_{{\bf P}_I,s'}\varphi^{\lambda_p}_{{\bf k},s'})^*}{E-\epsilon^{I,\lambda_I}_{{\bf P}_I}-\epsilon^{\lambda_p}_{{\bf k}}+\epsilon^{\lambda_h}_{{\bf q}}-Un_0}, \\
        &{\bf T}^e\equiv\left[{\bf T}^e(\mathbf{Q},E)\right]_{(s,{\bf q}, \lambda_h);(s',{\bf q}',\lambda_h')}=\Big(\big[\frac{1}{U} \delta_{s,s'}- \sum_{\stackrel{E_{\bf k}^{\lambda_p}>\mu}{\lambda_I=u,d}} \frac{\varphi^{\lambda_I}_{{\bf P}_I,s}\varphi^{\lambda_p}_{{\bf k},s}(\varphi^{\lambda_I}_{{\bf P}_I,s'}\varphi^{\lambda_p}_{{\bf k},s'})^*}{E+i\varepsilon-\epsilon^{I,\lambda_I}_{{\bf P}_I}-\epsilon^{\lambda_p}_{{\bf k}}+\epsilon^{\lambda_h}_{{\bf q}}-Un_0}  \big]_{s,s'}\Big)^{-1}\delta_{{\bf q},{\bf q}'}\delta_{\lambda_h,\lambda_h'},\label{eq:mobChevyTe}\\
        &{\bf T}^h\equiv\left[{\bf T}^h(\mathbf{Q},E)\right]_{(s,{\bf k}, \lambda_p);(s',{\bf k}',\lambda_p')}=\Big(\big[-\frac{1}{U} \delta_{s,s'}- \sum_{\stackrel{E_{\bf q}^{\lambda_h}<\mu}{\lambda_I=u,d}} \frac{\varphi^{\lambda_I}_{{\bf P}_I,s}(\varphi^{\lambda_h}_{{\bf q},s}\varphi^{\lambda_I}_{{\bf P}_I,s'})^*\varphi^{\lambda_h}_{{\bf q},s'}}{E+i\varepsilon-\epsilon^{I,\lambda_I}_{{\bf P}_I}-\epsilon^{\lambda_p}_{{\bf k}}+\epsilon^{\lambda_h}_{{\bf q}}-Un_0}  \big]_{s,s'}\Big)^{-1}\delta_{{\bf k},{\bf k}'}\delta_{\lambda_p,\lambda_p'}.
    \end{align}
\end{subequations}
Here, $\mathbf{P}_I=\mathbf{Q}-\mathbf{k}+\mathbf{q}$ is the impurity momentum required to conserve the total momentum $\mathbf{Q}$, and $\varepsilon>0$ is a positive infinitesimal. $U\,n_0=U\,\sum_{\epsilon_{\mathbf{q}}^\lambda<\mu} (1/2)$ is the mean field energy from the interaction potential $U$ of the impurity. 

We solve these linear equations numerically to extract the impurity self-energy $\boldsymbol{\Sigma}(\mathbf{Q}, E)$ and the spectral function $A(\mathbf{Q}, \omega)$. Notably, due to the independence of $\alpha_0^{\lambda_I}$ for $\lambda_I\in\{u, d\}$, the self-energy $\boldsymbol{\Sigma}$ is a matrix of dimension $2$. The spectral function $A(\mathbf{Q}, E) = (1/\pi) \text{Re} \int_0^\infty dt\, \langle\text{FS}|d_{\mathbf{Q},d}(t)d^\dagger_{\mathbf{Q},d}(0)|\text{FS}\rangle \exp(iEt-\varepsilon t)$, describes the response spectrum of the impurity in the lower band ($\lambda_I = d$),
\begin{subequations}
    \begin{align}
    &\left[\boldsymbol{\Sigma}(\mathbf{Q},E)\right]_{\lambda_I, \lambda_I'}=(\varphi_{{\bf Q},s}^{\lambda_I} \varphi_{{\bf q}, s}^{\lambda_h})^*_{(s,{\bf q}, \lambda_h)}\cdot \frac{\delta}{\delta \alpha_0^{\lambda_I'
}}(\boldsymbol{\chi}(\mathbf{Q},E))_{(s,\mathbf{q},\lambda_h)},\label{eq:slfeng}\\
&A(\mathbf{Q},E) = -\frac{1}{\pi}\text{Im}\left[\text{Tr}\left(P_d ( (E+i\varepsilon) \mathbf{I} - \boldsymbol{\epsilon}^I_{\mathbf{Q}}-\boldsymbol{\Sigma}(\mathbf{Q},E))^{-1}\right)\right]\label{eq:spec}
\end{align}
\end{subequations}
where $\mathbf{I}$ is the two-dimensional identity matrix, $\boldsymbol{\epsilon}^I_{\mathbf{Q}}=\text{diag}(\epsilon^{I,u}_{\mathbf{Q}}, \epsilon^{I,d}_{\mathbf{Q}})$ is the diagonal matrix of impurity dispersion energies, and $P_d$ projects onto the lower band $d$.

\section{Quasiparticle Weight and Lifetime}

\begin{figure}[ht!]
\includegraphics[width = \linewidth]{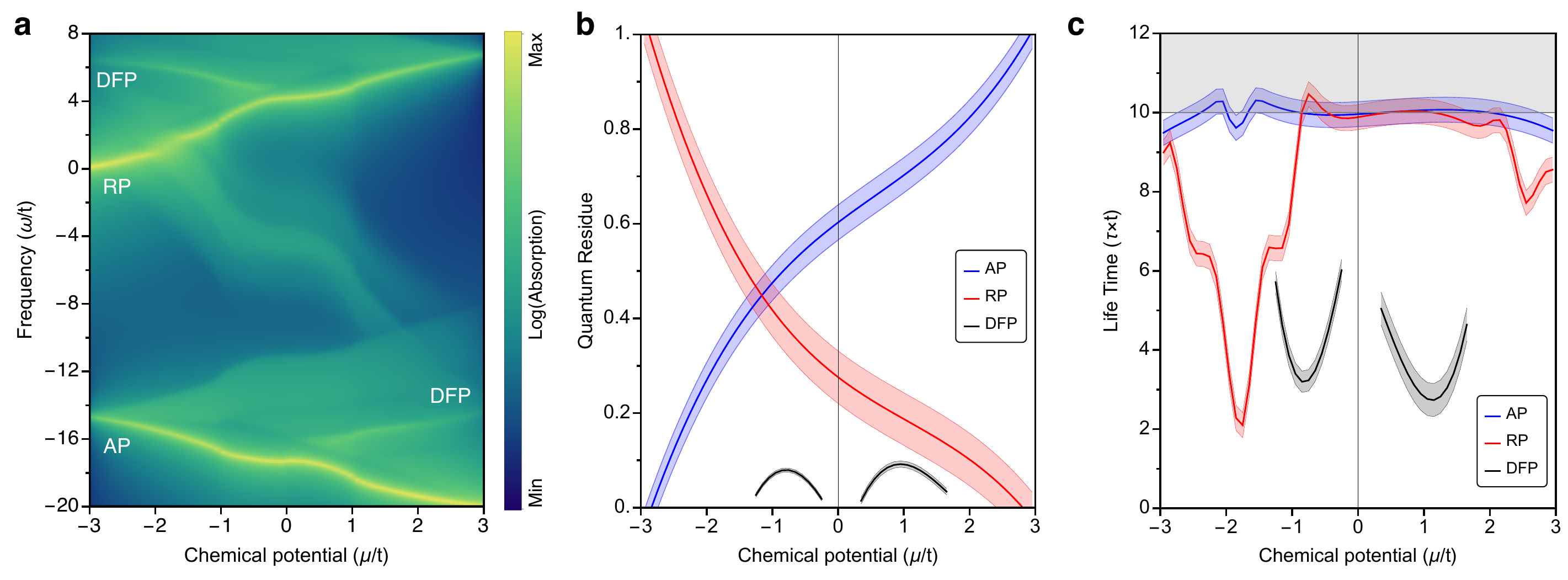}
\caption{Properties of the three polaron branches extracted from the Pad\'e approximant of the impurity spectral function $A(\mathbf{Q},E)$. All energies and chemical potentials are expressed in units of the bath fermion hopping parameter $t$. {\bf a} Logarithm of the impurity spectral function at $t_I=t$ and $U/t=-20$. {\bf b} Quasiparticle residues. The curves are obtained using a basis spline interpolation, with the bands indicating the error variance. {\bf c} Polaron lifetimes in units of $1/t$, interpolated similarly to {\bf b}. The lifetimes are capped at $10$, corresponding to the inverse of the artificial broadening $\varepsilon=0.1 \,t$ used in the numerical calculations. The gray shaded region indicates the domain where the lifetime exceeds this numerical bound.\label{fig:QR}}
\end{figure}

The variational framework allows us to investigate the spectral properties of the polaron branches in detail. Specifically, we focus on the polaron model for the graphene band structure, where a mass-balanced impurity (i.e., $t_I=t$) interacts with the fermionic bath at an interaction strength of $U/t=-20$. In \cref{fig:QR}{\bf a}, we plot the logarithm of the impurity spectral function, revealing distinct spectral features for the attractive, repulsive, and Dirac-Fermi polaron (DFP) branches. By analyzing the pole structure of the Green's function, we extract the quasiparticle residue $Z$ and lifetime $\tau$, see also the discussion below. The results are shown in \cref{fig:QR}{\bf b} and {\bf c}, respectively.

Crucially, the DFP exhibits a finite residue and a significant lifetime over a wide range of chemical potentials. These non-zero weights and long lifetimes confirm the robustness of the DFP, establishing it as a well-defined, coherent quasiparticle rather than a broad continuum feature.
The lifetimes in \cref{fig:QR}{\bf c} saturate at the inverse of the artificial broadening $\tau \sim 1/\varepsilon = 10/t$ (shaded region); this saturation indicates that the intrinsic lifetime of the quasiparticles exceeds the resolution limit of our numerical approach, further attesting to their stability.

\newpage
\section{Polaron self-energy}

\begin{figure}[ht!]
\includegraphics[width = \linewidth]{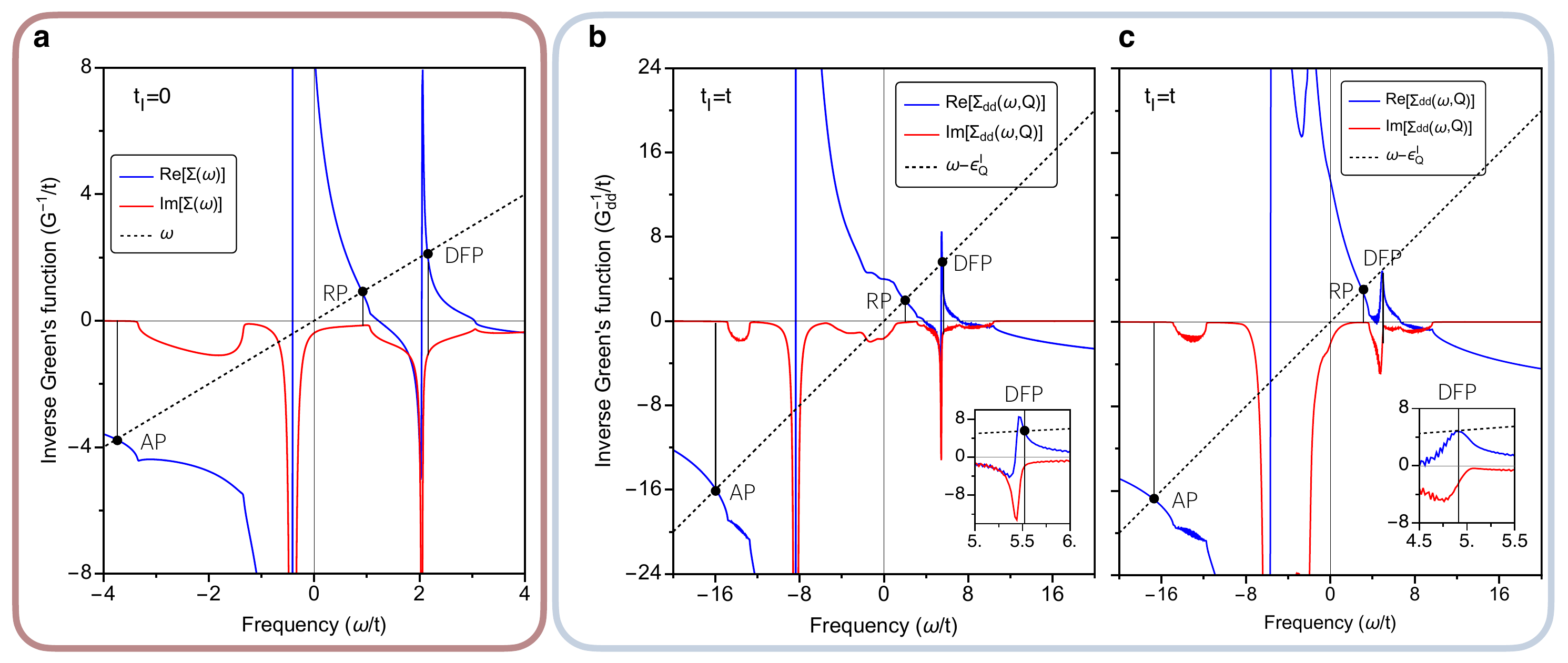}
\caption{Self-energy calculated from the variational ansatz at total momentum ${\bf Q}={\bf 0}$. {\bf a} Static impurity self-energy  $\boldsymbol{\Sigma}(\omega)$ evaluated at chemical potential $\mu=-\,t$ and artificial broadening $\varepsilon=0.01\,t$. Green dots mark the solutions to the quasiparticle dispersion equation, $\omega=\text{Re}[\boldsymbol{\Sigma}(\omega)]$, corresponding sequentially from left to right to the attractive polaron (AP), repulsive polaron (RP), and Dirac-Fermi polaron (DFP). {\bf b}, {\bf c} Lower-band ($d$) impurity self-energy $\boldsymbol{\Sigma}_{dd}(\omega)$ with broadening $\varepsilon=0.02\,t$, evaluated at a bath-fermion chemical potential of {\bf b} $\mu=-1.35\,t$ and {\bf c} $\mu=-0.85\,t$. In {\bf b}, intersections with the line $\omega -\epsilon_{\bf Q}^{I,d}$ indicate the emergence of the AP, RP, and DFP. In {\bf c}, specifically for the DFP branches, the real part $\text{Re}[\boldsymbol{\Sigma}_{dd}(\omega)]$ approaches the line $\omega-\epsilon_{\bf Q}^{I,d}$ before intersecting, indicating a precursor of the Dirac-Fermi polaron.\label{fig:SlfEng}}
\end{figure}

To confirm the DFP formation, we plot the lower-band impurity self-energy $\boldsymbol{\Sigma}_{dd}({\bf Q},\omega)$ at ${\bf Q}=0$ in \cref{fig:SlfEng}. While intrinsically a $2 \times 2$ matrix in the band basis, $\boldsymbol{\Sigma}$ is strictly diagonal at ${\bf Q}=0$ by spatial inversion symmetry. From \cref{eq:slfeng}, the self-energy takes the form $\boldsymbol{\Sigma}_{\lambda_I', \lambda_I} \propto (\varphi^{\lambda_I'}_{{\bf Q},s'})^{\dagger} M_{s',s} \varphi^{\lambda_I}_{{\bf Q},s}$, where $M$ represents the effective self-energy in the sublattice basis. Since the interaction preserves inversion symmetry, which exchanges the $a$ and $b$ sublattices and maps ${\bf Q} \to -{\bf Q}$, the matrix at the symmetric point ${\bf Q}=0$ must satisfy $M = \sigma_x M \sigma_x$. This restricts its form to $M = a I + b \sigma_x$. Projecting this restricted $M$ onto the ${\bf Q}=0$ Bloch wavefunctions  eliminates all off-diagonal mixing terms. Consequently, $\boldsymbol{\Sigma}({\bf Q}=0, \omega)$ is entirely diagonal, a result we have further verified numerically.

In \cref{fig:SlfEng}, we present the self-energy for both static [\cref{fig:SlfEng}{\bf a}] and mobile [\cref{fig:SlfEng}{\bf b}, {\bf c}] impurities.
For the static case, the variational ansatz explicitly captures the emergence of attractive, repulsive, and Dirac-Fermi polarons (DFP). While Anderson's orthogonality catastrophe typically dictates that polaron quasiparticle peaks vanish, \cref{fig:SlfEng}{\bf b} demonstrates that these peaks survive for the mobile impurity. Notably, much like the repulsive polaron~\cite{Scazza:2022bez}, the DFP can manifest as a well-defined quasiparticle embedded within the many-body scattering continuum, particularly when the interaction strength $U$ is sufficiently large and the bath chemical potential is tuned near the van Hove singularities. Finally, \cref{fig:SlfEng}{\bf c} illustrates a critical regime where $\text{Re}[\boldsymbol{\Sigma}_{dd}({\bf Q}, \omega)]$ closely approaches $\omega-\epsilon^{I,d}_{\bf Q}$, indicating a precursor of the DFP.

The stability and spectral prominence of these polaron features can be quantitatively understood through the analytic structure of the self-energy. Specifically, the emergence of a quasiparticle occurs at the polaron energy $\omega_*$, which is determined by the root of the Dyson equation,
\begin{align}
    \omega_*-\epsilon_{\bf Q}^{I,d}-\text{Re}[\boldsymbol{\Sigma}_{dd}(\omega_*,{\bf Q})]=0
\end{align}
At this resonance, the quasiparticle weight, or quantum residue, is given by $Z = 1 / (1 - \partial_{\omega_*} \text{Re}[\boldsymbol{\Sigma}_{dd}(\omega_*,{\bf Q})])$, while its corresponding lifetime is defined as $\tau = 1 / (Z \,\text{Im}[\boldsymbol{\Sigma}_{dd}(\omega_*,{\bf Q})])$. For the narrow resonances observed around the DFP, the intense dressing manifests as a steeply descending real part of the self-energy, which consequently yields a small quantum residue $Z$. Counterintuitively, this small residue plays a pivotal role in stabilizing the excitation, as it inversely prolongs the quasiparticle lifetime $\tau$. Furthermore, precisely at the resonance condition for both the quasiparticle and the broader resonance features, the imaginary part $\text{Im}[\boldsymbol{\Sigma}_{dd}]$ exhibits a pronounced drop from its scattering continuum peak. This simultaneous suppression of the decay rate minimizes spectral broadening, thereby leaving room for a robust, well-defined quasiparticle peak to form.

\section{$T$-matrix interpretation}

The formation of DFP quasiparticles can be understood at the single-particle level through the two-body $T$-matrix, which describes the scattering of an impurity with a fermion,
\begin{align}
    T^{-1}(\omega) = \frac{1}{U} - \sum_{{\bf k},\lambda} \frac{|\varphi^\lambda_{{\bf k},s}|^2}{\omega-\epsilon^\lambda_{\bf k}+i0^+} \,,
\end{align}
where $\lambda\in\{u,d\}$ is the band index and $\varphi^\lambda_{{\bf k},s}$ is the Bloch wave function component at the impurity site $s$.

\begin{figure}[ht!]
\includegraphics[width = \linewidth]{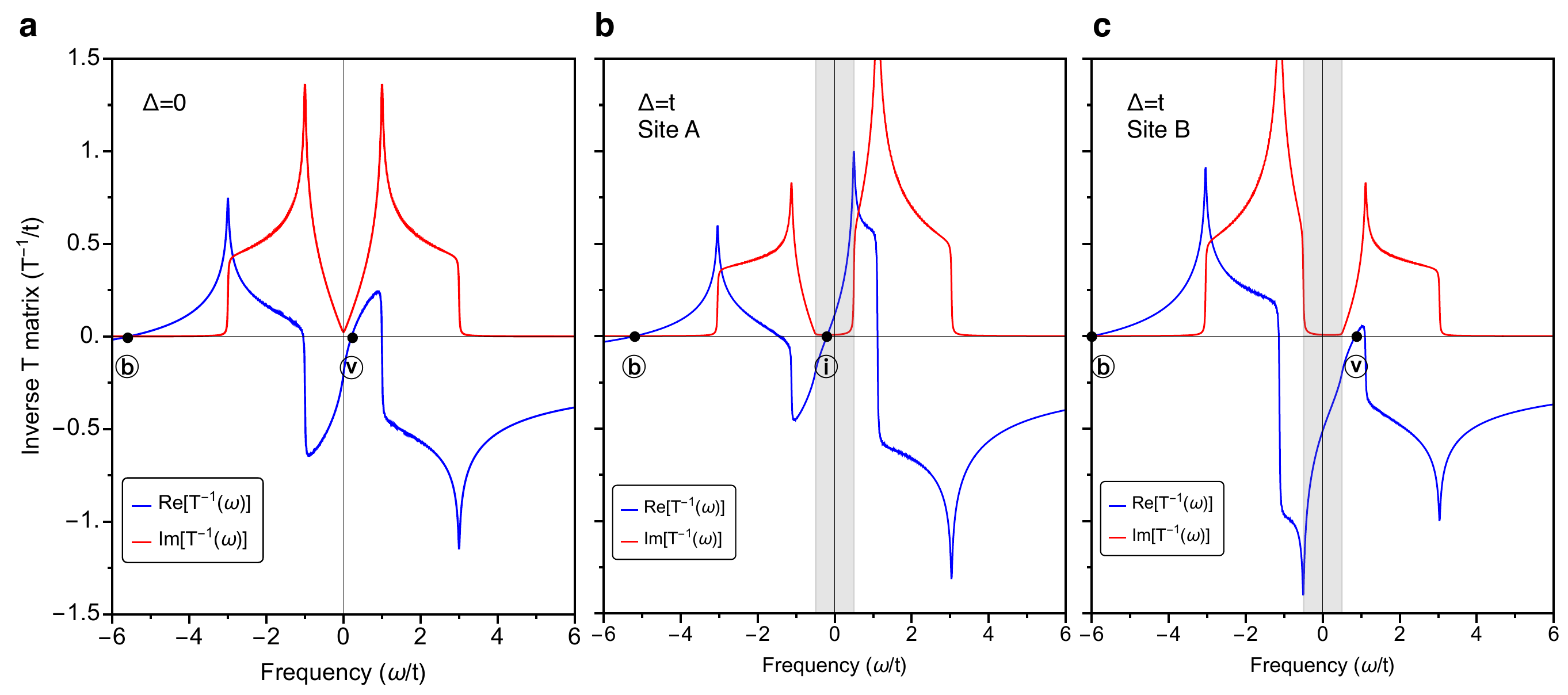}
\caption{Inverse two-body $T$-matrix for a static impurity with interaction strength $U=-5\,t$. {\bf a} Gapless single-layer graphene. The zero-crossings of the real part, $\text{Re}[T^{-1}]=0$, indicate the formation of (virtual) bound states. Here, \circleletter{b} denotes a true bound state, while \circleletter{v} marks a virtual bound state near the Dirac point. {\bf b}, {\bf c} Gapped honeycomb lattice ($\Delta=t$) with the impurity located at {\bf b} A-site and {\bf c} B-site. In {\bf b}, an in-gap bound state \circleletter{i} emerges alongside the bound state \circleletter{b}. In {\bf c}, \circleletter{b} and \circleletter{v} denote the true bound state and a virtual bound state in the upper band, respectively. \label{fig:Tinv}}
\end{figure}

As illustrated for a gapless graphene lattice in \cref{fig:Tinv}{\bf a}, the roots of the inverse $T$-matrix, $\text{Re}[T^{-1}]=0$, reveal the coexistence of two impurity-induced states: a strongly localized true bound state \circleletter{b} and a virtual bound state \circleletter{v} forming near the Dirac point. As the sublattice gap $\Delta$ increases, the zero-crossing associated with this virtual bound state eventually disappears. Crucially, these two-body bound states dictate the many-body scattering dynamics. They directly correspond to the two resonances that emerge in the impurity self-energy, which are clearly observable in the behavior of $\text{Re}[\boldsymbol{\Sigma}({\bf Q}, \omega)]$ (\cref{fig:SlfEng}), ultimately driving the formation of the distinct polaron branches.

\begin{figure}[ht!]
\includegraphics[width = 0.35\linewidth]{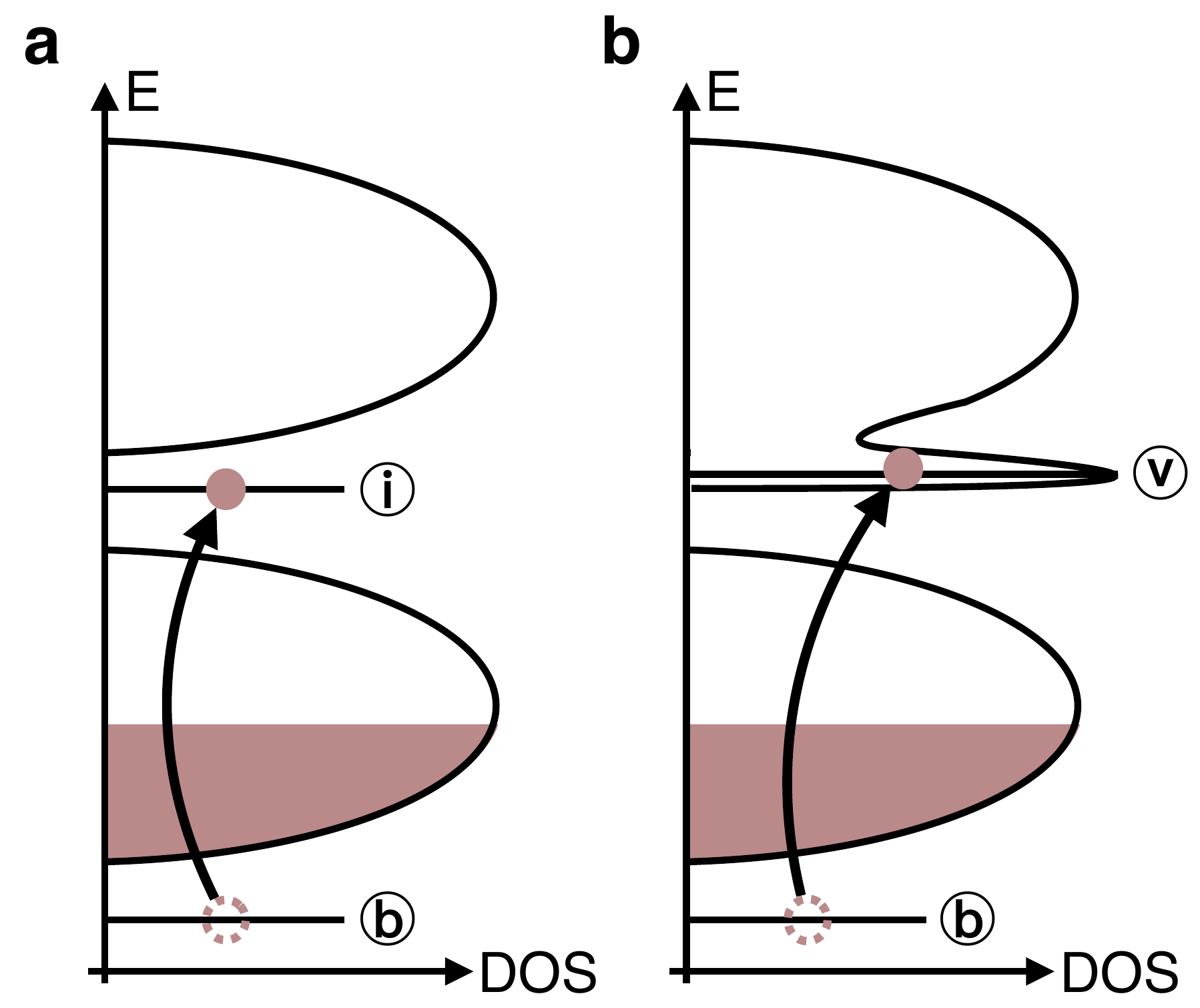}
\caption{Schematic illustration of the interacting density of states (DOS) governing the emergence of the Dirac-Fermi polaron (DFP). The interacting DOS accounts for the potential between the impurity and the fermion bath, directly revealing the formation of two-body bound states and scattering resonances. {\bf a} In a gapped band structure, interactions can induce a true in-gap bound state, denoted as \circleletter{i}, alongside the conventional lower-band bound state, denoted as \circleletter{b}. The DFP emerges as a distinct quasiparticle directly associated with the occupation of this in-gap state \circleletter{i}. {\bf b} Alternatively, the interaction may yield a virtual bound state, denoted as \circleletter{v}, which manifests as a pronounced resonance peak in the interacting DOS near the bottom of the upper band. In this case, the DFP corresponds to the excitation of the bound state \circleletter{b} to the virtual bound state \circleletter{v}.  \label{fig:illuTmat}}
\end{figure}

Building on this static potential scattering framework, the physical nature of the DFP can be intuitively deduced from the interacting density of states, as illustrated in \cref{fig:illuTmat}. As established in \cref{fig:Tinv}, the static scattering potential induces either a true in-gap bound state or a virtual bound state within the continuum, depending on the underlying band structure. In this context, the DFP arises when the fermion occupying the conventional bound state \circleletter{b} is excited to the in-gap state \circleletter{i} or the virtual bound state \circleletter{v}, see \cref{fig:illuTmat}.

\newpage

Crucially, this physical picture remains robust even when the impurity becomes mobile. For a mobile impurity, the two-body scattering $T$-matrix evaluated at zero total momentum (${\bf Q} = 0$) mathematically resembles the static scattering scenario. Because momentum conservation dictates that the impurity and the bath fermion must carry equal and opposite momenta, $-{\bf k}$ and ${\bf k}$, the total kinetic energy of the scattering pair is simply the sum of their individual single-particle dispersions. This combined energy landscape acts as an effective band structure that preserves the essential features of the original lattice, including the gapless Dirac cone. Consequently, the zero-momentum $T$-matrix for a mobile impurity maps directly onto the static impurity problem with a merely adjusted band structure, confirming that the emergence of the DFP is universally governed by the aforementioned in-gap and virtual bound state mechanisms.

\bibliographystyle{apsrev4-1}